\documentclass[11pt]{article}
\pdfoutput=1

\usepackage{jheppub} 

\usepackage{amsmath,amssymb,epsfig,amsfonts}
\usepackage{graphicx,subfigure}
\usepackage{epsf}
\usepackage{makeidx}
\usepackage[all]{xy}
\usepackage{slashed}
\usepackage{verbatim} %for comments
\usepackage{float}
\restylefloat{table}
\usepackage[all]{xy}
\usepackage{pdflscape}
\usepackage{tikz}
\usepackage{array}
\usepackage{youngtab}
\usepackage{multirow}
\usepackage{color}
\usepackage{ulem}
\usepackage{cleveref}
\usepackage{braket}
\usepackage{stmaryrd}
\usepackage{mathtools}
\usepackage{bm}

\usepackage{environ}
\usepackage{tikz-cd}

\usetikzlibrary{backgrounds}

\usepackage{tensor}

\newcommand{\bigslant}[2]{{\left.\raisebox{.2em}{$#1$}\right/\raisebox{-.2em}{$#2$}}}

\newcommand{\be}{\begin{equation}}
\newcommand{\ee}{\end{equation}}
\newcommand{\ba}{\begin{aligned}}
\newcommand{\ea}{\end{aligned}}

\newcommand{\bs}{\begin{split}}
\newcommand{\es}{\end{split}}

\def\tM3{\mathcal{M}_3}
\def\Sboundary{\Sigma}
\def\Graph{\Gamma}

\def\gau{\widetilde{G}}
\def\gut{G}

\def\Mgut{M_{\phi}}
\def\sph{\alpha}

\NewEnviron{eq}[1]{\begin{equation}\begin{split}\label{eq: #1} \BODY \end{split}\end{equation}}

\newcommand{\lb}{\left(}
\newcommand{\rb}{\right)}
\newcommand{\lbb}{\left[}
\newcommand{\rbb}{\right]}

\newcommand{\tn}[1]{\textnormal{#1}}

\def\hk{hyper-K\"ahler }
\newcommand{\s}{\mathbb{S}}

\newcommand\D{\mathcal{D}}

\newcommand{\Vol}{\mathrm{Vol}}

\newlength{\sswidth}

\newcommand{\tr}{\mathrm{Tr}}

\newcommand{\C}{\mathbb{C}}
\renewcommand{\P}{\mathbb{P}}

\newcommand{\bea}{\begin{eqnarray}}
\newcommand{\eea}{\end{eqnarray}}

\newcommand{\R}{{\mathbb R}}
\newcommand{\Z}{{\mathbb Z}}

\def\diag{\mathop{\mathrm{diag}}\nolimits}
\def\Im{\mathop{\mathrm{Im}}\nolimits}

\def\Re{\mathop{\mathrm{Re}}\nolimits}

\def\tr{\mathop{\mathrm{Tr}}\nolimits}

\def\unit{{1\kern-.65ex {\rm l}}}
\def\1{{1\kern-.65ex {\rm l}}}

\newcommand{\del}{\partial}

\def\CC{{\cal C}}
\def\CD{{\cal D}}

\def\CF{{\cal F}}

\def\CH{{\cal H}}

\def\CL{{\cal L}}
\def\CM{{\cal M}}
\def\CN{{\cal N}}
\def\CO{{\cal O}}

\def\CR{{\cal R}}
\def\CS{{\cal S}}

\newcount\hour \newcount\minute
\hour=\time \divide \hour by 60
\minute=\time
\def\now{%
\ifnum \hour<13
  \ifnum \hour=0 \advance \hour by 12 \number\hour:\else \number\hour:\fi%
     \ifnum \minute<10 0\fi%
     \number\minute%
\ A.M.%
\else \advance \hour by -12 \number\hour:%
  \ifnum \minute<10 0\fi%
  \number\minute%
  \ P.M.%
\fi%
}

\makeatother

\begin{document}

% format
\baselineskip=18pt  % a la harvmac
\numberwithin{equation}{section}  % make eq labels (sec.num)
%\allowdisplaybreaks  % allow page breaks in displayed eqs

%%%%%%%%%%%%%%%%%%%%%%%%%%%%%%%%%%%%%%%%%%%
%%%        TITLE BEGINS HERE
%%%%%%%%%%%%%%%%%%%%%%%%%%%%%%%%%%%%%%%%%%%

%% ========== title (note version) begins here ==========
%
%\vspace*{-1cm}
%\begin{center}
% {\Large\bf Title of the Document}
%\end{center}
%\vspace*{-.5cm}
%
%% ========== title (note version) ends here ==========

%% ========== title (paper version, a la harvmac) begins here ==========

\thispagestyle{empty}

% Report number
%\vspace*{-2cm} 
%\begin{flushright}
%{\tt }\\
%\end{flushright}

% title, authors, affiliation
\vspace*{1cm} 
\begin{center}
{\huge Higgs Bundles for  M-theory on $G_2$-Manifolds}\\

 \vspace*{2cm}
 
{Andreas P. Braun, Sebastjan Cizel, Max H\"ubner, and Sakura Sch\"afer-Nameki}\\

 \vspace*{1cm} 

{\it Mathematical Institute, University of Oxford, \\
Woodstock Road, Oxford, OX2 6GG, United Kingdom}\\
\bigskip

{\tt maths.ox.ac.uk:\,andreas.braun,\,cizels,\,hubner,\,sakura.schafer-nameki }
\vspace*{0.8cm}
\end{center}
\vspace*{.5cm}

% abstract
\noindent
M-theory compactified on $G_2$-holonomy manifolds results in 4d $\mathcal{N}=1$ supersymmetric gauge theories coupled to gravity. 
In this paper we focus on the gauge sector of such compactifications by studying the Higgs bundle obtained from a partially twisted 7d super Yang-Mills theory on a supersymmetric three-cycle $M_3$. We derive the BPS equations and find the massless spectrum for both
abelian and non-abelian gauge groups in 4d. The mathematical tool that allows us to  determine  the spectrum is Morse theory, and more generally Morse-Bott theory. 
The latter generalization allows us to make contact with twisted connected sum (TCS) $G_2$-manifolds, which form the largest class of examples of compact $G_2$-manifolds. M-theory on TCS $G_2$-manifolds is known to result in a non-chiral 4d spectrum. We determine the Higgs bundle for this class of $G_2$-manifolds and provide a prescription for how to engineer singular transitions to models that have chiral matter in 4d. 

\newpage
%%%%%%%%%%%%%%%%%%%%%%%%%%%%%%%%%%%%%%%%%%%
%%%           TITLE ENDS HERE
%%%%%%%%%%%%%%%%%%%%%%%%%%%%%%%%%%%%%%%%%%%

\tableofcontents
%%%%%%%%%%%%%%%%%%%%%%%%%%%%%%%%%%%%%%%%%%%
%%%     MAIN BODY OF THE TEXT BEGINS HERE
%%%%%%%%%%%%%%%%%%%%%%%%%%%%%%%%%%%%%%%%%%%

\section{Introduction}

Geometric engineering is at the heart of many applications of string theory, starting with model building for particle physics, the study of superconformal field theories, or sharpening the boundaries of the string theory landscape. For many of these applications F-theory has been the framework of choice in recent years, culminating for instance in the classification of 6d superconformal field theories \cite{Heckman:2013pva, Heckman:2015bfa}. Clearly a similarly robust and comprehensive analysis would be desirable for four-dimensional models with minimal supersymmetry. In F-theory, the defining data of 4d $\mathcal{N}=1$ theories is however not purely geometric, unlike the 6d setup, but includes a choice of $G_4$-flux. In particular, $G_4$-flux is crucial in order to get chiral 4d $\mathcal{N}=1$ theories from F-theory \cite{Donagi:2008ca,Braun:2011zm,Marsano:2011hv,Krause:2011xj,Grimm:2011fx}.

An alternative framework that yields minimal supersymmetry in 4d is obtained from M-theory on $G_2$-holonomy manifolds (for a review see \cite{Acharya:2004qe}). As is well known, the main challenge in this setup is the construction of {\it compact} $G_2$-holonomy manifolds\footnote{We shall often simply refer to these as $G_2$-manifolds.} with singularities, which yield both gauge (codimension 4) and chiral matter (codimension 7) degrees of freedom in 4d \cite{Acharya:1996ci,Acharya:2000gb,Atiyah:2000zz,Atiyah:2001qf,Witten:2001uq,Acharya:2001gy}. To this moment this is an open question. 

Until recently, the number of known compact $G_2$-manifolds was rather limited: the only concrete examples were the Joyce orbifolds given by resolutions of $T^7/\Gamma$ \cite{joyce1996I} and constructions based on orbifolds of a Calabi-Yau three-fold times $S^1$. Recently, a comparatively large class of examples (order millions) of compact $G_2$-manifolds was described in \cite{MR2024648,Corti:2012kd,MR3109862} as twisted connected sums (TCS). 

The physics of M-theory and string theory on TCS $G_2$-manifolds has been investigated in \cite{Halverson:2014tya, Halverson:2015vta, Braun:2016igl, Guio:2017zfn, Braun:2017ryx, Braun:2017uku, Braun:2017csz, Braun:2018fdp, Fiset:2018huv, Acharya:2018nbo}. One key property common to all TCS manifolds, which is a direct consequence of this particular construction, is that singularities will occur (if at all) in codimension 4 and 6, but not 7. From the standard geometric engineering dictionary for $G_2$-manifolds it then follows that the resulting models in 4d will not have chiral matter. An obvious question is then which type of deformations or singular transitions are required to remedy this limitation.  
The present paper will provide a setting which gives some answers to this question and explores how such transitions would be characterized in TCS geometries, by providing a local model description in terms of a Higgs bundle. To achieve this, we first refine and extend the local model framework of \cite{Pantev:2009de}, to incorporate the local limit of TCS $G_2$-manifolds, and then determine the type of deformations that are required.

The approach of using local Higgs bundle models and their spectral covers in F-theory \cite{Donagi:2008ca,Beasley:2008dc,Hayashi:2008ba,Beasley:2008kw,Hayashi:2009ge,Donagi:2009ra,Marsano:2009gv,Blumenhagen:2009yv,Marsano:2009wr,Hayashi:2010zp,Marsano:2011hv} has proven very successful in model building, and more importantly as a precursor to the study of compact F-theory models. The Higgs bundles characterize the gauge sector of a compactification in terms of the local geometry in the vicinity of an ADE-singularity. In F-theory, this was not only useful in making the geometric engineering dictionary precise, but also subsequently in the constructions of compact geometries with favorable 4d effective field theories. 
For $G_2$-manifolds the local structure close to conical singularities has been studied in \cite{Acharya:2001gy, Atiyah:2001qf}. Here we will take a slightly different approach, starting much like in F-theory with the statement that a local geometry that realizes in M-theory an ADE gauge group in 4d, will necessarily  have a description in terms of an ALE-fibration over a compact supersymmetric cycle. In F-theory the local Calabi-Yau is an ALE-fibration $\mathbb{C}^2/\Gamma_{ADE} \rightarrow M_4$, with $M_4$ a K\"ahler surface in the base of the elliptic Calabi-Yau four-fold. For a $G_2$-manifold, the local model is analogously given by a fibration 
\be\label{ALEFib}
\mathbb{C}^2/\Gamma_{ADE} \rightarrow M_3\,,
\ee
where $M_3$ is a supersymmetric three-cycle, i.e. an associative cycle, in the $G_2$-manifold.  This approach was advocated  in \cite{Pantev:2009de}, however much of the details of their paper remained somewhat ad hoc and more importantly, does not e.g. include the case of TCS $G_2$-manifolds as we shall explain. We will both provide an in depth exploration of the Higgs bundle associated to this model, that will in particular lend itself to generalizations. 

As M-theory compactified on an ALE space gives a 7d super Yang-Mills (SYM) theory with ADE gauge group, the effective 4d $\mathcal{N}=1$ theory of an ALE-fibration can be found by studying a topologically twisted 7d SYM-theory on a three-manifold $M_3$. The BPS equations then determine the field configurations along $M_3$ that ensure that $\mathcal{N}=1$ supersymmetry is preserved in 4d. They are given in terms of a Higgs bundle specified by an adjoint valued one-form Higgs field $\phi$ and a gauge connection $W$ along $M_3$. We will focus entirely on diagonalizable Higgs fields, which implies that the connection $W$ furthermore has to be flat. 
The diagonalizability implies that we can equivalently describe the Higgs bundle in terms of its eigenvalues or spectral data. 

The BPS equations imply that $d\phi =d^\dagger \phi=0$ and so $\phi =d f$, where $f$ is a harmonic function.
This in turn implies that $f$ is constant as long as we require $M_3$ to be compact and $f$ to be regular. To obtain interesting solutions we introduce `sources' or equivalently singularities for $f$, $\Delta f = \rho$. Alternatively, we may excise the loci where sources are located and study the corresponding $f$-twisted Laplace equation on the resulting three-manifold with boundary $\tM3$.

%%%%%%%%%%%%%%%%%%%%%%%%%%%%%%%%%%
%%%%%%%%%%%%%%%%%%%%%%%%%%%%%%%%%%

\begin{table}
	\begin{center}
		\begin{tabular}{|c|c|c|c|}\hline
		\multirow{2}{*}{ALE/$G_2$ Geometry} & \multirow{2}{*}{Higgs bundle} & \multirow{2}{*}{\shortstack{SQM/Morse-Bott\\ on $\tM3$}} &  \multirow{2}{*}{\shortstack{4d $\mathcal{N}=1$\\effective theory}}\cr
		& & & \cr \hline \hline
		\multirow{2}{*}{\shortstack{ALE-fibration for\\gauge group $G$}}   & \multirow{2}{*}{\shortstack{$\phi,W$ sections of \\ $ T^\ast(\tM3)\otimes \hbox{Ad} (G_\perp)$}} & \multirow{2}{*}{\shortstack{$\phi = df$\\Morse-Bott $f$}}  &  \multirow{2}{*}{\shortstack{Non-abelian\\gauge symmetry}} \cr
		 &   & & \cr \hline
		\multirow{2}{*}{\shortstack{Enhancement of \\ singularity}} & \multirow{2}{*}{$\phi =0$} & \multirow{2}{*}{Critical loci}  &  \multirow{2}{*}{Charged matter} \cr 
		& & & \cr\hline
		\multirow{2}{*}{\shortstack{Cycle\\decompactifies}} & \multirow{2}{*}{Singularities in $\phi$}	& \multirow{2}{*}{Location of charges $\rho$}  &  \multirow{2}{*}{--}\cr 
		 & & & \cr\hline
		\multirow{2}{*}{Associative $S^3$s} & \multirow{2}{*}{Gradient curves in $\mathcal{C}$} & \multirow{2}{*}{Gradient flow trees} & \multirow{2}{*}{Interactions} \cr 
		& & & \cr\hline 
		\multirow{2}{*}{Extra two-form}  & \multirow{2}{*}{Factored $\mathcal{C}$} & \multirow{2}{*}{Charges $\rho$ in Cartan} & \multirow{2}{*}{$U(1)$ symmetry}\cr
		& & & \cr\hline
		\end{tabular}
		\caption{Dictionary between  ALE-fibration/$G_2$ geometric data of the fibration over $\tM3$, Higgs bundle, Morse-Bott theory on $\mathcal{M}_3$ (alternatively SQM),  and the 4d $\mathcal{N}=1$ low energy effective theory. }
		\label{tab:SummaryDic}
	\end{center}	
\end{table}

%%%%%%%%%%%%%%%%%%%%%%%%%%%%%%%%%%
%%%%%%%%%%%%%%%%%%%%%%%%%%%%%%%%%%

In general the solutions to this zero-mode counting are difficult to determine. However, if we assume a fully factored spectral cover, the problem  of finding the zero mode spectrum and interactions maps to  Morse-Bott cohomology on $\tM3$. In this case the resulting 4d gauge theory has $U(1)$-gauge symmetries, which are determined by the number of factors of the spectral cover. 
The zero modes can then be computed in terms of relative cohomology of  $\tM3$ with respect to its boundary. 
The Higgs bundle spectral cover provides a construction of the three-cycles in the ALE-fibration, and determines the matter fields and couplings in 4d.  

If the spectral cover is not fully factored we only have a formal description of the spectrum in terms of the cohomology of the $f$-twisted complex. This may be somewhat surprising for the reader more familiar with the F-theory spectral cover description, see e.g. \cite{Donagi:2009ra, Marsano:2009gv, Marsano:2009wr, Marsano:2011hv}, where the factorization is usually achieved with some amount of tuning (in order to have extra $U(1)$ gauge symmetries) and the theories without this are usually simpler to describe. In the $G_2$-setting the factorization is paramount for even computing the 4d spectrum.

There is an alternative description -- again in the case of fully factored spectral covers -- in terms of supersymmetric quantum mechanics (SQM), whose grounds states can be computed using Morse (more generally Morse-Bott) theory as in Witten's classic work \cite{Witten:1982im}. 
This characterization in terms of SQM identifies matter and couplings in terms of gradient flow trees in $\tM3$. 
A summary of the dictionary between ALE-geometry, i.e. local $G_2$ geometry, the Higgs bundle, Morse-Bott theory on $\tM3$ or SQM, and the data of the 4d effective theory is provided in table \ref{tab:SummaryDic}.

This setup in particular allows modelling the local geometry of M-theory compactifications on TCS $G_2$-manifolds, which have an ALE-fibration over $S^3$ (e.g. as in \cite{Braun:2017uku}). Moreover it will allow us --  in the framework of the local Higgs bundle description of the geometry -- to make a concrete proposal for the types of deformations and transitions that the geometry needs to undergo.
Although we necessarily lose the concrete description of the geometry offered in terms of a twisted connected sum\footnote{Studying such transitions in a compact setting seems to go beyond the current tools available in geometry, as it can no longer be a TCS. However, see also the recent paper by Chen \cite{Chen}.} we may nevertheless track what happens to our model in the language of the local geometry, which may be useful in modifying/improving the TCS construction.

The plan of this paper is as follows: section \ref{sec:Cadburys} starts with a careful derivation of the partially topologically twisted 7d 
Super-Yang-Mills (SYM) theory on $M_3$, which in turn determines the BPS equations. We then discuss solutions in terms of Higgs bundles, which characterize the local geometry, and discuss the spectrum of gauge and bulk matter. The spectral cover approach for these Higgs bundles is set up in section \ref{sec:SpectralCovers} and localized matter is studied in Section \ref{sec:Haribo}. A description of abelian Higgs field backgrounds in terms of supersymmetric quantum mechanics and its connection with Morse and more generally Morse-Bott theory is given in section \ref{sec:BPSConfig}. This setup is then applied to the study of matter couplings in section \ref{sec:MatterInteractions}. Finally, in section \ref{sec:Snickers} we apply this framework to describe the local models for TCS $G_2$-manifolds and study the deformations of the associated local models. A summary of results useful for model building applications together with some concrete models is given in section \ref{sec:Manual}. The reader predominantly interested in the rules of how local $G_2$ Higgs bundles are constructed for practical purposes can focus almost entirely on this section.  We conclude with section \ref{sec:CD}, which furthermore contains a list of future research directions. A glossary of our notation and further technical details are relegated to the appendices.

%%%%%%%%%%%%%%%%%%%%%%%%%

\section{The Gauge Theory Sector of M-theory on $G_2$-manifolds}
\label{sec:Cadburys}

M-theory compactified  on a $G_2$-manifold gives rise to a 4d $\mathcal{N}=1$ supersymmetric gauge theory with matter fields, coupled to supergravity. 
In this paper we will be interested in the gauge theories obtained from such compactifications and therefore will decouple gravity. Gauge degrees of freedom in an M-theory compactification on a holonomy $G_2$-manifold are localized on codimension 4 subspaces, which are  associative (i.e. calibrated) three-cycles $M_3$.
Locally the geometry takes the form of an ALE-fibration over $M_3$ as in (\ref{ALEFib}).
A useful way to characterize the gauge sector is to think in terms of the 7d SYM-theory obtained from M-theory on the ALE-fiber: the gauge bosons in the Cartan subalgebra of the gauge group 
arise from dimensional reduction of the M-theory three-form $C_3$ on the two-forms in the ALE-fiber, and the remaining non-abelian gauge bosons arise from wrapped M2-branes.
 In an adiabatic approximation, where the ALE-fibration varies slowly over $M_3$, the 4d effective action can be obtained by dimensionally reducing this 7d SYM-theory on the three-cycle $M_3$, with a partial topological twist. 
 In this section we carry out this reduction and determine the spectrum of gauge and matter fields, which are determined by solutions of BPS equations along $M_3$ (see (\ref{BPS})). The solutions are given in terms of a  Higgs bundle over $M_3$, that is specified by a one-form Higgs field $\phi$ and an internal gauge field $W$.

\subsection{Partial Topological Twist and BPS Equations}
\label{sec:Twisted7d}

We start with 7d SYM with ADE gauge group $\gau$.
This theory can be obtained by dimensional reduction of the maximally supersymmetric 10d SYM on $\R^{1,6}\times T^3$. Our conventions are such that the 10d gauge multiplet consists of a (hermitian) gauge field $A$ and a Majorana-Weyl spinor $\lambda$ both valued in the adjoint representation of an ADE group $\gau$. The Lorentz group, and thereby the vector multiplet, reduce as follows
\be \ba \label{DimRed}
SO(1,9)_L &\quad \rightarrow \quad   SO(1,6)_L \times SO(3)_R  \\
A: \quad {\bf 10} & \quad \rightarrow \quad ({\bf 7},{ \bf 1}) \oplus ({\bf 1},{\bf 3})\equiv (A_M,\phi_i) \\
\lambda: \quad \bf{16} & \quad \rightarrow \quad (\bf{8},\bf{2})\equiv (\lambda_{\alpha\hat\alpha})\,,
\ea \ee
where the 10d vector indices are split into $M=0,\dots,6$ and $i=1,2,3$ and the spinor indices decompose as $\alpha=1,\dots,8$ and $\hat\alpha=1,2$, where we denote the R-symmetry indices with a hat. The 10d Majorana-condition descends to a 7d symplectic Majorana-condition\footnote{We refer  to appendix  \ref{app:Spinors} for our conventions with regards to spinors and supersymmetry.}.
%\be
%\lambda_{\alpha\hat\alpha}=(i\sigma^2)_{\hat\alpha}^{~\hat\beta}\lambda_{\alpha\hat\beta}^*\,,
%\ee
Denoting the gauge coupling in 7d by $g_7$ the action becomes 
\be\ba\label{Action7dSYM}
S_{7\tn{d}} =& \frac{1}{g_7^2}\int d^{7}x\lbb -\frac{1}{4} \tn{Tr}\lb F_{MN}F^{MN} \rb-\frac{1}{2} \tn{Tr} \lb D_{M}\phi_iD^{M}\phi^i \rb+ \frac{1}{4} \tn{Tr} \lb [\phi_i,\phi_j][\phi^i,\phi^j] \rb \rbb  \cr 
& + \frac{1}{g_7^2}\int d^7x \lbb +\frac{i}{2}\tn{Tr}\lb \bar{\lambda}^{\alpha\hat\alpha} (\hat{\gamma}^M)_{\alpha}^{\,~\beta}D_M\lambda_{\beta\hat\alpha} \rb -\frac{i}{2}\tn{Tr}\lb \bar{\lambda}^{\alpha\hat\alpha }(\sigma^i)_{\hat\alpha}^{\,~\hat\beta}[\phi_i,\lambda_{\alpha\hat\beta}] \rb \rbb \,,
\ea\ee
where $D_M=\del_M-i[A_M,\cdot\,]$ and $F$ is the field strength associated to $A$. The supersymmetry variations  are
\be\ba
\delta A_M &= +\frac{i}{2} \bar{\epsilon}^{\,\alpha\hat\alpha} (\hat\gamma_M)_\alpha^{\,~\beta} \lambda_{\beta\hat\alpha}  \cr
\delta \phi_i &= +\frac{1}{2} \bar{\epsilon}^{\,\alpha\hat\alpha} (\sigma_i)_{\hat\alpha}^{\,~\hat\beta} \lambda_{\alpha\hat\beta} \cr
\delta \lambda_{\alpha\hat\alpha} &= -\frac{1}{4} F_{MN}(\hat\gamma^{MN})_\alpha^{\,~\beta}\epsilon_{\beta\hat\alpha}  + \frac{i}{2}  D_M\phi_i (\hat\gamma^M)_\alpha^{\,~\beta}(\sigma^i)_{\hat\alpha}^{\,~\hat\beta}\epsilon_{\beta\hat\beta}  -\frac{1}{4}  [\phi_i,\phi_j]\epsilon^{ij}_{~~k}(\sigma^k)_{\hat\alpha}^{\,~\hat\beta}\epsilon_{\alpha\hat\beta}   \,,
\ea\ee 
where $\hat\gamma$ denotes the 7d gamma matrices.

This 7d SYM theory is the starting point for the analysis of gauge degrees of freedom in a local $G_2$-holonomy compactification of M-theory. For a given ALE-fiber, the singularity determines the 7d gauge group $\gau$. We now reduce this theory further on an associative three-cycle $M_3$. Since this will be generically curved with holonomy group $SO(3)$, the 4d theory will in turn only retain supersymmetry if we partially topologically twist the local Lorentz group $SO(3)_M$ with the R-symmetry $SU(2)_R$.
Upon compactification on $M_3$ the local Lorentz symmetry is broken to
\be\ba\label{UntwistedReduction}
SO(1,6)_L \times SU(2)_R &\quad \rightarrow \quad  SO(1,3)_L \times SO(3)_M \times SU(2)_R \cr 
A: \quad ({\bf 7},{\bf 1}) & \quad \rightarrow \quad ({\bf 2}, {\bf 2}; {\bf 1},{\bf 1}) \oplus ({\bf 1},{\bf 1};{\bf 3},{\bf 1}) \equiv ({A_\mu},W_{\underline{i}})\cr 
\phi: \quad ({\bf 1},{\bf 3}) & \quad \rightarrow \quad ({\bf 1}, {\bf 1};{\bf 1},{\bf 3})\equiv (\phi_{\hat\imath})\cr  
\epsilon, \ \lambda: \quad ({\bf 8},{\bf 2}) & \quad \rightarrow \quad ({\bf 2},{\bf 1};{\bf 2},{\bf 2})\oplus ({\bf 1},{\bf 2};{\bf 2},{\bf 2})\equiv (\lambda_{\alpha\underline{\alpha} \hat\alpha},\bar{\lambda}_{\dot{\alpha}\underline{\alpha} \hat\alpha})\,, 
\ea\ee
where the vector indices split as $\mu=0,\dots,3$ and $\underline{i},\hat\imath=1,2,3$ and the spinor indices are  $\alpha,\dot\alpha,\underline{\alpha},\hat\alpha=1,2$. The fermions $\lambda_{\alpha\underline{\alpha} \hat\alpha}$ satisfy a Majorana-condition as described in the appendix in \eqref{untwistedMajo}.

The supersymmetry parameter $\epsilon$  transforms non-trivially under $SO(3)_M$, so that to preserve supersymmetry in 4d, we redefine the local Lorentz group $SO(3)_M$ by an R-symmetry transformation\footnote{We will be slightly casual here and in the following, in that the twist involves the Lie algebras, rather than the groups.}
\be
SU(2)_{\rm twist}  = \diag (SO(3)_M, SU(2)_R)\,,
\ee
with generators $(\Sigma_M)_i +(\Sigma_R)_i$, where $\Sigma$ denotes the generators of the respective algebras. 
The field content and supersymmetry parameters transform under the partially twisted Lorentz group as follows
\begin{equation}\begin{split}\label{TwistedFieldcontent}
SO(1,3)_L \times SU(2)_M \times SU(2)_R  &\quad \rightarrow \quad   SO(1,3)_L \times SU(2)_\tn{twist}  \cr 
A:\quad ({\bf 2},{\bf 2};{\bf 1};{\bf 1}) & \quad \rightarrow \quad ({\bf 2},{\bf 2};{\bf 1}) \equiv (A_\mu)\cr 
W: \quad ({\bf 1},{\bf 1};{\bf 3},{\bf 1}) & \quad \rightarrow \quad ({\bf 1},{\bf 1};{\bf 3}) \equiv (W_i)\cr 
\phi:\quad ({\bf 1}, {\bf 1};{\bf 1},{\bf 3}) & \quad \rightarrow \quad ({\bf 1}, {\bf 1};{\bf 3})\equiv (\phi_i)\cr 
\epsilon,\ \lambda: \quad ({\bf 2},{\bf 1};{\bf 2},{\bf 2}) & \quad \rightarrow \quad ({\bf 2},{\bf 1};{\bf 1}) \oplus ({\bf 2},{\bf 1};{\bf 3})\equiv (\chi_\alpha,\psi_{i\alpha})\cr 
\bar\epsilon, \ \bar{\lambda}: \quad ({\bf 1},{\bf 2};{\bf 2},{\bf 2}) & \quad \rightarrow \quad ({\bf 1},{\bf 2};{\bf 1}) \oplus ({\bf 1},{\bf 2};{\bf 3})\equiv (\bar{\chi}^{\dot\alpha},\bar{\psi}_{i}^{~\dot\alpha})\,.
\end{split}\end{equation}
It follows that there are four real supercharges, as required for 4d $\mathcal{N}=1$ supersymmetry, 
\be
\epsilon_\alpha = ({\bf 2},{\bf 1}; {\bf 1}) \,,\qquad 
\bar\epsilon_{\dot\alpha} = ({\bf 1}, {\bf 2}; {\bf 1})\,.
\ee
That this supersymmetry is indeed preserved is shown in appendix \ref{app:SUSY}.
After the twist the fermions $\chi$ and $\psi$ transform as singlets and triplets of the twisted Lorentz group and are identified with 
0- and 1-forms on $M_3$ valued in $\tn{ad} (P)$, i.e. 
\be\label{FormIdentification}
\ba
\chi\in& \ \Omega^0(M_3,\tn{ad}(P))\otimes \C \cr 
 \psi \in& \ \Omega^1(M_3,\tn{ad}(P))\otimes \C\,,
\ea \ee
where $P$ is a $\gau$-principal bundle.
We denote the field strengths associated to the gauge fields $A_\mu$ and the Wilson lines $W_i$ by  $F_{\mu\nu}$ and $(F_W)_{ij}$, respectively, and their associated covariant derivatives as $D_\mu$ and $D_i$. 
The latter can be combined with the scalars $\phi_i$, which both transform as a ${\bf 3}$ of  $SU(2)_{\rm twist}$, into a complex 1-form
\be\label{CoVDiV}
\varphi_i=\phi_i+iW_i\,,\qquad \bar{\varphi}_i=\phi_i-iW_i\,, \qquad \D_i = \del_i +[\varphi_i,\cdot\,]\,, \qquad \bar{\D}_i = \del_i -[\bar{\varphi}_i,\cdot\,]\,.
\ee
Note that $\varphi,\bar{\varphi}$ and $\CD,\bar{\CD}$ are related by conjugation in the gauge algebra. We further introduce  
\be
(\CF_\varphi)_{ij} = [\CD_i,\CD_j]\,, \qquad (\CF_\varphi)_{\mu i}=[D_\mu,\CD_i]\,, \qquad (\CF_{\bar{\varphi}})_{\mu i}=[D_\mu,\bar{\CD}_i] \,,
\ee
and its conjugate $\CF_{\bar{\varphi}}=\CF_{\varphi}^\dagger$. We assume that the 4d  gauge fields $A_\mu$ are independent of the internal coordinates along $M_3$, so that the latter two expressions become standard space-time derivatives of the complex scalars $\varphi,\bar{\varphi}$
\be \label{KinTerm}
(\CF_\varphi)_{\mu i}=D_\mu\varphi_i\,, \qquad (\CF_{\bar{\varphi}})_{\mu i}=D_\mu\bar{\varphi}_i\,. 
\ee
Define the interaction term 
\be\ba
I_{\varphi,\bar{\varphi}} \equiv 2D_i\phi^i  =\del_i\varphi^i+\del_i\bar{\varphi}^i+[\varphi_i,\bar{\varphi}^i]\,.
\ea\ee
The partially twisted 7d SYM action is then
\be
\begin{alignedat}{3} \label{TwistedActionMain}
	S_{F,\tn{\,twist}} &=  \frac{1}{g_7^2}\int d^7x \bigg\{ &&\tn{\,Tr}\Big[  -i\bar{\chi} \bar{\sigma}^\mu D_\mu \chi -i  \bar{\psi}^i \bar{\sigma}^\mu D_\mu \psi_{i}+\sqrt{2}i\bar{\chi} \D^i \bar{\psi}_i- \sqrt{2}i\chi \bar{\D}^i \psi_{i} \\ &&&~~~~\:\, + \frac{i}{\sqrt{2}}\epsilon^{ijk} \bar{\psi}_{i} \bar{\D}_j \bar{\psi}_k -\frac{i}{\sqrt{2}}\epsilon^{ijk} \psi_i \D_j \psi_{k} \Big]  \bigg\}\\
	S_{B,\tn{\,twist}} &= \frac{1}{g_7^2}\int d^7x \bigg\{&&-\frac{1}{4}\tn{\,Tr}\Big[ F_{\mu\nu}F^{\mu\nu}\Big]- \tn{\,Tr}\Big[(\CF_{\varphi})_{\mu i} (\CF_{\bar{\varphi}})^{\mu i}\Big]-\tn{\,Tr}\Big[(\CF_{\varphi})_{ij}(\CF_{\bar{\varphi}})^{ij}\Big] \\  &&& -\frac{1}{2}\tn{\,Tr}\Big[  I_{\varphi,\bar{\varphi}}^2\Big]\bigg\}\,.
\end{alignedat}
\ee
%The fermions $\chi,\psi_i$ of this action are left handed. 
The supersymmetry variations for the bosonic fields are 
\be\begin{alignedat}{3}\label{bosonicVar}
	\delta A_\mu &=i\epsilon \sigma_\mu\bar{\chi}    \,, \qquad  &&\bar{\delta}  A_\mu &&= -i\bar{\epsilon}\bar{\sigma}_\mu \chi    \\
	\delta W_i &= -\frac{i}{\sqrt{2}}\epsilon \psi_i \,, \qquad &&\bar{\delta} W_i &&= \frac{i}{\sqrt{2}}\bar{\epsilon}\bar{\psi}_i   \\
	\delta \phi_i &=\frac{1}{\sqrt{2}} \epsilon\psi_i \,, \qquad &&\bar{\delta} \phi_i &&= \frac{1}{\sqrt{2}} \bar{\epsilon}\bar{\psi}_i   \\
	\delta \varphi_i &= \sqrt{2}\epsilon \psi_i \,, \qquad &&\bar{\delta} \varphi_i &&= 0 \cr 
	\delta \bar{\varphi}_i &= 0 \,, \qquad && \bar{\delta} \bar{\varphi}_i &&= \sqrt{2} \bar{\epsilon}\bar{\psi}_{i} \,,  
\end{alignedat}\ee
and for the fermionic ones we find 
\be
\begin{alignedat}{3}\label{fermionicVar}
	\delta \chi &= F_{\mu\nu}\sigma^{\mu\nu}\epsilon+iI_{\varphi,\bar{\varphi}\,}\epsilon \,, \qquad && \bar{\delta} \chi &&= 0 \cr  
	\delta \bar{\chi} &= 0 \,, \qquad && \bar{\delta} \bar{\chi} &&= F_{\mu\nu}\bar{\epsilon}\bar{\sigma}^{\mu\nu} -iI_{\varphi,\bar{\varphi}\,}\bar{\epsilon} \cr  
	\delta \psi^k  &=  i(\CF_{\bar{\varphi}})_{ij}\epsilon^{ijk} \epsilon  \,, \qquad && \bar{\delta} \psi^k &&= \sqrt{2}i (\CF_{\varphi})_{\mu}^{~\,k}\sigma^\mu\bar{\epsilon}  \cr 
	\delta \bar{\psi}^{k}  &= \sqrt{2}i (\CF_{\bar\varphi})_{\mu}^{~\,k} \bar{\sigma}^\mu\epsilon  \,, \qquad && \bar{\delta} \bar{\psi}^{k}  &&= -i(\CF_{\varphi})_{ij}\epsilon^{ijk} \bar{\epsilon}\,.
\end{alignedat}
\ee
To obtain a 4d supersymmetric theory upon twisted dimensional reduction, the field configuration along $M_3$ needs to preserve supersymmetry. We further require the background to enjoy 4d Poincar\'e-invariance and therefore require it to be independent of the coordinates along $\mathbb{R}^{1,3}$
\be
(\CF_{\varphi})_{\mu i}=0 \,, \qquad  (\CF_{\bar{\varphi}})_{\mu i}=0\,.
\ee
The BPS equations are then obtained by setting $\braket{\delta\lambda}=0$ and are 
\be\label{CBPS}
I_{\varphi,\bar{\varphi}}= \del_i\varphi^i+\del_i\bar{\varphi}^i+[\varphi_i,\bar{\varphi}^i]=0\,, \qquad (\CF_\varphi)_{ij}=0\,, \qquad (\CF_{\bar{\varphi}})_{ij}=0\,,
\ee
where the first equation is obtained by setting the real and imaginary parts of $\delta \chi$ to zero separately. 4d  Poincar\'e invariance requires $\langle F_{\mu\nu} \rangle=0$.
Rewriting \eqref{CBPS} with respect to the notation in (\ref{TwistedFieldcontent}) the BPS equations become the F- and D-term equations \be\label{BPS}
\boxed{\ba
\quad 0&= F_W-i[\phi,\phi]\cr 
 0& = D_W\phi\cr 
0&= D_W^\dagger \phi\,.\quad 
\ea}\ee
Background values for the Higgs field $\phi$ and gauge field $W$ along $M_3$ that solve these equations will determine the effective field theory in 4d\footnote{Note that we have chosen Hermitian representatives for the gauge algebra. Transitioning to anti-Hermitian representatives we recover the results of \cite{Pantev:2009de}.}.
In components the BPS equations are 
\be\ba
0&= \del_i W_j-\del_j W_i+i[W_i,W_j]-i[\phi_i,\phi_j]\cr 
0&=\del_i\phi_j+i[W_i,\phi_j]- \del_j\phi_i-i[W_j,\phi_i]\cr  0&=g^{ij}\lb\del_i\phi_j+i[W_i,\phi_j]\rb\,.
\ea\ee

Depending on the topology of $M_3$ there are various solutions to these equations. 
The simplest  set of solutions are obtained for commuting Higgs fields 
\be
[\phi, \phi]=0\,, \qquad F_W=0 \,.
\ee  
We will generally assume this to be the case. The remaining equations are $D_W\phi = D_W^\dagger \phi =0$. 
If $M_3$ is a compact three-manifold without boundaries and $\phi$ is regular, there are two cases to consider: 
\be
\ba
\pi_1 (M_3) = 0 \quad  &\Rightarrow \quad W=0\,, \quad  d\phi = d^\dagger \phi =0 \quad \Rightarrow \quad \phi=0 \cr 
\pi_1 (M_3) \not=0 \quad &\Rightarrow \quad D_W\phi = D_W^\dagger \phi =0  \,.
\ea\ee
In the first case $\phi$ has to be a harmonic 1-form and thus must be trivial, in the second case it can be non-trivial. 

We will be interested in simply-connected three-manifolds (i.e. a homology three-spheres) in the following. To nevertheless have non-trivial solutions we relax the assumption  that $\phi$ is regular, which can be achieved by including sources into the D-term equations. Writing $\phi= df$, the function $f$ is then required to satisfy Poisson's equation 
\be\label{Fish}
\phi= df\,,\qquad \Delta f = \rho\,,
\ee
where $\rho$ models the sources supported on a closed subset $\Graph$.
This maps the solution of the BPS equations to an electrostatics problem with the identification
\be
\ba
f &=\hbox{electrostatic potential}\cr 
\rho &= \hbox{charge density, supported on  }\Graph \subset M_3 \,.
\ea
\ee
%%%%%%%%%%%%%%%%%%%
\begin{figure}
  \centering
  \includegraphics[width=7cm]{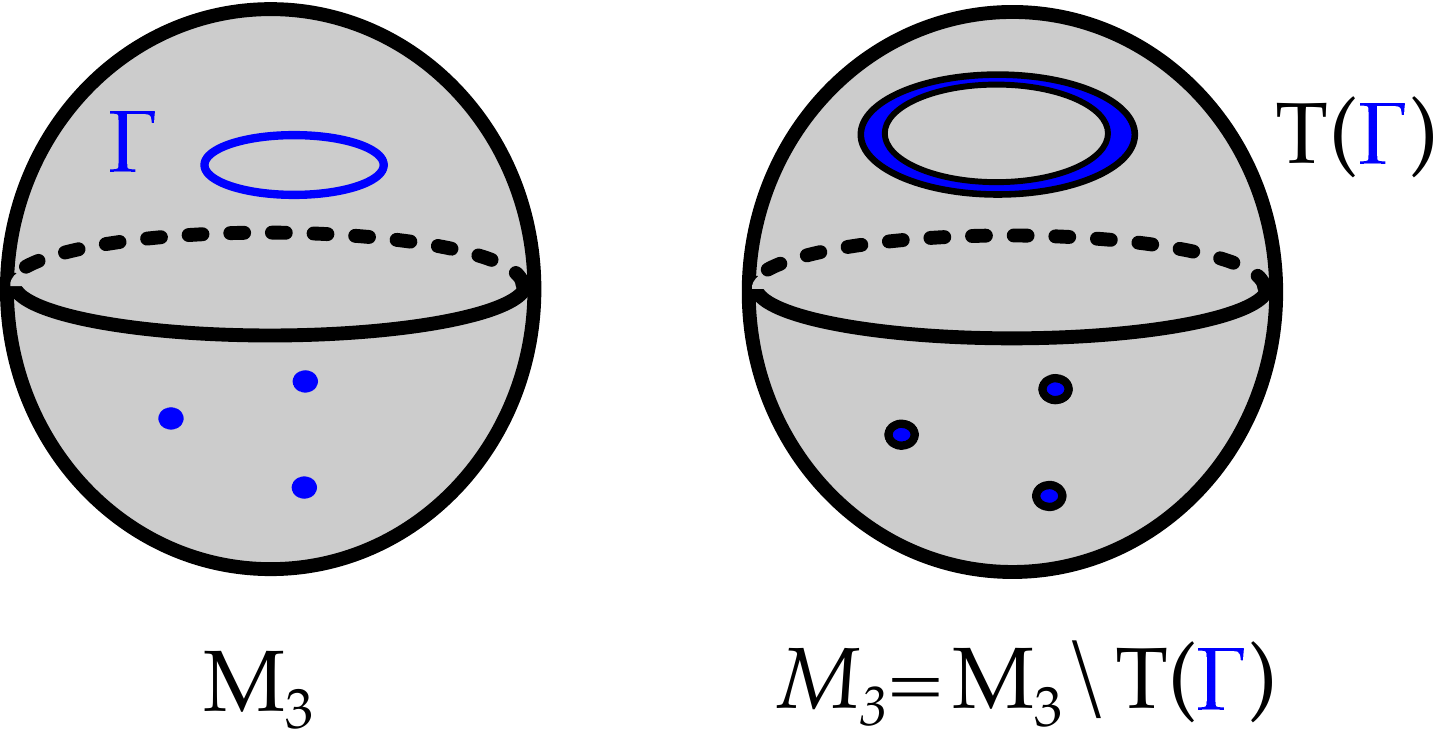}
    \caption{On the left hand side the three-cycle $M_3$ is shown, with the charge distribution $\rho$ that is located along $\Gamma$. On the right hand side, a  tubular neighborhood $T(\Gamma)$ is excised and the resulting manifold is $\tM3$.}
 \label{fig:M3M3T}
\end{figure}
%%%%%%%%%%%%%%%%%%

Alternatively this system can be described by excising a tubular neighborhood $T(\Graph)$ of the charge support $\Graph$, and studying the problem of finding solutions on $\tM3= M_3 \setminus T(\Graph)$ -- see figure \ref{fig:M3M3T}.  In this case 
$\phi$ needs to be regular, with suitable boundary conditions along $\partial\tM3$.
In summary we are going to consider the following setup
\be\label{Milka}
\boxed{\hbox{\ $\phi \hbox{ regular }, \quad \phi = df \,, \quad \Delta f= 0\,, \quad \partial \tM3  = T(\Graph) \not= \emptyset$\ }}
\ee
which will be used in section \ref{sec:Haribo} to determine physically interesting solutions to the BPS equations including localized matter. 
Localized matter is characterized by the vanishing of $\phi$. When $f$ is Morse {(i.e. it has no degenerate critical points)} these are isolated points and we will discuss this setup in section \ref{sec:Haribo}. By relaxing the constraint of $f$ only having isolated critical points this can be generalized to situations where $f$ is a Morse-Bott function and higher-dimensional matter loci can be included as well. We will discuss this in section \ref{subsec:Lindt} and apply it to TCS $G_2$-manifolds in section \ref{sec:Snickers}.

In the remainder of this section, we assume that $\phi$ is non-trivial and regular, but make no further assumptions on the details of the loci $\phi = 0$.

\subsection{Higgs Bundles}
\label{sec:HiggsALE}

Before studying the low energy effective theory, let us briefly recall the relation between the Higgs bundle and the local ALE-fibration. 
The BPS equations (in the absence of sources) (\ref{BPS})
 are in fact precisely the odd dimensional analogue of the Hitchin equations for the Higgs field $\phi$ giving rise to the data of a Higgs bundle. In the case $[\phi,\phi]= 0$ there is an elegant geometric description of the Higgs field $\phi$ in terms of an ALE-fibration over $M_3$, which we now summarize \cite{Pantev:2009de}. This construction is analogous to the one in F-theory, where the Higgs field specifies the unfolding (a complex structure deformation) of the ALE singularity and is closely connected to the compact Calabi-Yau underlying the F-theory compactification \cite{Beasley:2008dc, Donagi:2009ra, Marsano:2011hv}. Recently this was developed also for Spin(7) manifolds \cite{Heckman:2018mxl}.
 In our case the Higgs field describes the deformations of the full \hk structure of an ALE fiber.

Recall that $\phi$ is an adjoint valued 1-form $\Omega^1(\tM3)$ or a section of $T^*(\tM3)$, and we take it to be non-trivial along the commutant $G_\perp$ of the 4d gauge group $G$ in
\be
\gau \ \rightarrow \ G_{\perp} \times G \,.
\ee
The Higgs field is 
\be\label{phidef}
\phi \in  \Gamma(T^\ast (\tM3) \otimes \text{Ad} (G_\perp)) \,,
\ee
i.e. $\phi$ lives in a local geometry in the vicinity of $\tM3$ which is the total space of the cotangent bundle $T^*(\tM3)$. This is a local Calabi-Yau threefold. 
Since $[\phi,\phi]= 0$, we can diagonalize the Higgs field to obtain $n$ 1-forms $\phi_j$, where $n$ is the rank of the Lie algebra $\mathfrak{g}_\perp$ of $G_\perp$.
To locally recover the ALE-fibration over $\tM3$ associated to this Higgs field, we use the Kronheimer construction \cite{Kronheimer,Acharya:2001gy}.  Every ALE-space is of the form $\mathbb{C}^2/\Graph_{\rm{ADE}}$, where $\Gamma_{\rm{ADE}}$ is a finite subgroup of $SU(2)$, which are classified by the corresponding ADE Dynkin diagrams. The second homology of the resolution of singularities of $\mathbb{C}^2/\Gamma_{\rm{ADE}}$ is isomorphic to $\mathfrak{g}$ and we can think of the components $\phi_j$ as measuring the periods of the \hk structure forms. More explicitly, over a local patch of $M_3$ we can write the fibration as $\R^3\times \mathbb{C}^2/\Gamma_{\rm{ADE}}$. We chose a basis $\sigma_j$ of $H_2(\mathbb{C}^2/\Gamma_{\rm{ADE}},\Z)$ and fix a \hk triple $(\omega_I,\omega_J,\omega_K)$. The 1-form $\phi_j$ can be written as
\be
\label{eq:higgscomponents}
  \phi_j = \phi_{j,I} dx^1+\phi_{j,J} dx^2+\phi_{j,K} dx^3 \,,
\ee
where we identify 
\be\label{eq:higgscomponents_periods}
  \phi_{j,I} = \int_{\sigma_j} \omega_I,\qquad \phi_{j,J} = \int_{\sigma_j} \omega_J,\qquad\phi_{j,K} = \int_{\sigma_j} \omega_K \,.
\ee
This uniquely defines the \hk structure on each fiber. Observe that the Higgs field has an $SO(3)$ symmetry arising from the $SO(3)$ acting on $\omega_I$, $\omega_J$ and $\omega_K$. 

In geometric terms we can describe our situation as follows. For simplicity, assume that we have a $G_\perp= U(1)$-valued Higgs field $\phi$. We are considering a local model for a $G_2$-manifold with ADE-singularities located along an associative submanifold $\tM3$, which physically means that gauge degrees of freedom are localized along $\tM3$ and the gauge group is given by the ADE type of the singularity. Consider the gauge group $\gau$, which by turning on a non-trivial background vev for $\phi$ generically higgses to 
$\gau \ \rightarrow \  \gut \times U(1)$.
This means that the ALE fiber over a generic point of $\tM3$ will have the singularity corresponding to $\gut$ via the ADE correspondence and there will be a two-cycle in the $U(1)$ direction with non-zero volume, given by $\phi$. Over the points where $\phi=0$, the two-cycle collapses and the ALE singularity worsens; equivalently the gauge group enhances from $\gut$ to $\gau$. We will elaborate this point in section \ref{sec:MatterInteractions}.

We can in fact make the local geometry of the gauge enhancement fairly explicit. For the moment let us restrict our attention to the case where $\gau= SU(2)$ which corresponds to a $\C^2/\Z_2$ singularity over $\tM3$. Giving a non-trivial background vev for $\phi$ corresponds to deforming the generic fiber to a smooth Eguchi-Hanson space. More precisely, consider the generator $\sigma$ of $H_2(\widetilde{\C^2/\Z_2},\Z)$ of the resolved geometry. Recall that $\sigma$ is topologically a two-sphere. From  \eqref{eq:higgscomponents} and \eqref{eq:higgscomponents_periods} we see that  at a generic point $x\in\tM3$ (which for this purpose is approxated locally by $\mathbb{R}^3$) we have
\be
	\Vol(\sigma) = |\phi(x)|\,,
\ee 
by which we mean the volume of $\sigma$ in the Eguchi-Hanson space above $x$. Consider now a neighborhood of a non-degenerate zero of $\phi$, which we can assume to be at $0\in \R^3$. We can locally write $\phi = df$, where 
\be
	f(x_1,x_2,x_3) = f(0) +\frac{1}{2}\sum_{i=1}^3 \pm x_i^2 \,.
\ee
The signs depend on the eigenvalue of the Hessian at $0$. The Higgs field $\phi$ now has an isolated zero at the origin. The explicit local description of the ALE-fibration is given by
\be
	X = \left\{(z_1,z_2,z_3),\, (x_1,x_2,x_3)\,\left|\, z_1^2+z_2^2+z_3^2=\sum_{i=1}^3 x_i^2\right.\right\}\subset \C^3\times \R^3\,.
\ee
Viewing $X$ as a fibration over $\R^3$ all of the fibers are smooth apart from the fiber over $(0,0,0)$ i.e. the zero of $\phi$. Moreover, $X$ is a cone in $\C^3\times\R^3$ with the apex at the origin. The link of the cone can be found by intersecting $X$ with the unit sphere in $\C^3\times\R^3$ and is in fact $\P^3$ realized as the twistor bundle over $S^4$. The approximate $G_2$-metric on $X$ is given by 
\be
	\Phi = dx_1\wedge dx_2\wedge dx_3 + dx_1\wedge\omega_I+dx_2\wedge\omega_J+dx_3\wedge\omega_K + \phi\wedge\eta\,,
\ee
where $\eta$ is the 2-form dual to the two-cycle $\sigma$.

This can be generalized to arbitrary ALE-fibrations. The local geometry is of the form $\C^2/\Gamma_{G}\times \R^3$, with a $\C^2/\Gamma_{\widetilde{G}}$ fiber over the origin. We again work with $\widetilde{G} = SU(n+1)$. For the deformations of other ADE singularities see \cite{Katz:1996xe}. The topology in a neighborhood of an isolated zero is 
\be
	X = \left\{(z_1,z_2,z_3),\, (x_1,x_2,x_3)\,\left|\, z_1^2+z_2^2+z_3^n\left(z_3-\sum_{i=1}^3 x_i^2\right) = 0 \right.\right\}\subset \C^3\times \R^3\,.
\ee
This describes a family of $SU(n)$ singularities, with enhancement to $SU(n+1)$ at the origin (note that we again write $\phi=df$ as above). There are also explicit deformations for other ADE groups. Topologically $X$ is now a cone over the weighted projective space $\P^3_{n,n,1,1}$ with coordinates $(y_1,y_2,y_3,y_4)$ \cite{Acharya:2001gy}. In the link, there is a family of $SU(n)$ singularities along an $S^2$ given by $y_3=y_4=0$. In the ambient space, the location of the singularities is a cone $\R^+\times S^2=\R^3$, which is identified in our context with a local patch of the base $\R^3$ of $X$. As before, the apex of the cone is where the cycle $\sigma$ collapses to zero volume. 

This therefore establishes a key piece of the dictionary between properties of $\phi$ and the ambient $G_2$-geometry. The isolated zeroes of $\phi$ give rise to conical singularities of the ALE fibered $G_2$-manifold. As we show in section \ref{sec:BPSConfig}, this fits together nicely with the physics side as zeroes of $\phi$ which occur at codimension 7 are precisely the loci where chiral fermions are localized.

\subsection{Massless Spectrum}
\label{sec:ZeroModes}

Given a solution to the BPS equations (\ref{BPS}) with regular Higgs field we can ask what the spectrum of the 4d gauge theory is. The equations of motion of the fermions follow from (\ref{TwistedActionMain}) to be 
\be\ba \label{EOM}
0&=\bar{\sigma}^\mu D_\mu \chi -  \sqrt{2}\CD_i \bar{\psi}^{i}   \\ 
0&=\bar{\sigma}^\mu D_\mu \psi^i+\sqrt{2}\CD^i \bar{\chi}-\sqrt{2}\epsilon^{ijk} \bar{\CD}_j \bar{\psi}_k\,,
\ea\ee
which are equivalent to the decoupled equations
\be\ba\label{EOM2}
0&=D_\mu D^\mu\chi+2\CD_i\bar{\CD}^i\chi \cr 
0&=D_\mu D^\mu\psi_i+2[\CD_i,\bar{\CD}_j]\psi^j+2\bar{\CD}_j\CD^j\psi_i\,.
\ea\ee 
So far we have not imposed $[\phi,\phi]=0$. 
Define the twisted exterior derivative and Laplace operator
\be\label{DiffOps}
\CD=d+[\varphi\wedge \cdot\,]\,,\qquad \bar{\CD}=d-[\bar{\varphi}\wedge \cdot\,]\,,\qquad 
\Delta=\CD^\dagger\CD+\CD\CD^\dagger\,, \qquad \bar{\Delta}=\bar{\CD}^\dagger\bar{\CD}+\bar{\CD}\bar{\CD}^\dagger\,,
\ee
where the adjoint is taken with respect to the Hermitian inner product
\be \label{InnerProd}\ba
\braket{\,\cdot\,,\cdot\,}: \Omega^p(M_3,\tn{ad}(P))\times\Omega^p(M_3,\tn{ad}(P))\  &\rightarrow\  \C\cr 
(\alpha, \beta) \ &\rightarrow \ 
\braket{\alpha,\beta}=\int_{M_3}\tn{Tr}\lb \bar{\alpha} \wedge * \beta \rb\,.
\ea\ee
Acting on functions $g \in \Omega^0(M_3,\tn{ad}\,P)$ and written in coordinates, e.g. the operator $\bar{\Delta}$ becomes 
\be
\bar{\Delta}g=\bar{\CD}^\dagger \bar{\CD}g=\bar{\CD}^\dagger (\bar{\CD}_mg\,dx^m) = \CD^m\bar{\CD}_m g\,,
\ee
where we pick up a conjugation due to the inner product. We find that \eqref{EOM2} may be rewritten as
\be\label{EOMmain}\ba
0&=D_\mu D^\mu\chi+2\bar{\Delta} \chi \cr 
0&=D_\mu D^\mu\psi +2\Delta \psi\,,
\ea\ee
where by (\ref{FormIdentification}), $\chi$ and $\psi$ are 0- and 1-forms, respectively. Massless modes are therefore described by the kernels of the Laplacians $\Delta,\bar{\Delta}$ or equivalently by closed and co-closed forms with respect to the operators in \eqref{DiffOps}
\be\ba\label{CoClosed}
\bar{\CD} \chi&=0\,, \qquad \bar{\CD}^\dagger \chi=0\cr 
 \CD \psi&=0 \,, \qquad \CD^\dagger \psi=0\,.  
\ea\ee
By the BPS equations the co-boundary operators $\CD,\bar{\CD}$ and their adjoints close $\CD^2=\bar{\CD}^2=0$ and $(\CD^\dagger)^2=(\bar{\CD}^\dagger)^2=0$, and via the Hodge correspondence for elliptic complexes we can describe the zero-modes equivalently as cohomology groups. The non-vanishing background value of $\phi$ or $W$ oriented along a subgroup $\gut_\perp$ of $\gau$ breaks the gauge group to its commutant $\gut \subset\gau$. The adjoint fermions $\psi,\chi$ will decompose accordingly to give matter valued in irreducible representation. In this higgsed theory the fermions are sections of the associated gauge bundles, $E$. The action of $\CD$ restricts to each of these subbundles allowing us to make the identification
\be\ba\label{IniHom}
\chi_\alpha&\in H_{\bar{\CD}}^0(M_3,E)\,,\qquad \bar{\chi}_{\dot\alpha} \in H_{\CD}^0(M_3,E) \cr  \psi_\alpha&\in H_{\CD}^1(M_3,E)\,, \qquad \bar{\psi}_{\dot\alpha}\in H_{\bar{\CD}}^1(M_3,E)\,.
\ea\ee
We next rewrite these cohomology groups with respect to the same co-boundary operator by dualising  $H^0_{\bar{\CD}},H^1_{\bar{\CD}}$ with the Hodge star. Note that by \eqref{InnerProd} we have $\CD^\dagger = *\,\bar{\CD}\, *$ and $\bar{\CD}^\dagger = *\,\CD\, *$ so that taking $\chi_\alpha\in H^0_{\bar{\CD}}(M_3,E)$ for example we find that $*\chi_\alpha$ is annihilated by the operators $\CD,\CD^\dagger$
\be
\CD^\dagger(*\chi_\alpha)=*\bar{\CD} \chi_\alpha=0\,, \qquad \CD*\chi_\alpha=* \bar{\CD}^\dagger \chi=0\,.
\ee
This precisely states that $*\chi_\alpha\in H^3_{\CD}(M_3,E)$, i.e. we have mapped from $\bar{\CD}$-cohomology to $\CD$-cohomology using the Hodge star. The same observations hold true for $\bar{\psi}_\alpha$.
The Hodge star  relates
\be
H^0_{\bar{\CD}}(M_3,E) \cong H^3_{\CD}(M_3,E)\,,\qquad H^1_{\bar{\CD}}(M_3,E) \cong H^2_{\CD}(M_3,E)\,.
\ee
This allows us to make the following identifications
\be\label{Finally}
\boxed{
\ba
&\chi_\alpha\in H_{\CD}^3(M_3,E)\,,\qquad \bar{\chi}_{\dot\alpha}\in H_{\CD}^0(M_3,E)\,,\cr
&\psi_\alpha\in H_{\CD}^1(M_3,E)\,,\qquad \bar{\psi}_{\dot\alpha}\in H_{\CD}^2(M_3,E)\,,
\ea}\ee
where now all cohomologies are with respect to $\CD$ and forms of all degrees are employed. Note that the $\Z_2$-grading of the exterior algebra aligns with the 4d chirality of the fermionic zero-modes. The Hodge star depends on the metric of $M_3$ which itself is induced from the metric of $G_2$-holonomy of the ambient 7d manifold.

Since $M_3$ is associative and so calibrated with respect to $\Phi_{ijk}$ we equivalently could have used the $G_2$ 3-form $\Phi_{ijk}$ to dualize since it restricts to a volume form of $M_3$. Contracting elements of $H^0_{\bar{\CD}}$ and $H^1_{\bar{\CD}}$ with the 3-form $\Phi_{ijk}$ is then exactly the same as taking their Hodge dual.

\subsection{Bulk Matter}
\label{sec:Twix}

The first type of matter we will discuss arises from a background Higgs bundle, where  $\langle \phi \rangle =0$, which solves the BPS equations, but  $W\not=0$ with $F_W =0$. This will be referred to as bulk matter, as the modes will not be localized. We will see that for $\pi_1(\tM3)=0$
there is no chiral index for this matter type. It may be interesting to extend this to non-trivial $\pi_1$ setups, which we relegate to future work, and also has been discussed in earlier works from a different point of view (see e.g. \cite{Witten:2001bf}).

Turning on a flat gauge field along a subgroup $\gut_\perp \subset \gau$ the gauge group $\gau$ is Higgsed to the commutant $\gut$ of $\gut_\perp$ in $\gau$ and the adjoint representation of $\gau$ decomposes as
\be\label{Decomp}  
\ba
\gau \ & \rightarrow \ \gut \times \gut_\perp\cr 
\text{Ad} (\gau)  \ & \rightarrow \   \left(\text{Ad}(\gut)\otimes {\bf 1} \right) \oplus  
\left({\bf 1} \otimes \text{Ad} (\gut_\perp) \right) \oplus \bigoplus_n {\bf R}_n \otimes {\bf S}_{n} \,.
\ea\ee
For the fields of the theory this decomposition is lifted to the bundle level, where $\text{Ad}(P)$ decomposes into the vector bundles 
$\CR_n \otimes \CS_n$
in the representations ${\bf R}_{n}$ and ${\bf S}_{n}$ of $\gut$ and $\gut_\perp$, respectively. The chiral and conjugate-chiral zero modes transforming in ${\bf R}_n$ are then counted by the cohomology groups
\be
\tn{Chiral\,}:\qquad 
\ba\label{Groundstates}
(\chi_{{\bf R}_n})_\alpha &\in H^0_{\bar{\CD}}(M_3,\CS_n)  \cr 
(\psi_{{\bf R}_n})_{\alpha} &\in H^1_{\CD}(M_3,\CS_n)
\ea
\qquad 
\tn{Conjugate-chiral\,}: \qquad \ba 
(\bar{\chi}_{{\bf R}_n})_{\dot\alpha}&\in  H^0_{\CD}(M_3,\CS_n) \cr 
(\bar{\psi}_{{\bf R}_n})_{\dot\alpha}&\in  H^1_{\bar{\CD}}(M_3,\CS_n)\,.
\ea\ee
Their CPT-conjugate zero modes in $\overline{\bf{R}}_n$ are obtained by Hermitian conjugation in the gauge algebra or equivalently from \eqref{IniHom} with $E= \bar{\CS}$.
In order to rewrite these cohomology groups with respect to the same boundary operator $\CD$ we again dualise $H^0_{\bar{\CD}},H^1_{\bar{\CD}}$ using the Hodge star and obtain 
\be
\tn{Chiral\,}:\qquad 
\ba\label{Groundstates2}
(\chi_{{\bf R}_n})_\alpha &\in H^3_{\CD}(M_3,\CS_n)  \cr 
(\psi_{{\bf R}_n})_{\alpha} &\in H^1_{\CD}(M_3,\CS_n)
\ea
\qquad 
\tn{Conjugate-chiral\,}: \qquad \ba 
(\bar{\chi}_{{\bf R}_n})_{\dot\alpha}&\in  H^0_{\CD}(M_3,\CS_n) \cr 
(\bar{\psi}_{{\bf R}_n})_{\dot\alpha}&\in  H^2_{\CD}(M_3,\CS_n)\,.
\ea\ee
These cohomology groups completely determine the chiral and conjugate-chiral spectrum in 4d transforming in ${\bf R}_n$ of the remnant gauge symmetry $\gut$
\be\ba
\tn{Chiral fermion zero-modes}\, &:\qquad  H_{\CD}^3(M_3,\CS_n)\oplus H_{\CD}^1(M_3,\CS_n) \,,\\
\tn{Conjugate-chiral fermion zero-modes}\, &:\qquad  H_{\CD}^0(M_3,\CS_n)\oplus  H_{\CD}^2(M_3,\CS_n)\,.
\ea\ee
The chiral index of the representation ${\bf R}_n$ is
\be
\label{eq:ChiralInd}
\chi(M_3,{\bf{R}}_n,\CD)=\sum_{i=0}^3(-1)^i \dim_{\C} H^i_{\CD}(M_3,\CS_n)\,,
\ee
which is nothing other than the Euler characteristic of the $\CD$-complex.
In the case of trivial fundamental group $\pi_1(M_3)$, there is no flat bundle to break the gauge group, and  
$\dim H^i_{\CD}(M_3,\CS_n)= b^i(M_3,\CD)$ reduce to the Betti numbers of the de Rham complex on $M_3$.
The chiral index is then given by the usual Euler characteristic, which vanishes for odd dimensional closed manifolds
\be
\pi_1(M_3) =0:\qquad \chi(M_3,{\bf{R}}_n,\CD) =0  \,.
\ee
This concludes our discussion of `bulk' matter. In the following we will focus our attention on localized matter modes, which arise from 
non-trivial $\phi$ background values. Since these are best characterized in terms of spectral covers we will first develop the framework for that. 
We will briefly discuss interactions between bulk matter and localized matter fields later on.

\subsection{Defect Description of Matter}

Thus far our discussion was based on starting with a 7d SYM theory on $M_3$ with a gauge group $\gau$ which is generically broken to a smaller subgroup by the Higgs background. Over the zero locus of the Higgs field some of the gauge symmetry is restored. An equivalent description  starts with a bulk 7d SYM with gauge group $G\times U(1)$, and an unhiggsing to $\gau$  by inserting defects at points in $M_3$. We discuss this mechanism for rank 1 enhancements.

The starting point is the off-shell formulation of the 7d SYM given in appendix \ref{app:OffShell}, now with gauge group $\gut\times U(1)$. We take the Higgs field to have a background turned on along the abelian directions as $\phi=df\mathfrak{t}$ where $\mathfrak{t}$ denotes the $U(1)$ generator. The permitted configurations for $f$ are again determined by the BPS equations, including sources. Whenever the Higgs field vanishes the gauge symmetry could potentially enhance. For this we need to extend the field content by the required degrees of freedom.  We therefore add defects, coupled to the bulk fields as
\be\ba\label{eq:DefectActionFree}
I_{\tn{defect}}^{(\pm)}&=\int_{\R^{(1,3)}\times M_3} \lb  \Lambda^\dagger e^{\pm2qV} \Lambda \rb \Big|_{\theta\theta\bar\theta\bar\theta}\,,
\ea\ee
where we have introduce a chiral multiplet
\be\ba
\Lambda =\sigma +\sqrt{2} \theta \lambda + \theta \theta K + \cdots\,,
\ea\ee
valued in the representation ${\bf R}_q$ or $\overline{\bf R}_{-q}$ of $\gut\times U(1)$ depending on the choice of sign $+,-$ in \eqref{eq:DefectActionFree} respectively. Here $K$ is the auxiliary field of the chiral multiplet. The multiplet $\Lambda$ has a fixed profile along $M_3$ which is determined by the condition that $\Lambda$ descends to a massless 4d field upon reduction and will be subject of section \ref{sec:BPSConfig}. This also fixes the choice of sign in \eqref{eq:DefectActionFree}. These fields provide the additional degrees of freedom to enhance the gauge symmetry from $\gut\times U(1)\rightarrow \gau$ at the critical points of $f$. The action of \eqref{eq:DefectActionFree} is not the most general and can be extended by superpotential terms. The superpotential derived from M-theory in the picture where we Higgs to the gauge group $\gut \times U(1)$ is discussed in section \ref{sec:MatterInteractions}.

\section{Spectral Covers}
\label{sec:SpectralCovers}

\subsection{Spectral Cover for the Higgs Field}

For the case when a higher rank Higgs bundle is turned on but the Higgs field commutes, it is useful to describe the solution to the BPS equations in terms of the spectral data of the Higgs field. 
This framework is of course very familiar from F-theory spectral covers, see e.g. \cite{Donagi:2009ra, Marsano:2009gv, Marsano:2009wr, Marsano:2011hv}, and for the Lagrangians in Calabi-Yau threefolds and the associated $G_2$-manifolds with pointlike singularities was touched upon in \cite{Pantev:2009de}. Here we will prepare the setup to also account for more general Higgs field configurations, with the goal to apply it to the TCS-manifolds.

\begin{figure}
  \centering
  \includegraphics[width=8cm]{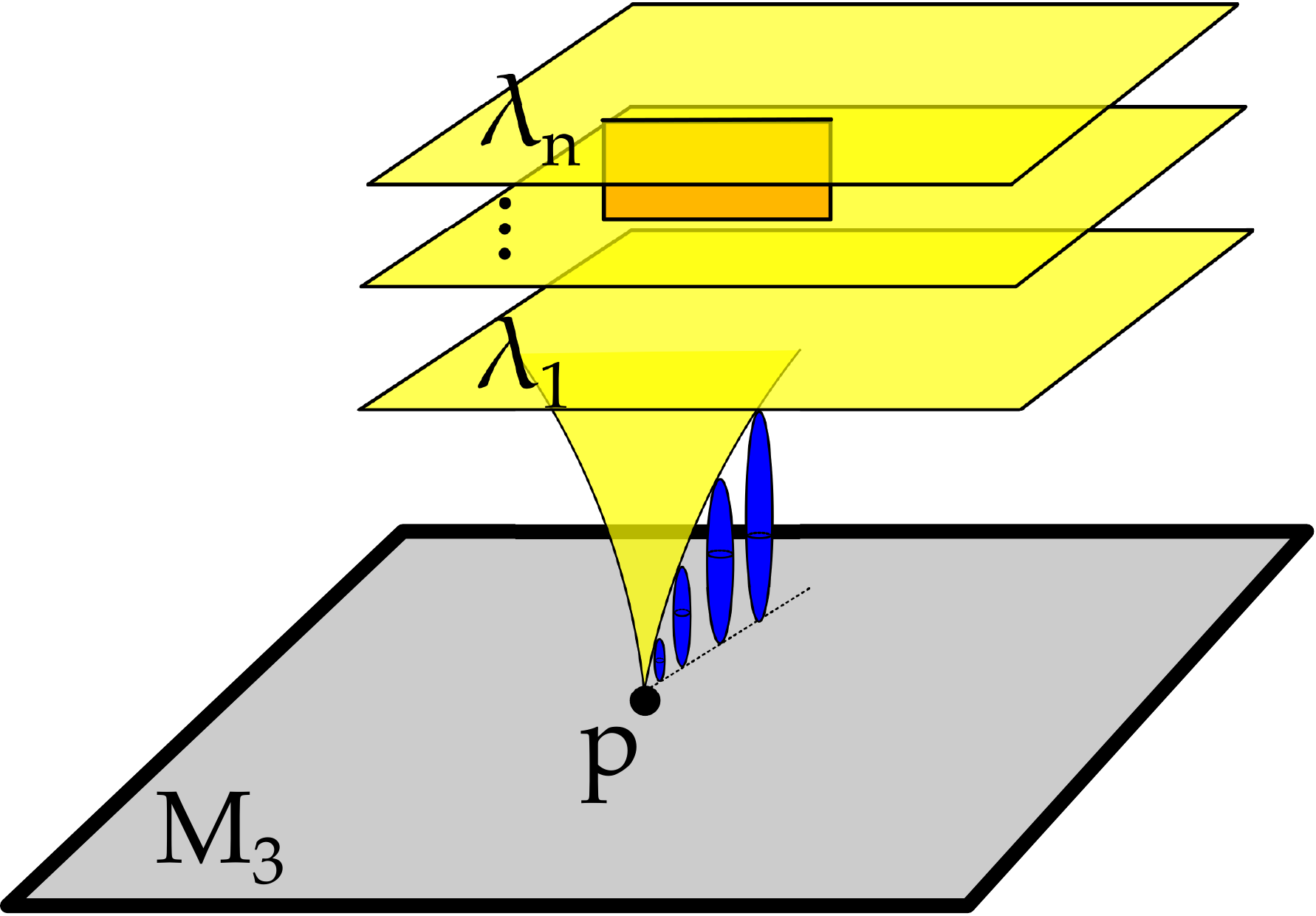}
    \caption{Higgs bundle spectral cover over $\tM3$. Each sheet is labeled by an eigenvalue $\lambda_i$ of $G_\perp$. Each $\lambda_i$ is a one-form and their vanishing thus implies that they intersect the `zero section', i.e. the locus of the ADE singularity of type $G$ present in every ALE fiber, over points $p_k$ on $M_3$. Those points $p_i$ are precisely the loci where matter is localized. In the generic case of non-factored spectral covers, the sheets are furthermore connected by branch-cuts (orange). As we will see later on: In case the point $p$ is connected by a flow line in $M_3$ to another critical point, there is a corresponding associative three-cycle which is built by fibering the collapsing $S^2$ (blue) over the flow line. The resulting contribution to the superpotential gives a mass term for the localized states.}
  \label{fig:Spec}
\end{figure}

Recall that $\phi$ is a section of $\Omega^1(\tM3)\otimes \hbox{Ad}(G_\perp)$. For concreteness let $G_\perp= SU(n)$.
For a commuting Higgs field we can choose to diagonalize it and study the resulting spectral cover
\be
\mathcal{C}:\qquad 0 =  \det (\phi - s ) = \sum_{i=0}^n b_{n-i}  s^i =  b_0   \prod_{i=1}^n (s - \lambda_i)  \,,
\ee
where  $b_i$ are symmetric polynomials in the eigenvalues of $\phi$ and for $SU(n)$, $b_1=0$. The eigenvalues $\lambda_i$ are one-forms which give rise to an $n$-sheeted cover of $M_3$ and 
\be 
b_i \in S^i (T^*\tM3) \,,
\ee
where $b_0$ is the zero-section. A cartoon of this is given in figure \ref{fig:Spec}.
Using the spectral cover the associated ALE-fibration is simply 
\be
y^2 = x^2 +  \mathcal{C}(s) \,.
\ee
Each $\lambda_i$ parametrizes the volumes of a corresponding two-sphere in the $G_\perp$-ALE-fiber. The gauge symmetry $\gau$ is generically higgsed to $G$, except at the loci 
\be
{b_{n}}= b_0 \prod_{i=1}^n \lambda_i =0 \,,
\ee
 when the gauge symmetry enhances. {Since $\lambda_i$ is a  one-form, this condition implies that this happens generically over points in $\tM3$, though we will encounter other situations as well.}
The relation between the eigenvalues $\lambda_i$ and coefficients in the ALE-fibration $b_i$ is not linear, and generically the sheets of the spectral cover will be connected by branch-cuts. This effect implies in particular that the $U(1)$-symmetries associated to the Higgs bundle are not actually present in the low energy effective theory. 

The classic example for spectral cover models starts with an $E_8 \rightarrow SU(5) \times SU(5)_\perp$. The spectral cover is five-sheeted and $\lambda_i=0$ characterizes the location of ${\bf 10}$ matter representations (we refer the reader to the F-theory literature where these models have been studied in depth \cite{Donagi:2009ra, Marsano:2009gv, Marsano:2009wr, Marsano:2011hv}). We will construct an example of this kind in detail at the end of this paper.

\subsection{$U(1)$ Symmetries}

Generically the sheets of the cover are connected by branch-cuts and therefore, although locally it may appear otherwise, the independent gauge group is $G$, the commutant of $G_\perp= SU(n)$ in $\gau$. If however the coefficients of the spectral cover are tuned such that it globally factors over $\tM3$ 
\be
\label{eq:SpectralDecomp}
\mathcal{C} (s)= \prod_{k=1}^{N+1 }\mathcal{C}^{(k)}(s) \,,
\ee
then this corresponds to $N$ independent $U(1)$ factors in the 4d effective theory \cite{Marsano:2009wr}.
The possibilities of factorization depend on the monodromy group that acts on the spectral cover, which for $SU(n)$ covers is $S_n$. If the group acts transitively on the $n$ sheets then there is no additional $U(1)$ symmetry. If it has $N+1$ orbits then there are $N$ globally defined two-forms, which define $U(1)$ symmetries. To see this, we consider the difference between the factored cover components  $\mathcal{C}^{(k)} - \mathcal{C}^{(l)}$. Fibered over $\tM3$, the associated two-cycles define a non-trivial five-cycle in both the local model and in the compact $G_2$ manifold $J$. The Poincar\'e dual two-form to this five-cycle then gives a $U(1)$ gauge boson in the Kaluza-Klein reduction of the three-form $C_3$ in compactification of M-theory. This can be also be seen concretely in the context of TCS $G_2$, see section \ref{sec:Snickers}. 

Perhaps the only important difference to the F-theory spectral cover story is that here it will be paramount that the spectral cover factors, in order to determine the spectrum of the 4d theory. Although the general twisted cohomology will continue to compute the zero mode spectrum, we do not have a computational tool to determine these cohomologies, unless the spectral cover factors completely. 

%%%%%%%%%%%%%%%%%%%%%%%%%%%%%%%%%%%%%%%%%%%%%%%%%%%%%%%%%%%%%%%%%%%%%%%%%%%%%%%%%%%%%%%%%%%%%%%%%%%%%%%%%%%%%%%%%%%%%%%%%%%%%%%%%%%%%%%%%%

\section{Localised Matter}
\label{sec:Haribo}

We will now study the more interesting and richer class of matter fields, localized on points or one-dimensional loci of $\tM3$. 
So far in section \ref{sec:Twix} we considered only flat gauge fields on along $\tM3$, which corresponds to bulk matter.
Turning on vevs for the Higgs fields $\phi$ will enlarge the possible matter structure and will allow us to engineer spectra with non-trivial chiral index.
The simplest case is an abelian Higgs field configuration
\be\label{Splitting}
\ba
\gau \ &\rightarrow \  \gut\times U(1)_{\perp} \cr 
{\rm Ad}(\gau) \ &\rightarrow\  {\rm Ad}(\gut) \oplus {\rm Ad} (U(1)) \oplus {\bf R}_{+q} \oplus \overline{\bf R}_{-q} \,,
\ea
\ee
where $G$ is the 4d gauge group and $U(1)_\perp$ the commutant, along which the Higgs fields is turned on. 
The expectation is that since $\phi$ is in the ${\bf 3}$ of $SO(3)_{\rm twist}$, the condition for local gauge enhancement to $G$ occurs at codimension 3 in the base $M_3$, i.e. codimension 7 in the $G_2$-manifold $J$. This is also suggested by the earlier spectral cover discussion. We will discuss this case of codimension 7 localized matter first. In less generic situations, such as the twisted connected sums, however, enhancement occurs at codimension 6 loci. 

The solution to the BPS equations (\ref{Milka}) on $M_3$ will be constructed by excising a tubular neighborhood $T(\Gamma)$ of a graph $\Gamma$, with boundary conditions, which we will discuss in detail. The central question is how the zero modes in ${\bf R}_{+q}$ and $\overline{\bf R}_{-q}$ are counted. 
In this section we provide the cohomological answer to this question, which applies to both codimension 6 and 7 gauge enhancements. In the next section we will provide specific solutions to the BPS-equations, to which the general analysis in this section can be applied, thereby computing the  zero mode spectrum.

\subsection{Zero Modes from Relative Cohomology}
\label{sec:MilkyWay}

We now turn on a background value for the Higgs field $\phi$, which to begin with is $U(1)$-valued.
As explained in section \ref{sec:Cadburys}, 
we now set out to solve the   D-term equation \eqref{BPS} for $\phi = df$ with sources, i.e. the Poisson equation
\be
\label{eq:Poisson}
\Delta f = \rho \,,
\ee
where the charge density $\rho$ satisfies charge conservation
\be
\int_{M_3}\rho = 0  \,.
\ee
 We take $\rho$ to be localized on links $\Graph_i$ in $M_3$ of definite signs of the charges, $\Graph_{\pm}$,
\be
\Graph =  \Graph_+\cup\Graph_-\,.
\ee
Both the Higgs field $\phi$ and $f$ diverge along $\Graph$. 
We again excise a tubular neighborhood as in section \ref{sec:Cadburys}. 
The boundary $\del\tM3$ splits into connected components $\Sboundary_i$, which correspond to the connected components of the underlying links $\Graph_i$ and correspondingly the boundary splits as 
\be\label{eq:SplitOfBoundary}
\del\tM3 =\bigcup_i \Sboundary_i = \Sboundary_+\cup\Sboundary_- \,.
\ee
 The normal derivatives of $f$, which are computed with respect to the outward pointing unit normal vector fields, have to be positive (resp. negative) restricted to $\Sboundary_+$ (resp. $\Sboundary_-$). 

The zero modes of the fields in the representation ${\bf R}_q$ and $\overline{\bf R}_{-q}$ in the presence of a background Higgs vev $\phi = df$ are obtained from the twisted Laplacian 
\be\label{eq:WittenLap}
\Delta_f = \CD\CD^\dagger+ \CD^\dagger\CD = \lb d^\dagger d+dd^\dagger \rb + q^2|df|^2 + q\sum_{i,j=1}^3 \lb H_f\rb_{ij} \lbb (a^i)^\dagger, a^j\rbb  \,,
\ee
where 
\be
\label{eq:Doperator}
\CD = d+ q df\wedge\,, \qquad \CD^\dagger = d^\dagger + q \, \iota_{\tn{grad} f}\,,
\ee
and  $H_f$ is the Hessian of $f$. Furthermore we defined the raising/lowering operators
\be
(a^i)^\dagger=dx^i\wedge \,,\qquad 
a^i=\iota_{\del_i}\,.
\ee
Note that $\CD^\dagger$ is not necessarily adjoint to $\CD$ on manifolds with boundary $\Sigma$ as
\be\ba
\braket{\CD\alpha,\beta}- \braket{\alpha,\CD^\dagger\beta}=\int_{\Sboundary} \bar{\alpha}\wedge*   \beta\,.
\ea\ee
Requiring appropriate boundary conditions fixes this problem. Consider a form $\alpha$ split into its tangential and normal component to the boundary 
\be
\alpha = \alpha_t+\alpha_n \,.
\ee
The tangential part $\alpha_t$ is defined as the pullback of $\alpha$ to the boundary and the normal part as $\alpha_n = \alpha-\alpha_t$. The boundary contribution is sensitive only to the tangential components i.e.
\be
\int_{\Sboundary} \bar\alpha \wedge\ast\beta = \int_{\Sboundary} \bar\alpha_t\wedge\ast\beta_n = \sum_i \int_{\Sboundary_i} \bar\alpha_t\wedge\ast\beta_n\,,
\ee
where we have used the fact that $(\ast \alpha)_t = \ast\alpha_n$. 
The two types of boundary conditions are
\be\ba
\label{eq:boundaryconditions}
\hbox{Dirichlet}:\qquad &   \ \  \left.\alpha_t\right|_{\Sboundary_i} = 0\cr 
\hbox{Neumann}:\qquad & \left. \ast\alpha_n\right|_{\Sboundary_i} = 0 \,,
\ea\ee
which can be imposed on every boundary component $\Sboundary_i$ independently. Choosing one of the above boundary conditions for every $\Sboundary_i$ amounts to restricting the domains of the operators $\CD$ and $\CD^\dagger$ to an appropriate subspace of forms. Within the restricted domains, the operators then become adjoints to each other. Moreover, by restricting the domain of $\Delta_f$ to make it self-adjoint, we can identify the zero-modes of $\Delta_f$ with the elements of cohomology groups $H^\ast_\CD(\tM3)$ using Hodge theory. We supply more details on the boundary conditions in appendix \ref{sec:BoundaryAndHodge}.

A natural choice is to split the boundary conditions according to whether the normal derivative $\del_n f$ is inward or outward pointing at a particular component of the boundary. This is the unique choice of boundary conditions, which precludes localization of zero-modes on the boundary $\Sboundary$. The relevance of this choice will become clear in section \ref{subsec:Lindt}.
Extending the set-up in \cite{chang1995} we first restrict the domains of $d$ and $d^\dagger$ to 
\be
\ba
\label{eq:ChoiceofBC}
D(d) & := \left\{\alpha\in\Omega^p(\tM3)\,\left|\, \alpha_t|_{\Sboundary_-}= 0 \ (\hbox{Dirichlet})\right. \right\} \cr 
D(d^\dagger) &:= \left\{\alpha\in\Omega^p(\tM3)\,\left|\,\ast\alpha_n|_{\Sboundary_+}= 0 \ (\hbox{Neumann})\right. \right\} \,,
\ea
\ee
i.e. we are imposing Neumann conditions on the positive boundary and Dirichlet conditions on the negative. Moreover, we define the domains of $\CD$ and $\CD^\dagger$ to be $D(\CD) = D(d)$ and $D(\CD^\dagger) = D(d^\dagger)$. 
The corresponding boundary conditions on the metric Laplace operator are given as
\be\label{DomainLaplace}
D^{\rm matter}(\Delta) = \left\{\alpha\in\Omega^p(\tM3)\,\left|\, \alpha_t|_{\Sboundary_-}= (d^\dagger\alpha)_t|_{\Sboundary_-}=0\quad \tn{and}\quad \ast\alpha_n|_{\Sboundary_+}= \ast(d\alpha)_n|_{\Sboundary_+}=0  \right.\right\}\, ,
\ee
where we set again $D^{\rm matter}(\Delta_f) = D^{\rm matter}(\Delta)$.  Note that the $d$-complex and $\CD$-complex are isomorphic (cf. appendix \ref{sec:BoundaryAndHodge}), so they have isomorphic cohomology groups. In this case, the $d$-complex is restricted to forms which vanish on $\Sboundary_-$. This computes the relative cohomology of the pair $(\tM3, \Sboundary_-)$ \cite{ElementsTopAlg} so we get
\be
\label{eq:SpectrumCohomology}
H^p_\CD(\tM3)= H^p (\tM3,\Sboundary_-)\,.
\ee
The sign of the $U(1)$-charge $q$ is important. Changing it amounts to changing the sign of $f$, which inverts the signs of normal derivatives and consequently exchanges the boundary conditions imposed on the positive and negative boundaries, and we obtain  the cohomology groups with respect to the positive boundary. In terms of the operators defined in section \ref{sec:ZeroModes} changing the sign of $q$ corresponds to computing the cohomology with respect to the operator $\bar\CD$, which is isomorphic to the $\CD$-cohomology but this time with respect to the conjugate representation $\overline{\bf{R}}$. 

Returning to the analysis of the spectrum above, we have seen that it is computed by the relative cohomology with respect to the negatively charged boundary components. Clearly, $H^0(\tM3,\Sboundary_-)$ vanishes since any constant function which vanishes on the boundary is identically zero. Moreover, by Lefschetz duality $H^3(\tM3,\Sboundary_-)$ also vanishes. Therefore, the discussion from section \ref{sec:ZeroModes} shows that the chiral fermions are counted by $H^1(\tM3,\Sboundary_-)$, while the conjugate-chiral fermions are counted by $H^2(\tM3,\Sboundary_-)$
\be
\ba
\hbox{chiral}:& \qquad H^1(\tM3,\Sboundary_-) \cr 
\hbox{conjugate-chiral}: &\qquad H^2(\tM3,\Sboundary_-) \,.
\ea
\ee
The net amount of chiral matter transforming in the representation $\mathbf{R}$ is therefore given by the relative Euler characteristic
\be
\chi(\tM3,\Sboundary_-) = b^2(\tM3,\Sboundary_-)-b^1(\tM3,\Sboundary_-),
\ee
where $b^1(\tM3,\Sboundary_-)$ and $b^2(\tM3,\Sboundary_-)$ are the dimensions of the respective cohomology groups. The Hodge star induces the isomorphism
$H^\ast(\tM3, \Sboundary_-) =  H^{3-\ast}(\tM3, \Sboundary_+)$, 
so that 
\be
\chi(\tM3,\Sboundary_-) %=b^2(\tM3,\Sboundary_-)-b^1(\tM3,\Sboundary_-) = b^1(\tM3,\Sboundary_+)-b^2(\tM3,\Sboundary_+) 
= - \chi(\tM3,\Sboundary_+)\,.
\ee

%Before we move to a generalization, we should briefly discuss what happens to the gauge fields for $G\times U(1)$. 
We have seen that for an $M_3$ without boundary there is a 4d vector multiplet in the spectrum. Once we introduce sources along $\Gamma$ and excise a tubular neighborhood around them, we need to check that the vector multiplets remain in the spectrum. 
Since these adjoint fields are uncharged under the $U(1)$, the associated forms cannot have any tangential boundary conditions, and we  
impose purely normal boundary conditions. In this case the domain of the relevant Laplace operator becomes
\be\label{DomainLaplace2}
D^{\rm gauge}(\Delta) := \left\{\alpha\in\Omega^p(\tM3)\,\left|\,\ast\alpha_n|_{\Sboundary}= \ast(d\alpha)_n|_{\Sboundary}=0  \right.\right\}.
\ee
The kernel is then isomorphic to the de Rham cohomology groups \cite{chang1995} and we obtain the required zero modes for the vector multiplets in 4d.

\subsection{Higher Rank Higgs bundles}
\label{sec:U1SymMod}

Next we generalize to higher rank Higgs bundles in $G_\perp$. 
We still assume that $[\phi,\phi]=0$. If we cannot diagonalize the Higgs bundle globally, (i.e. in the spectral cover language the spectral cover does not fully factor) then we still have a local description in terms of the Higgs field along the CSA, but not globally:
\begin{alignat}{4}\label{eq:GeneralBackgroundLocal}
\tn{globally on }M_3&:\qquad&&\phi=df\,, \qquad &&\Delta f=\rho\,, \qquad &&\int_{M_3}\rho=0 \,, \\
\tn{locally on }M_3&:\qquad  &&\phi=\mathfrak{t}^idf_i\,,\qquad&&\rho=\mathfrak{t}^i\rho_i \qquad &&\Delta f_i=\rho_i\,,
\end{alignat}
i.e. we can only diagonalise locally in a patch $U$ of $M_3$. Here $f_i,\rho_i:U\rightarrow\R$ are functions, $n=\tn{rk}\,G_\perp$ and $\mathfrak{t}^i$ the generators of the CSA. 
Locally this background breaks the gauge symmetry into 
\be\ba\label{eq:Multibreaking}
\gau \quad &\rightarrow \quad \gut \times U(1)^n\,, \\
\tn{Ad}\, \gau \quad &\rightarrow \quad \tn{Ad}\, \gut \oplus \tn{Ad} (U(1)^n)\oplus \bigoplus_{Q= (q_1,\dots, q_n)} \tn{\textbf{R}}_{Q  }\,,
\ea\ee 
where $Q=(q_1,\dots,q_n)$ denotes a vector of $U(1)$-charges. If the spectral cover has $N+1$ irreducible components (as in \eqref{eq:SpectralDecomp}), $N$ of these $n$ $U(1)$ factors descend to the gauge group of the 4d effective theory.
The operator $\CD$ defined in \eqref{DiffOps} acts on ${\bf R}_Q$ by
\be\ba\label{eq:CalDMulti}
\CD|_{{\bf R}_Q}&=\CD_Q=d+\lb q_1df_1+\cdots+q_ndf_n  \rb\wedge\,,\\
\CD|_{{\bf R}_Q}^\dagger &=\CD_Q^\dagger =d^\dagger+ \iota_{\tn{grad}(q_1f_1+\dots+q_nf_n)}\,.
\ea\ee
Let us introduce 
\be\label{eq:EffMorsefunction}
f_Q=q_1f_1+\cdots+q_nf_n\,.
\ee
The zero-modes are counted by \eqref{Finally} where $E=\tn{Ad}\,G_\perp$. If the spectral cover does not factor, i.e. the sheets mix under monodromy, the cohomologies of the operator $\CD$ cannot be rewritten in terms of e.g. de Rham cohomologies. For the case of rank 1 Higgs bundles the isomorphism given between the corresponding complexes was given by conjugation with $e^{qf}$ (this is explained in greater detail in appendix \ref{sec:BoundaryAndHodge}). This required a globally defined function $f$ whose role for fully reducible Higgs bundles is played by $f_Q$ as we will explain in the next section. This isomorphism cannot be adapted in a straightforward manner to general Higgs bundles.

Restricting $\CD$ to $\tn{Ad}\,\gut$ or  $\tn{Ad} (U(1)^n)$, it is reduced to the exterior derivative 
\be
\CD|_{\tn{Ad}\,\gut}=\CD|_{{U(1)}}=d\,,\qquad \CD|_{\tn{Ad}\,\gut}^\dagger=\CD|_{{U(1)}}^\dagger=d^\dagger\,.
\ee
Vector and chiral multiplets transforming in these representations are thus simply counted by the zeroth and first Betti numbers of $\tM3$, respectively.

%%%%%%%%%%%%%%%%%%%%%%%
\begin{table}
	\begin{center}
		\begin{tabular}{|l||*{5}{c|}}\hline
			&\makebox[3em]{$\tn{Ad}\,\gut$}&\makebox[3em]{$\tn{Ad}\,U(1)^n$}&\makebox[3em]{${\bf R}_Q$}
			&\makebox[3em]{$\overline{\bf R}_{-Q}$}\\\hline\hline
			Vector multiplets & $1$ & $1$ & $0$ & $0$ \\\hline
			Chiral multiplets &$b^1(\tM3)$ & $ b^1(\tM3)$ & $b^1(\tM3,{\Sboundary}^-_Q)$ & $b^2(\tM3,{\Sboundary}^-_Q)$ \\\hline
		\end{tabular}
		\caption{The 4d $\mathcal{N}=1$ matter content for a background given by a $U(1)^n$ valued Higgs bundle whose spectral cover is fully factored. Here ${\Sboundary}^-_Q$ denotes the negative boundary of $\tM3$ with respect to the function $f_Q$. Note $b^1(\tM3)=b^2(\tM3)$ and $b^1(\tM3,\Sboundary_Q^\mp)=b^2(\tM3,{\Sboundary}_{Q}^\pm)$.}
		\label{tab:MatterContent}
	\end{center}	
\end{table}
%%%%%%%%%%%%%%%%%%%%%%%

However, if the Higgs bundle diagonalizes globally, i.e. if we have rank $G_\perp$ many $U(1)$ symmetries, then a simple generalization of the rank one case applies. 
The zero modes are counted with respect to
\be
\CD=d+df_Q\wedge\,,
\ee
where $f_Q$ is globally well-defined and a function. 
 As a consequence the results of section \ref{sec:MilkyWay} carry over upon making the replacement $qf\rightarrow f_Q$.  $\tM3$ is obtained by excising the singularities of all the $f_i$ and the boundary decomposes again into positive and negative parts 
\be\label{eq:DecompBoundary}
\Sigma=\Sigma^+_Q\cup \Sigma^-_Q\,,
\ee
depending on whether $f_Q\rightarrow \pm\infty$ when approaching the excised charge. By \eqref{eq:EffMorsefunction} the charge vector can flip the sign of a boundary as seen by the individual functions $f_i$ used to define $f_Q$, i.e. for differently charged representation ${\bf R}_Q$ each zero mode counting requires an alternate decomposition of the boundary. We therefore find the fermionic zero-mode spectrum in the representation ${\bf R}_Q$ to be enumerated by the relative Betti numbers
\be\ba
b^1(\tM3,\Sigma_Q^-)&=\tn{\,chiral zero-modes in\,}{\bf R}_Q\,, \\
b^2(\tM3,\Sigma_Q^-)&=\tn{\,conjugate-chiral zero-modes in\,}{\bf R}_Q\,.
\ea\ee
This parallels the identification of cohomologies as in \eqref{eq:SpectrumCohomology}. Each of these fermionic zero-modes contributes to a chiral multiplet upon reduction to 4d by supersymmetry. The CPT conjugate of the fermionic zero-modes enumerated by $b^2(\tM3,\Sigma_Q^-)$ will be of positive chirality in 4d and contribute to a chiral multiplet valued in ${\overline{\bf R}_{-Q}}$. 

For the representations uncharged under any of the factors of $U(1)$ we have $\CD=d$ and their boundary conditions on $\tM3$ are chosen purely normal as in \eqref{DomainLaplace2}, and they are counted by de Rham cohomology. The complete  4d spectrum is summarized in table \ref{tab:MatterContent}.

%%%%%%%%%%%%%%%%%%%%%%%%%%%%%%%%%%%%%%%%%%%%%%

\begin{figure}
	\centering
	\includegraphics[width=12cm]{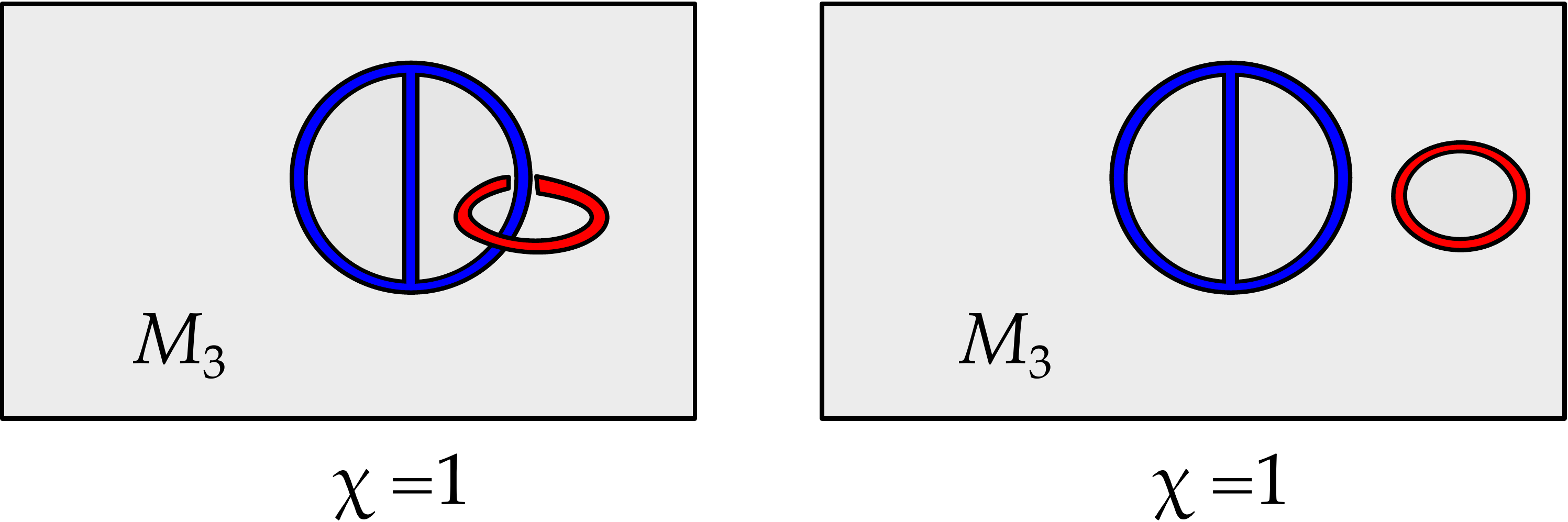}
	\caption{Examples of charged graphs in $\tM3$. Positive, negative charges are coloured red, blue respectively. Both charge distribution give rise to the same chiral index but a different number of zero-modes.}
	\label{fig:chrageexa2}
\end{figure}

%%%%%%%%%%%%%%%%%%%%%%%%%%%%%%%%%%%%%%%%%%%%%%

\subsection{Example 1: Wires in $S^3$}
\label{sec:Ex1}

We now turn to describing concrete charge configurations -- these configurations were studied in \cite{Pantev:2009de} and we revisit them here. 
Let $M_3=S^3$ and embed charges in $S^3$ which are localized on a graph $\Graph$. The positively and negatively charged components of the graph are  disjoint $\Graph=\Graph_+\,\cup\,\Graph_-\,$. We denote by $n_+,n_-$ the number of components and by $\ell_+,\ell_-$ the number of loops of $\Graph_+,\Graph_-$ respectively. The total charge on $\Graph_\pm$ is again constrained to vanish. Excising tubular neighbourhoods of $\Graph_\pm$ we obtain $\tM3$ with associated boundaries $\Sigma_\pm$. By \eqref{eq:SpectrumCohomology} the number of non perturbative chiral and conjugate-chiral zero-modes are then given by the Betti numbers $b^i(\tM3,\Sigma_-)$ for $i=1,2$ respectively. The top and bottom cohomologies vanish as discussed in section \ref{sec:MilkyWay}. The first and second cohomology are
\be\ba
\label{eq:chiralindexcharges}
b^1(\tM3,\Sigma_-) &= \ell_++n_-- r -1\cr 
b^2(\tM3,\Sigma_-)& = \ell_-+n_+- r -1\,,
\ea\ee
where $r$ counts the number of negative loops which are independent in homology when embedded in $M_3\setminus \Graph_+$. The chiral index is then computed to be
\be
\chi(\tM3,{\bf R}_{q})= (n_+ - \ell_+)  - (n_- - \ell_-)\,.
\ee 
It solely depends on the charge configuration $\Graph$ and is independent of the number $r$. A chiral spectrum is therefore easily generated. Multiple charged graphs will give rise to the same spectrum.
Another point to note here is that a non-trivial chiral index will only arise if for some sign of the charge, the number of loops and components is different, i.e. the charge distribution is not localized solely on a disjoint union of circles. This will later on give hints as to how to deform the Higgs bundles for TCS $G_2$-manifolds.

%%%%%%%%%%%%%

%%%%%%%%%%%%%

\section{BPS-Configurations, Super-QM and Morse-Bott Theory}
\label{sec:BPSConfig}

In the discussion above we were not interested in any particular features of the harmonic function $f$ on $\tM3$ and the computation of the spectrum is in fact valid for any such $f$. In this section we first specialize to the case where the Higgs field $\phi=df$ has isolated, non-degenerate zeros, which is the same as requiring $f$ to be a Morse function -- this is the case already studied in \cite{Pantev:2009de}. We show that massless chiral matter is localized at the zeros of the Higgs field. We then generalize this to the case where $f$ can have critical loci of dimension one, in which case it is Morse-Bott. The latter will be essential for the TCS geometries. 
The main tool here is reformulating the problem of finding the kernel of the Laplacian $\Delta_f$ in terms of a supersymmetric quantum mechanics and Morse theory. This approach is useful as it lends itself to the generalized  Morse-Bott setup that we are interested in. 

\subsection{Matter, Morse and (Super-Quantum-) Mechanics}
\label{sec:WKB}

Let us consider again the abelian case where $\phi=df$ with $f$ harmonic in the decomposition (\ref{Splitting}), which counts the  fermionic zero-modes transforming in the representation $\mathbf{R}_q$, that are in the kernel of 
\be\label{WittensLaplacian}
\Delta_f = \CD\CD^\dagger+\CD^\dagger\CD=\lb d^\dagger d+dd^\dagger \rb + q^2 |df|^2 + q\{d,\iota_{\tn{grad}\,f}\}+q \{d^\dagger, df\wedge \} \,.
\ee
The twisted Laplacian $\Delta_f$ can be interpreted as the Hamiltonian of a supersymmetric quantum mechanics (SQM) with the target space $\tM3$ where the supercharges are given by the operators $\CD$ and $\bar{\CD}$ \cite{Witten:1982im}.  In section \ref{sec:ZeroModes} we have shown that (due to the partial topological twist) the state space is identified with the space of differential forms on $\tM3$. However, since $\tM3$ is now a manifold with boundary, we have to restrict the state space to forms satisfying the boundary conditions given in \eqref{eq:ChoiceofBC}, which we denote by $\CH = \oplus_i\Omega^i_b(\tM3)$. The subscript $b$ indicates that the forms satisfy the boundary conditions. The function $f$ now plays the role of a superpotential and the kernel of $\Delta_f$ characterizing the true zero modes in the reduction to 4d is now enumerating the supersymmetric ground states of the SQM \cite{Witten:1982im}. In summary:
\begin{center}
	\begin{tabular}{|c|c|}
	\hline
		4d Effective Theory &   SQM  \cr \hline
		Matter fields & State Space \cr
		$\CD$, $\CD^\dagger$ & Supercharges \cr
		$\Delta_f$ & Hamiltonian \cr 
		Higgs field $\phi = df$ & $f=\tn{Superpotential}$ \cr 
		Matter zero modes & Ground states \cr \hline
	\end{tabular}
\end{center}

As in Witten's analysis, we can now use perturbation theory to compute the zero mode spectrum. To compute the perturbative kernel of $\Delta_f$, rescale $f\mapsto tf$. In terms of the electrostatics problem \eqref{eq:Poisson}, this amounts to rescaling the charges globally by a factor of $t$, which does not alter the overall ground state count. The term $q^2|df|^2$ in \eqref{WittensLaplacian} scales quadratically in $t$. Hence, for large $t$, the solutions of the equation $\Delta_{tf}\psi= 0$ are localized at the points where $df=0$ i.e. the zeros of the Higgs field $\phi$.

In this discussion we focus on harmonic functions $f$ which are Morse. The local physics will then be given by a supersymmetric harmonic oscillator. Before continuing with the computation we recall the definition of a Morse function.  
A smooth function $f: \, \tM3\rightarrow\R $ is called Morse if its set of critical points 
\be
	N=\{p\in \tM3: df(p)=0\}
\ee 
is discrete and all points $p\in N$ are non-degenerate. A critical point $p\in N$ is called non-degenerate if its Hessian 
$H_f(p)$ is non-degenerate as a bilinear map. In this case $p\in N$ is assigned a number $\mu(p)$ called the Morse index given by the number of negative eigenvalues of $H(p)$
\be
	p\in N:\qquad \mu(p)= \left| \{c = \hbox{eigenvalue of }H_f(p); \  c  <0 \} \right| \,.
\ee
In the case of manifolds with boundary, we further assume that there are no critical points of $f$ on $\del\tM3$. Note that this is true in our case, since the normal derivative of $f$ at the boundary is non-zero (see section \ref{sec:MilkyWay}). For more details on Morse theory we refer the reader to \cite{Hori:2000kt, milnormorse}.

We can choose a normal coordinate system in which $f$ and the metric $g$ on $\tM3$ take the form 
\be\ba\label{eq:expansions}
f(x)&=f(0)+\frac{1}{2}\sum_{i=1}^3 c_i(x^i)^2+ \CO ((x^i)^3)\,,\\
g_{ij}(x)&=\delta_{ij}+\CO ((x^i)^2)\,,
\ea\ee
where we assumed that $p= 0$ and 
$c_i$ are  the eigenvalues of the Hessian, which due to the harmonicity of $f$ sum to zero. This means that only points with Morse index 1 and 2 can occur. 
Expanded in these coordinates $\Delta_{t f}$ reduces to the Hamiltonian of a supersymmetric harmonic oscillator with 
\be\label{eq:LaplacianExpanded}
\Delta_{tf} = \sum_{i=1}^3 \lb -\frac{\del^2}{\del (x^i)^2} + q^2t^2c_i^2(x^i)^2 + qt c_i[dx^i, \iota_{\del /\del x^i}] \rb + \CO((x^i)^{3})  \,.
\ee
Solving for the ground states of the harmonic oscillator locally, near a critical point $p$ of Morse index $\mu(p)$, we find a unique solution given by a differential form of degree $\mu(p)$. The zero modes of $\psi$, which are identified with $1$-forms in \eqref{Finally},  localize at critical points of Morse index $1$. For  $c_i$ with signature $(-,+,+)$, the solution to leading order is
\be
\label{SolutiontoWitten}
\mu(p)=1: \quad \psi = \psi_{(p,q)} \exp \lb -qt\sum_{i=1}^3 |c_i|(x^i)^2 \rb dx^1\,.
\ee
In other words the form part is oriented along the negative eigenspaces of the Hessian of the function $f$. Here we have decomposed the 7d spinor $\psi$ into a Weyl-spinor $\psi_{(p,q)}$ carrying the anti-commuting, gauge and 4d spinor structure and its internal profile along $\tM3$. The index $(p,q)$ indicates the point $p$, where the correspondicng perturbative ground state localizes and $q$ keeps track of the charge of ${\bf{R}}_q$. The boundary conditions we described in section \ref{sec:MilkyWay} are exactly such that the solutions of \eqref{SolutiontoWitten} collected from all critical points of $f$ of Morse index 1 span the complete perturbative kernel of $\Delta_f$ at degree 1 \cite{chang1995}.

If $p$ has Morse index $2$, the ground state localized near $p$ is of degree 2 and letting $c_i$ have signature $(-,-,+)$, the solution is
\be
\label{eq:groundstatedeg2}
\mu(p)=2: \quad \bar\psi = \bar\psi_{(p,q)} \exp \lb -qt\sum_{i=1}^3 |c_i|(x^i)^2 \rb dx^1\wedge dx^2\,.
\ee
Likewise the fermions in $\overline{\bf R}_{-q}$ are obviously counted by replacing $f$ with $-f$.
% The critical points of $f$ and $-f$ are identical and critical points with a Morse index $\mu$ with respect to $f$ have a Morse index of $3-\mu$ with respect to $-f$. This identifies the solutions of degree $2$ in \eqref{eq:groundstatedeg2} as the CPT conjugates of the fermionic grounds states transforming in $\overline{\bf R}_{-q}$ localizing at critical points of $-f$ of Morse index 1. Each of these ground states contributes to a chiral multiplet in 4d. We therefore obtain a chiral multiplet transforming in ${\bf R}_q,\overline{\bf R}_{-q}$ for every critical point of Morse index 1 with respect to the functions $f,-f$ respectively.

\subsection{Exact Spectrum from SQM}
\label{sec:MassTermsExactSpec}

The perturbative calculation in the previous section does not necessarily give the exact spectrum of the full theory. On the SQM side this is due to the fact that quantum mechanical instanton corrections  can cause perturbative ground states to acquire a mass and be lifted in the full theory \cite{Witten:1982im,Hori:2000kt}. We now complete the dictionary between the 4d effective theory of 7d SYM and SQM by showing that masses of perturbative zero modes in the 4d theory arise precisely from instanton corrections on the SQM side.

We start our analysis with the off-shell action in \eqref{eq:OffShellAction} and split the complex $1$-form $\varphi=\varphi_0+\delta\varphi$ into its background $\varphi_0=tdf$ and fluctuations $\delta\varphi$. The 7d fields are expanded in terms of a basis of perturbative ground states of the twisted Laplacian as
\be\ba\label{eq:expansion}
\psi(x,y)&=\psi_{(a,q)}(x)\psi^{(a,q)}(y)\,, \\ \varphi(x,y)&=tdf(y)+\delta\varphi(x,y)=tdf(y)+\delta\varphi_{(a,q)}(x)\delta\varphi^{(a,q)}(y)\,,
\ea\ee
where $(x,y)\in\R^{1,3}\times \tM3$. Here the sum runs over the charged representations, ${\bf R}_q$ and $\overline{\bf R}_{-q}$, and all critical points $p_a$ of Morse index $1$ with respect to the relevant Morse function, $f$ and $-f$ respectively. The fermionic field $\psi_{(a,q)}(x)$ carries the anti-commuting, gauge and 4d spinor structure while $\psi^{(a,q)}(y)$ is a $1$-form on $\tM3$ annihilated by the twisted Laplacian in perturbation theory. In leading order in $t$ these are \eqref{SolutiontoWitten} or the CPT conjugate of \eqref{eq:groundstatedeg2}. 
The decompositions for $\delta\varphi$ are of analogous structure.

A mass term in 4d descends from the 7d interaction 
\be \label{eq:InterestingTerm}
\tn{Tr}\lbb \psi \wedge \CD \psi \rbb =\tn{Tr}\lbb \psi\wedge \lb d\psi +[\varphi\wedge,\psi]\rb\rbb  \,,
\ee
which for an abelian Higgs background yields the mass matrix
\be\label{eq:MassMatrix}
M^{ab}=\int_{\tM3} \psi^{(a,-q)} \wedge (d+tqdf\wedge)\psi^{(b,q)}=\int_{\tM3} \bar\psi^{(a,q)} \wedge \ast(d+tqdf\wedge)\psi^{(b,q)}\,.
\ee
This precisely computes the instanton corrections between the perturbative ground states in SQM theory and is simply the matrix element 
\be
\label{eq:MatrixElementSQM}
	M^{ab} = \braket{\psi^{(a,q)}|\CD\psi^{(b,q)}}\,.
\ee
Let us briefly summarize the classic results on these instanton corrections, see \cite{Witten:1982im,Hori:2000kt} for a detailed treatment. 
The (Euclidean) action of the SQM with  target space $\tM3$ is given by a standard sigma-model action
\be
\ba
\label{eq:SQMaction}
	 S_{\rm SQM} = \int_{\mathbb{R}} ds&\left(
	 	\frac{1}{2}g_{ij}\frac{d\gamma^i}{ds}\frac{d\gamma^j}{ds} +\frac{q^2t^2}{2}g^{ij}\del_i f\del_j f \right. \\
	 	&\left. \qquad +g_{ij}\bar\eta^i D_s\eta^j + qt D_i\del_j f\bar\eta^i\bar\eta^j+\frac{1}{2}R_{ijkl}\eta^i\bar\eta^j\eta^k\bar\eta^l
	 \right)\,,
\ea
\ee
where $g_{ij}$ is the metric on $\tM3$, $D$ the covariant derivative and $R_{ijkl}$ the curvature tensor. Canonically quantizing this action, one gets the SQM
we have described in the previous section \cite{Witten:1982im}. The matrix element \eqref{eq:MatrixElementSQM} now has the the following path integral expression
\be\ba
\label{eq:pathintegralexpression}
	\braket{\psi^{(a,q)}|\CD\psi^{(b,q)}}&=\frac{1}{qf(p_a)-qf(p_b)+O(1/t)}\lim_{T\rightarrow \infty} \braket{\psi^{(a,q)}|e^{T\Delta_{tf}}[\CD,f]e^{-T\Delta_{tf}}\psi^{(b,q)}}\\
	&=\frac{1}{qf(p_a)-qf(p_b)+O(1/t)}\int_{\substack{\gamma(+\infty)=p_a\\ \gamma(-\infty)=p_b}}D\gamma D \eta D\bar{\eta}\,[\CD,f]e^{-S_{\tn{SQM}}}\,,
\ea\ee
which is valid to leading order in $1/t$. The path integral is taken over the space of all trajectories $\gamma$ connecting the critical point $p_b$ to $p_a$, where $\mu(p_b)=1$ and $\mu(p_a)=2$. The integrand $[\CD,f]$ is $\CD$-exact and hence the path integral receives contributions only from fixed points of the fermionic variations generated by the corresponding supercharge $\mathcal{D}$. Such fixed points are given by trajectories $\gamma$

\be
\frac{d\gamma^i}{ds}=tq g^{ij}\del_j f\,,
\ee
which is the gradient flow equation. With this the mass matrix is evaluated in \cite{Hori:2000kt} to leading order in $1/t$ as
\be\label{eq:Massmatrix}
M^{ab}\, =\sum_{\gamma} n_\gamma e^{-tq(f(p_a)-f(p_b))}\,.
\ee
Here the sum runs over all ascending gradient flow lines $\gamma$ starting at $p_b$ and ending at $p_a$. The contribution from a flow line $\gamma$ is weighted by a sign $n_\gamma=\pm 1$, which arises from a choice of orientation on the moduli space of gradient trajectories. The precise derivation from the SQM context is intricate and is given in \cite[Appendix F]{Gaiotto:2015aoa}. The main takeaway is that perturbative ground states form a complex, where the coboundary operator is given by 
\be
\CD \psi^{(b,q)} = \sum_{a} M^{ab}\bar\psi^{(a,q)} \,.
\ee
This is exactly the Morse-Witten complex for the Morse function $f$. Massless states are counted by the cohomology of this complex and can be found by diagonalising $M^{ab}$.  Recall from section \ref{sec:MilkyWay} that $f$ is a solution of an electrostatics problem and satisfies $\del_n f<0$ (resp. $\del_n f>0$)  on $\Sboundary_-$ (resp. $\Sboundary_-$). The Morse-Witten complex therefore recovers the relative cohomology of a pair $(\tM3,\Sboundary_-)$ \cite{dur4050}. In 4d these give rise to $b^1(\tM3,\Sboundary_-)$ chiral multiplets valued in ${\bf R}_{q}$ and $b^2(\tM3,\Sboundary_-)$  chiral multiplets valued in $\overline{\bf R}_{-q}$.

It is possible that the boundary operator of the Morse-Witten complex is trivial. This is equivalent to a vanishing of the mass matrix $M^{ab}=0$, i.e. all perturbative ground states are true ground states. In this case the Morse function $f$ is called perfect. This is precisely the case when $f$ has $b^i(\tM3,\Sboundary_-)$  critical points of Morse index $i$, for $i=1,2$.

%%%%%%%%%%%%%%%%%%%%
\begin{figure}
	\centering
	\includegraphics[width= 6cm]{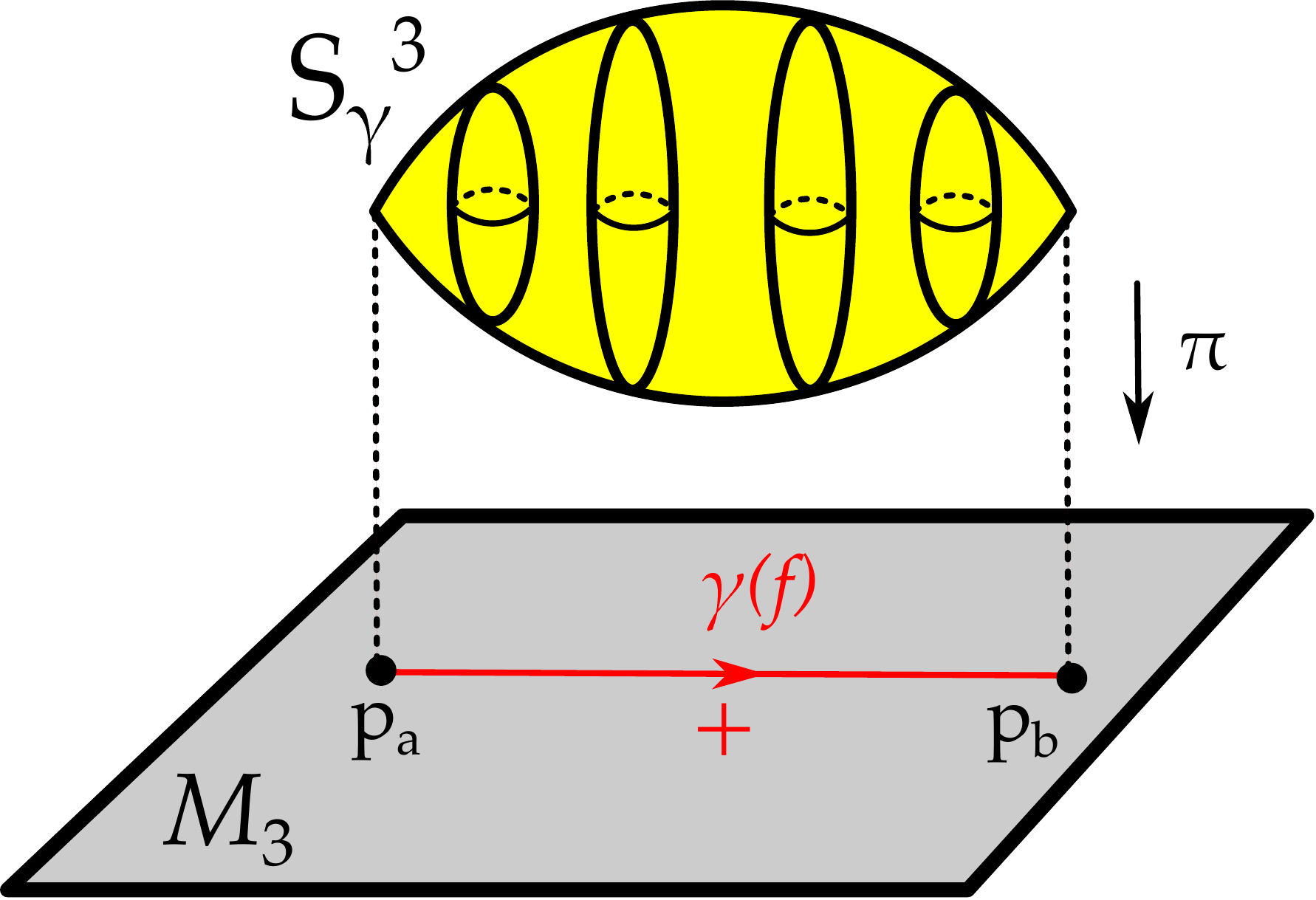}
	\caption{The supersymmetric three-cycle responsible for mass terms. 
	The two critical points $p_a$ and $ p_b$ of the function Morse $f$ in the base of the ALE-fibration $\tM3$ are connected by a gradient flow line $\gamma (f)$. Above each point along this path there is a  two-spheres  in the ALE fiber. Traversing along a gradient flow line of $f$ a 3-sphere $S^3_\gamma$ is traced out. }
	\label{fig:3Sphere}
\end{figure}
%%%%%%%%%%%%%%%%%%%

We can consider these mass terms also in the M-theory picture. In section \ref{sec:HiggsALE} we have interpreted the Higgs field $\phi= df$ as measuring the periods of the vanishing cycle in an ALE-fibration, with respect to a reference \hk structure. For an abelian Higgs field there is exactly one such vanishing cycle which is of finite volume through-out $\tM3$ and collapses precisely at the critical points of $f$. As this vanishing cycle is a two-sphere, paths connecting two critical points lifts to a 3-sphere in the ALE geometry. This 3-sphere is of minimal volume whenever it projects to a gradient flow line in $\tM3$. This is depicted in figure \ref{fig:3Sphere}. 

M2-instanton wrapped on such a three-sphere $S^3_\gamma$ reduces to SQM \cite{Harvey:1999as,Beasley:2003fx}. The stationary points of the M2-brane action, which correspond to associative three-cycles, hence become a fibration of the vanishing cycle of the ALE-fiber over the gradient trajectories $\gamma(f)$ determined by the Morse function $f$. These associatives then give a non-perturbative correction to the superpotential \cite{Harvey:1999as,Beasley:2003fx} which is of the form
\be\label{eq:S3SuperPotContribution}
\Delta W = n_\gamma \exp \lb i \int_{S^3_\gamma}(C+ \, i\Phi) \rb \,.
\ee
In particular, the coefficients originating from a one-loop determinant in the M2-brane action are the same as the those computed in the supersymmetric quantum mechanics and hence give the same coefficients $n_\gamma=\pm 1 $ as those appearing in the Morse theory analysis. In the case of several flow lines connecting the same critical loci $p_a,p_b$, the corresponding associatives are homologous  and there can hence be cancellations among the different contributions depending on the relative orientation.

%%%%%%%

\begin{figure}
	\centering
	\includegraphics[width=14cm]{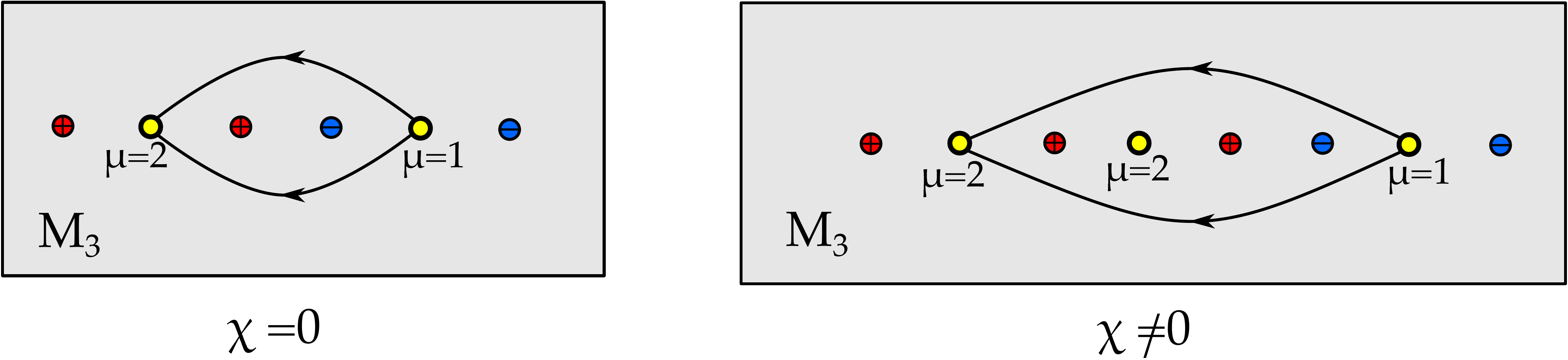}
	\caption{Examples of point like charge configurations in $\tM3$. Depicted are positive (red) and negative (blue) charges, critical points (yellow) and flow lines starting and ending at critical points. The critical points have Morse index $\mu$. The contributions of the flow lines cancels and for generic set-ups each critical point will give rise to a ground state of positive, negative chirality if $\mu=1,2$ respectively. The LHS thus has an equal number of chiral and conjugate-chiral ground states, the chiral index vanishes. For the same reasons the chiral index does not vanish on the RHS.}
	\label{fig:chargeexa}
\end{figure}

%%%%%%

\subsection{Example 2: $n_++n_-$ Point Charges in $S^3$}
\label{sec:Ex2}
We apply the analysis of section \ref{sec:WKB} and \ref{sec:MassTermsExactSpec} to  point charges on the three-sphere. Example configurations are shown in figure \ref{fig:chargeexa}.
Let $M_3=S^3$ and $\gau=SU(n+1)$. Consider  $n_\pm$ positive/negative point charges with the total  charge vanishing. The function $f:M_3\rightarrow \R$ is the electrostatic potential generated by these charges. This function gives rise to a singular abelian Higgs field background on $S^3$ via $\phi=df$ which breaks 
\be\label{SUnU1}
\tn{Ad}\,SU(n+1) \rightarrow \tn{Ad}\,SU(n) \oplus {\tn{Ad}\, U(1) }\oplus {\bf n}_{q} \oplus {\bf \overline{n}}_{-q} \,,
\ee
Perturbative ground states localize at the critical points of the harmonic function $f$. Let $n_\mu$ be the number of points with Morse index $\mu$, then there are $n_1$ chiral fermions $\psi$ and $n_2$ conjugate-chiral fermions $\bar{\psi}$ transforming in ${\bf n}_{q}$. The harmonicity of $f$ forbids points of Morse index $0$ or $3$ as these are minima or maxima respectively. The chiral index as defined \eqref{eq:ChiralInd}
is given by the difference
\be\label{eq:ChiralIndPlump}
\chi\lb S^3,{\bf n}_{q}\rb =n_2-n_1\,,
\ee
as perturbative ground states are lifted by M2-brane corrections in pairs leaving the difference of ground states of positive and negative chirality unchanged.

Next smear out the charges to small balls so that the singularities of $f$ are removed without altering $f$ away from the support of the charge distribution. In this case $\tn{grad}\,f$ becomes a smooth vector field on $M_3$ and the Poincar\'e-Hopf theorem can be applied. We denote the critical points of $f$ by $x_i$, then the topological index $I(x_i,f)$ of $\tn{grad}\,f$ at $x_i$ is determined by the topological index of the map
\be
\frac{\tn{grad}\,f}{|\tn{grad}\,f|}:\quad S^2_{x_i}\rightarrow S^2\,,
\ee
where $S^2_{x_i}$ is a small ball containing the critical point $x_i$. The Poincar\'e-Hopf theorem asserts that the sum of all indices is the Euler characteristic of $M_3=S^3$
\be\label{PoincareHopf}
\sum_i I(x_i,f)=\chi\lb S^3 \rb =0\,.
\ee 
Note that $I(x_i,f)=(-1)^{\mu(x_i)}$ for all critical points $x_i$ and that each charge contributes one maximum or minimum upon smearing it out, whereby \eqref{PoincareHopf} simplifies to
\be
0=n_--n_1+n_2-n_+\,.
\ee 
Combining this result with \eqref{eq:ChiralIndPlump} we find the chiral index to be determined solely by the composition of the initial charge configuration
\be\label{eq:ChiralIndexPoints}
\chi\lb S^3,{\bf n}_{q}\rb =n_+-n_-\,.
\ee
We thus find a rather simple criterion to determine whether the true ground state spectrum of the theory is chiral or not:
\be
n_+\neq n_-\quad\leftrightarrow\quad\tn{chiral spectrum}\,.
\ee 
Two examples are shown in figure \ref{fig:chargeexa}.
This result is of course recovered from the more general charge distributions discussed in section \ref{sec:Ex1} upon setting the number of loops $l_+$ and $l_-$ to zero. In particular for generic placements of the $n_++n_-$ charges one has
\be\label{eq:Chiralcount}
n_1=n_--1\,, \qquad n_2=n_+-1\,.
\ee  
Each critical point thus constitutes a true ground state and we recover \eqref{eq:chiralindexcharges}. This is made explicit in figure \ref{fig:chargeexa}. If flow lines between critical points exist, they always do so in pairs with $n_\gamma=\pm1$. Hence the corresponding ground states are not lifted.

\subsection{Generalized Critical Loci and Morse-Bott Theory}
\label{subsec:Lindt}

\begin{figure}
	\centering
	\includegraphics[width=5cm]{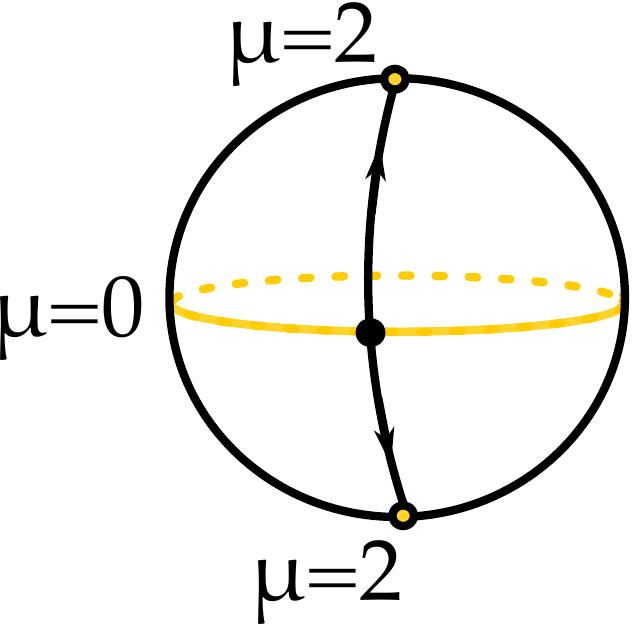}
	\caption{$S^2$ with the Morse-Bott function given by $f(x,y,z)= z^2$. The critical locus is colored in yellow and consists of two critical points of index 2 (north ant south pole) and a critical circle of index 0 (the equatorial circle). The gradient curves are depicted in black. Note that $\CM(N_0,N_2)= \s^1\coprod\s^1$. These two circles parametrize the gradient trajectories in the upper and lower hemisphere.}
	\label{fig:morsebottflow}
\end{figure}

The setup studied in \cite{Pantev:2009de} and in the last section assumes that the critical loci of the function $f$ are isolated points. Although this is the generic situation, it will be important to relax this assumption and consider the generalized setup in which the critical locus of $f$ can be one-dimensional, which happens for the recent TCS constructions of $G_2$-manifolds. Functions $f$ with critical loci of dimension greater than zero whose Hessian at its critical closed submanifold is non-degenerate in the normal direction are called Morse-Bott functions. An example is given in figure \ref{fig:morsebottflow}. For further background on this see \cite{Hori:2000kt,FloerMemo}.

The starting point is once more an abelian Higgs field $\phi=df$ as in section \ref{sec:WKB} where now $f$ is taken to be a harmonic Morse-Bott function. We are again interested in the fermionic zero modes transforming in the representation ${\bf R}_q$ which are in the kernel of the twisted Laplacian \eqref{WittensLaplacian}. As before, rescaling $f\rightarrow tf$ these localize on the critical loci of $f$ and we can solve for the zero mode solutions locally. However, $f$ now has higher dimensional critical loci and our previous analysis needs to be adapted. We begin by analyzing the critical loci of $f$. 

The local analysis of the perturbative ground states is now the same as in section \ref{sec:WKB}, although some extra care is required to keep track of the critical loci of different dimensions. The critical locus of $f$ splits into connected components all of which are compact closed submanifolds of $\tM3$. Let $N$ denote a single connected component. The normal bundle $\nu N$ splits into the positive and negative eigenspace of the Hessian $H_f$ of $f$
\be
\nu N = \nu_+N\oplus\nu_- N
\ee
and the Morse index of $N$ is defined as the rank of $\nu_- N$. In our context the Morse-Bott function $f$ is also harmonic. This precludes critical submanifolds of dimension 2 since harmonicity of $f$ implies that 
$\tr H_f = 0$, 
which would mean that $H_f$ is degenerate in the normal direction, which is not possible since $f$ is Morse-Bott by assumption. For harmonic Morse-Bott functions on a three-manifold,  
$N$ can thus only be a point or a circle. 
Moreover, if $N= \s^1$, it can only have index 1. This is again due to the requirement that $\tn{Tr}\,H_f$ vanishes everywhere. 
The case where $N$ is a point has been analyzed in section \ref{sec:WKB}. 

If $N = \s^1$ we can proceed analogously. As $N$ has index 1, $f$ is locally of the form
\be\label{eq:AdaptedCoords}
f(x)=f(0)-\frac{c}{2}\lb (x^1)^2-(x^2)^2\rb +O((x^i)^3)\,,
\ee
in a suitable normal coordinate chart centered at a point $p\in N$. In this coordinate system $x^3$ is the coordinate tangential to $N$ and the Hessian $H_f$ is diagonalized with the eigenvalues $c$ and $-c$. In these coordinates the twisted Laplacian  \eqref{WittensLaplacian} now takes the form
\be\ba
\Delta_{tf} &= \lb \Delta_{tf}\rb _{\perp}+\lb \Delta_{tf}\rb _{\parallel}+ \CO((x^i)^3)\,, \\
\lb \Delta_{tf}\rb _{\perp}&= \sum_{i=1}^2 \left(-\frac{\del^2}{\del (x^i)^2} + q^2t^2c^2(x^i)^2\right) - qt c[dx^1, \iota_{\del /\del x^1}] + qt c[dx^2, \iota_{\del /\del x^2}]\,,\\
\lb \Delta_{tf}\rb _{\parallel}&= -\frac{\del^2}{\del (x^3)^2}\,.
\ea\ee

The analysis of perturbative ground states thus splits into normal and tangential parts relative to $N$. In the normal direction we get a single 1-form solution $\psi_\perp$ given by
\be\label{eq:psiperp}
\psi_\perp = \psi_{(n,q)}\exp\lb-qtc\lb (x^1)^2+(x^2)^2\rb\rb dx^1 \,.
\ee
Here we have split $\psi_\perp$ into a 4d Weyl spinor $\psi_{(n,q)}$ carrying the anti-commuting, gauge and spinor structure and its internal profile normal to $N$. In principle $\psi_\perp$ is defined only locally on $N$. However, observe that $\psi_\perp$ is a volume form on the fiber of $\nu_- N$. Hence, assuming that the negative eigenbundle $\nu_- N$ is orientable, the local solutions can be patched together to a global form on $N$. Since $f$ is constant on $N$ the tangential equation reduces to a Laplace equation on $\s^1$. Let the coordinate on the circle be $\theta$. Then we obtain two solutions
\be
\label{eq:MorseBottPerturbative}
\psi_1 = \psi_\perp\,,\qquad \psi_2 = d\theta \wedge\psi_\perp \,.
\ee
For every circle $N$ contributing to the perturbative spectrum we therefore obtain a pair of states consisting of a 1- and 2-form. From \eqref{Finally} we know that the degree of the ground state correlates with the 4d chirality of fermions, i.e. the state described by a 1,2-form has positive, negative chirality upon a reduction to 4d. These fermionic states again contribute to chiral multiplets in 4d.

As in the case of Morse functions, perturbative zero modes for $\chi,\bar\chi$ transforming in ${\bf R}_q$ are absent as $f$ is harmonic. To conclude we again remark that the analysis above extends to fermionic ground states transforming in $\overline{\bf R}_{-q}$ by replacing $f$ with $-f$. The function $-f$ now exhibits the same critical loci. A critical point of Morse index $\mu$ with respect to $f$ has a Morse index of $\mu-3$ with respect to $-f$, however critical circles exhibit an unchanged Morse index of $1$ with respect to both $f$ and $-f$. The modes localising on the critical circles of $-f$ transforming in $\overline{\bf R}_{-q}$ are CPT conjugate to the solutions found in \eqref{eq:MorseBottPerturbative}. As a consequence we find the localized perturbative ground states on every critical circle contributing to the perturbative spectrum to assemble to two chiral multiplets transforming in ${\bf{R}}_q$ and $\overline{\bf{R}}_{-q}$. 

\subsection{Generalized Critical Loci and SQM}
\label{sec:1dmatter}

We now turn to the computation of the exact spectrum from the perturbative solutions in the Morse-Bott case, where the critical loci of $f$ consist of points and circles. While it is possible to compute the SQM instanton correction in much greater generality \cite{Hori:2000kt,FloerMemo}, the applications for TCS local models allow us to consider only the set-up with this restriction. The instanton calculation in this case effectively reduces to the one considered in section \ref{sec:MassTermsExactSpec}.

To find the exact spectrum, we again want to compute the matrix element \eqref{eq:Massmatrix} between perturbative zero modes localized at critical submanifolds we use the analogous SQM computation. Let $N_m$ denote the disjoint union of critical submanifolds of Morse-Bott index $m$ (recall that this is the dimension of the negative eigenspace of the Hessian matrix). In our case, $m$ can take the values $1$ or $2$. For $m=2$, all of the components of $N_2$ must be points, whereas $N_1$ can contain points as well as circles. 

Recall that among the ground states localized at critical circles there are chiral multiplets transforming in the representation ${\bf{R}}_{q}$ and $\overline{\bf{R}}_{-q}$. As already discussed in section \ref{subsec:Lindt}, this is because perturbative ground states are of the form 
\be\label{eq:MorseBottGS}
  \psi = \alpha\wedge\psi_\perp,
\ee 
with $\deg(\psi_\perp)= 1$ and $\alpha$ a harmonic form on $N_1$. When $N_1$ is a circle, $\alpha$ can be a function or a one-form. 
Consider again the matrix element 
\be
\label{eq:MEMB}
M^{ab}=\int_{\tM3} \bar\psi^{(a,q)} \wedge \ast(d+tqdf\wedge)\psi^{(b,q)}\, .
\ee
Here we again use the indices $a$ and $b$ to enumerate all the perturbative ground states of total degree $2$ and $1$ respectively. However, note that for Morse-Bott functions the index is no longer in one-to-one correspondence with critical loci since there are two perturbative ground states localized at each critical $\s^1\subset N_1$. For the following we will require the assumption that there are no ascending gradient flow lines between connected components in $N_1$.\footnote{In this case $f$ is said to be weakly self-indexing. This assumption can be avoided at a cost of making the exposition much more technical \cite{FloerMemo}.}

To compute $M^{ab}$ we need to consider three cases. First, both $\bar\psi^{(a,q)}$ and $\psi^{(b,q)}$ may be localized at points in which case the discussion of section \ref{sec:MassTermsExactSpec} applies verbatim. We now turn to the second possibility, where the ground states are both localized at the same circle critical circle $\s^1\subset N_1$. The matrix element is then given by the integral
\be
\label{eq:noinstantons}
	\int_{\tM3} d\theta\wedge\psi_\perp \wedge\ast(d+tqdf\wedge)\psi_\perp\,,
\ee
where we have used the explicit expression of for such ground states given in \eqref{eq:MorseBottPerturbative}. Using the expression for $\psi_\perp$ in \eqref{eq:psiperp} one can see that $d\theta\wedge \ast (df\wedge \psi_\perp) = 0$ and also $d\theta\wedge\psi_\perp\wedge\ast d\psi_\perp=0$. This implies that the matrix element $M^{ab}$ is zero, if $\bar\psi^{(a,q)}$ and $\psi^{(b,q)}$ are both localized at the same circle. 

The third possibility is that $\bar\psi^{(a,q)}$ is localized at a point $p_a$ in $N_2$ and $\psi^{(b,q)}$ is localized at a circle $\s^1_b\in N_1$.  To keep track of all of the gradient curves between critical loci of $f$, we introduce the moduli space of gradient trajectories between $N_m$ and $N_n$
\be
\CM(N_m,N_n) = \bigslant{\left\{\gamma: \R\rightarrow M \left| \,\,\lim\limits_{t\rightarrow -\infty}\gamma(t)\in N_m\,,\quad\lim\limits_{t\rightarrow \infty}\gamma(t)\in N_n\,, \quad \frac{d\gamma^i}{ds} = tq g^{ij}\del_j f\right. \right\}}{\R}.
\ee
Here, the quotient is taken with respect to the remaining reparametrization invariance of the gradient flow: $\gamma(t)\mapsto \gamma'(t) = \gamma(t+\delta t)$. The moduli space $\CM(N_m, N_n)$ is a smooth manifold, and it follows from simple dimensional analysis that its dimension is $m-n-1$. An illustrative example is given by $S^2$ with the Morse-Bott function $f(x,y,z)=z^2$, see figure \ref{fig:morsebottflow}.

For our purposes, the only relevant case is $m=1$ and $n=2$ in which case the moduli space is a finite set of points. This means that there are finitely many gradient trajectories connecting $N_1$ and $N_2$ and there are finitely many ascending gradient flow lines connecting $\s^1_b$ and $p_a$. We can now continue with the computation. In terms of the SQM path integral we have the expression 
\be
M^{ab}=\braket{\psi^{(a,q)}|\CD\psi^{(b,q)}}=\frac{1}{qf(p_a)-qf(p_b)+O(1/t)}\int_{\substack{\gamma(+\infty)=p_a\\ \gamma(-\infty)\in \s^1_b}}D\gamma D \eta D\bar{\eta}\,[\CD,f]e^{-S_{\tn{SQM}}}\,,
\ee
where $p_b$ is an arbitrary point in $\s^1_b$ (note that $f$ is constant along $\s^1_b$). This is nearly the same expression as in \eqref{eq:pathintegralexpression}, with the only difference being that we integrate over all curves with $\gamma(-\infty)\in \s^1_b$. However, the same localization argument as before applies. As we have seen above, the number of gradient trajectories is still finite and the result of the path integral computation has exactly the same form as for points, i.e. (\ref{eq:Massmatrix}).
The expression for the operator $\CD$ also remains unchanged
\be
\CD \psi^{(b,q)} = \sum_{a} M^{ab}\bar\psi^{(a,q)} \,.
\ee
The exact spectrum is given as the cohomology of $\CD$, which acts on the following complex
\be
  C^1 = \Omega^0(N_1)\,, \qquad  C^2 = \Omega^1(N_1)\oplus \Omega^0(N_2)\, .
\ee
This complex is a convenient way to arrange all the perturbative ground states of degree $p$ in $C^p$. It is a specific instance of a Morse-Bott complex for $f$, which can be defined for $f$ with critical loci of arbitrary dimension \cite{FloerMemo}. If $f$ is a solution to the electrostatics problem in section \ref{sec:MilkyWay}, the Morse-Bott cohomology again recovers the relative cohomology of a pair $(\tM3,\Sboundary_-)$.

\subsection{Chiral Index from Spectral Covers}
\label{sec:ChiralIndSpecCover}

We close this section by introducing yet another picture for counting the perturbative zero modes, namely using the spectral cover introduced in section \ref{sec:SpectralCovers}. For certain configurations it is possible to read off the exact spectrum using the spectral cover, this was already observed for the $U(1)$ case in \cite{Pantev:2009de}. 

For simplicity let us begin by recalling the statement for the rank 1 Higgsing in \eqref{SUnU1} where $\gau=SU(n+1)$. There we turned on a single abelian Higgs background parametrised by the Morse function $f$ via $\phi= df$. The spectral cover $\mathcal{C}$ in this case is simply the graph of $\phi$. The intersection number of $\mathcal{C}$ with the zero section $b_0=0$ (i.e. $\tM3$) at a critical point $p$ is denoted by $n_p$. 
This can be identified with  the degree of the vector field $\tn{grad}\,f$ at the critical point $p$. In a coordinate system where the Hessian $H_f$ is diagonal it follows immediately that the degree is determined by the Morse index $\mu(p)$ of $f$ at $p$ as $n_p=(-1)^{\mu(p)}$. We can therefore recast the counting of perturbative ground states as
\be\ba\label{eq:ZeroModesSpecCover}
|(\CC\cap \tM3)_-|&=\tn{perturbative zero modes in }{\bf R}_q \cr
|(\CC\cap \tM3)_+|&=\tn{perturbative zero modes in }\overline{{\bf R}}_{-q}\,,
\ea\ee
where  $(\CC\cap \tM3)_\pm$ counts the number of critical points $p$ with $n_p=\pm1$. The chiral index is thus simply given by the signed count of all points of intersection
\be\label{eq:CountModeSpecCov}
\chi (\tM3 ,{\bf R}_{q}) = \mathcal{C} \cap \tM3 = \sum_{p\in\tM3\,:\,df(p)=0}n_p= (\CC\cap \tM3)_+ - (\CC\cap \tM3)_-\,.
\ee

The above carries over straightforwardly to higher rank Higgs bundles if their corresponding spectral cover factors completely. We start from the set-up in which we have broken the gauge symmetry to $\gut \times U(1)^n$ by turning on sources for the Higgs field along the CSA of $\gau$ as in section \ref{sec:U1SymMod}. The representation $\tn{Ad}\,\gau$ decomposes into irreducible representation ${\bf R}_Q$ of $\gut \times U(1)^n$ where $Q$ denotes a vector of $U(1)$ charges. 
Generically the representation $\tn{Ad}\,\gau$ decomposes into irreducible representation of $\gut\times\gut_\perp$ with the weights $\lambda_i$ of the representation of $G_\perp$ determining the different spectral covers. Due to the special choice of background the representations of $G_\perp$ have decomposed into representations of $U(1)^n$ and to construct the spectral cover we must group the representations ${\bf R}_Q$ according to this decomposition. This grouping depends on $\gau$ but the weights will always be determined by the corresponding effective Morse functions as $\lambda_i=df_{Q_i}$ where $i=1,\dots,N$. The effective Morse function $f_{Q_i}$ was defined in \eqref{eq:EffMorsefunction} and $N$ denotes the rank of the spectral cover. A spectral cover is thus the union of graphs of multiple $df_{Q_i}$ and an $N$-fold covering of $\tM3$. The matter loci are as before the critical points of $f_{Q_i}$, i.e. the intersection of the spectral cover with the zero section. This is just $b_0=0$ in the language of section \ref{sec:SpectralCovers}.

To compute the perturbative spectrum we thus just need to count the intersections of the different sheets with their signs as in the rank 1 case above. Let $\CC_i\subset \CC$ denote the sheet of a spectral cover $\CC$ with $\tn{Graph}(df_{Q_i})=\CC_i$ then
\be\ba
|(\CC_i\cap \tM3)_-|&=\tn{perturbative zero modes in }{\bf R}_{Q_i}\cr
|(\CC_i\cap \tM3)_+|&=\tn{perturbative zero modes in }\overline{{\bf R}}{_{-Q_i}}\,, \\
\ea\ee
where the notation is as in \eqref{eq:ZeroModesSpecCover}. Similarly we compute the chiral index to 
\be
\chi (\tM3 ,{\bf R}_{Q_i}) = \mathcal{C}_i \cap \tM3 =(\CC_i\cap \tM3)_+ -(\CC_i\cap \tM3)_- \,.
\ee
Perturbative zero modes transforming a representation ${\bf R}_{Q_i}$ which is not associated by $\lambda_i=df_{Q_i}$ to a sheet of this spectral cover are enumerated by the intersection of the different sheets
\be\ba\label{eq:IntersectionChiralInd}
|(\CC_i\cap \CC_j)_-|&=\tn{perturbative zero modes in }{\bf R}_{Q_i-Q_j}\cr
|(\CC_i\cap \CC_j)_+|&=\tn{perturbative zero modes in }\overline{{\bf R}}_{-(Q_i-Q_j)}\,. \\
\ea\ee 
The chiral index again given by the difference
\be
\chi (\tM3 ,{\bf R}_{Q_i-Q_j}) = \mathcal{C}_i \cap  \mathcal{C}_j =(\CC_i\cap  \mathcal{C}_j)_+ -(\CC_i\cap  \mathcal{C}_j)_- \,.
\ee
This is pictorially most clear in the case of $A_n$ singularities. In this case the ALE-fiber is given by a circle fibration over $\R^3$ and the eigenvalues $\lambda_i$, which are characterised by the sheets of the spectral cover, correspond to the points at which the circle collapses. A vanishing sphere is stretched between any pair of these points and collapses whenever they come together, i.e. when the sheets intersect. This enhances the spectrum and constitutes an additional ground state.

%%%%%%%%%%%%%%%%%%%%%%%%%%%%%%%

%%%%%%%%%%%%%%%%%%%%%%%%%%%%%%%

\section{Yukawa Couplings and Higher-Point Interactions}
\label{sec:MatterInteractions}

In this section we discuss the interactions of bulk and localized matter. It will be useful to  consider the case of a fully factored spectral cover, in which case we can compute the zero-modes. In M-theory interactions between localized matter fields come from M2-instantons wrapped on calibrated $3$-spheres of the local ALE-fibration. This is simply a generalisation of the results of section \ref{sec:MassTermsExactSpec}, where we interpreted non-perturbative mass terms as arising from M2-instantons wrapping three-cycles which connect two critical points over a gradient flow line. For higher point interactions these three-cycles project to gradient flow trees on $\tM3$ and studying the moduli space of these constrains the corresponding interactions in 4d. Corrections to these couplings are obtained from integrating out states with masses induced by M2-instantons as discussed in section \ref{sec:MassTermsExactSpec}. 

Consider again the background as in section \ref{sec:U1SymMod}
\be\ba\label{eq:GeneralBackgroundLocal2}
\tn{globally on }M_3:\qquad \langle \phi \rangle = \diag (\lambda_1, \,\cdots, \lambda_n)=\sum_{i=1}^n\mathfrak{t}^idf_i\,,\quad \Delta f_i=\rho_i\,, \quad \int_{M_3}\rho_i=0\,.
\ea\ee
The matter content is summarized in table  \ref{tab:MatterContent}.

\subsection{Bulk--Localized-Matter Interactions}
\label{sec:BulkInt}

First consider the bulk-localized-localized interactions. These interactions are present for rank 1 abelian Higgs background. The contribution to the 
$\CN=1$ 4d superpotential is canonically derived by expanding the partially twisted  7d action
\be\ba\label{eq:OffShellAction1}
I_{7d}&=\frac{1}{g_{7d}^2} \int_{\R^{1,3}}\int_{\tM3}d^{7}x\bigg[ -\frac{1}{4}F_{\mu\nu}F^{\mu\nu}-D_\mu \bar{\varphi}_k D^\mu\varphi^k+i D_\mu \chi \sigma^\mu\bar{\chi} -i D_\mu \psi_k \sigma^\mu\bar{\psi}^k\\
&\qquad\quad   +\frac{1}{2}D^2+H^k\bar{H}_k-DI_{\varphi, \bar{\varphi}}+\frac{i}{\sqrt{2}}(F_{\bar{\varphi}})_{ij}\epsilon^{ijk}\bar{H}_k-\frac{i}{\sqrt{2}}(F_{\varphi})_{ij}\epsilon^{ijk}H_k \\
&\qquad\quad  -\frac{i}{\sqrt{2}}\epsilon^{ijk}\psi_i\CD_j\psi_{k}+\frac{i}{\sqrt{2}}\epsilon^{ijk}\bar{\psi}_i\bar{\CD}_j\bar{\psi}_{k}+\sqrt{2}i\chi \bar{\CD}_i \psi^i-\sqrt{2}i\bar{\chi} \CD_i \bar{\psi}^i\bigg] \,,
\ea\ee
in perturbative zero modes, see  appendix \ref{app:OffShell}. We include the light modes whose masses are induced by M2-instantons in order to discuss corrections to the couplings between the true zero modes at a later point. The gauge symmetry constrains the interaction to $
{\bf R}_Q \otimes \overline{\bf R}_{-Q}\otimes \lb \tn{Ad}\,\gut \oplus \tn{Ad}\,U(1)^n \rb$, 
and so we focus on a pair of conjugate representations in the subsequent analysis. A more detailed discussion is found in appendix \ref{sec:EffAction}. The interactions are deteremined by overlap integrals  
\be\label{eq:Zukawa1U1}
Z^{(ab,k)}=\int_{\tM3} \varphi^{(a)}\wedge \varphi^{(b)}\wedge h^{(k)}\,,
\ee
where the 1-forms $\varphi^a,\varphi^b$ describe the profile of the bosonic ground states along $\tM3$ localized at the critical points $p_a$ and $p_b$ of the function $f_Q$ and $f_{-Q}$, that transformin the representations ${\bf R}_{Q}$ and $\overline{\bf R}_{-Q}$, respectively. The 1-forms $h^{(k)}$ with $k=1,\dots,b^1(\tM3)$ form an harmonic basis. 

Reduction to  4d yields at every critical point $p_a$ of $f_Q$ of Morse index $\mu(p_a)=1$ a chiral multiplet in ${\bf R}_{Q}$. We denote the multiplets corresponding to $\varphi^{(a)}$ by $\Phi_a$ respectively. In addition there are $b^1(\tM3)$ chiral multiplets valued in $\tn{Ad}\,\gut$ and $\tn{Ad}\,U(1)^n$ obtained from expansions in ordinary harmonics of the bulk fields, denoted by $\Phi_k',\tilde{\Phi}_k$. The interaction term is then
\be
\CL_{4d,{\tn{int}}}=  \tn{Tr}\lbb \frac{i}{2\sqrt{2}} \sum_{ab,k}Z^{(ab,k)}\lb \Phi_a\Phi_k'\Phi_b \rb\Big|_{\theta\theta} - \frac{i}{2\sqrt{2}} \sum_{ab,k}Z^{(ab,k)} \lb \Phi_a\tilde{\Phi}_k\Phi_b \rb \Big|_{\theta\theta} + \tn{h.c.} \rbb \,.
\ee
We now turn to interactions of the localized matter fields only.

%%%%%%%%%%%%%%%

\begin{figure}
	\centering
	\includegraphics[width=6cm]{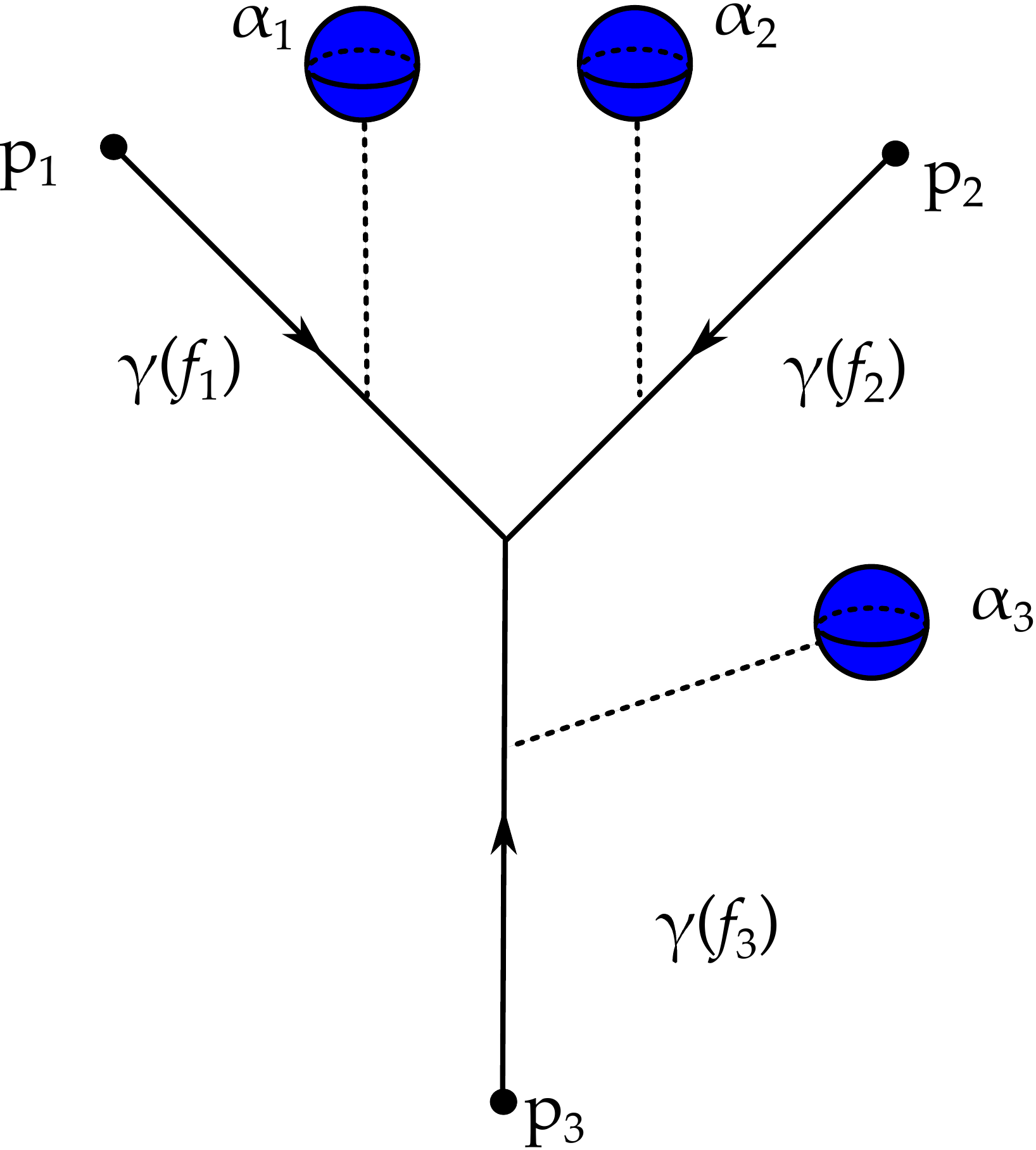}
	\caption{Gradient flow tree for Yukawa couplings. The picture shows three critical points $p_i$ of the functions $f_i$ of Morse index $\mu_{f_i}(p_i)=1$. The gradient flow lines $\gamma(f_i)$ of the $f_i$ are marked by arrows. Every $f_i$ controls the size of a 2-cycle $\alpha_i$ which has the topology of a two-sphere and collapses over the points $p_i$. The three three-chains formed by fibering the two-spheres $\alpha_i$ over the segments $\gamma(f_i)$ can be joined at their meeting point as $\alpha_1 + \alpha_2 + \alpha_3 = 0$, and the resulting three-cycle is an associative.}
	\label{fig:3cycle}
\end{figure}
%%%%%%%%%%%%%%%

\subsection{Yukawa Couplings}
\label{sec:MassYukawa}

For Yukawa couplings we need a rank $n=2$ Higgs bundle (or higher).
There are two Morse functions $f_1$ and $f_2$ and the combination $f_Q=q_1f_1+q_2f_2$.
From the effective field theory we obtain this coupling by expanding the action \eqref{eq:OffShellAction1} in perturbative zero modes 
\be\label{eq:Yukawa}
Y^{abc}_{pqr}=\int_{\tM3}\psi^{(a,p)}\wedge \varphi^{(b,q)}\wedge \psi^{(c,r)}\,, \qquad Q_p+Q_q+Q_r=0\,,
\ee
where  $(a,p)$ refers to the internal profile of the perturbative zero mode localized at the critical point $p_a$ transforming in ${\bf R}_{Q_p}$.
The Yukawa couplings arise from M2-instantons wrapping associative three-cycles. 
To characterize the three-cycles consider the Morse functions
\begin{alignat}{5}\label{eq:Charges}
Q_1&=(1,0)\,, \qquad &&Q_2&&=(0,-1)\,, \qquad &&Q_3&&=(-1,1)\cr 
f_{Q_1}&=f_1\,, \qquad &&f_{Q_2}&&=-f_2\,, \qquad &&f_{Q_3}&&=-f_1+f_2=f_3\,,
\end{alignat}
which describe an $SU(3)$ ALE-fibration over the base $\tM3$. Each of the functions $f_i$ controls the volume of a corresponding two-sphere $\sph_i$ in the ALE fiber, which satisfy 
\be
\label{eq:homrelation}
	\sph_1+\sph_2+\sph_3 = 0
\ee
in the homology of every fiber. Recall that $\sph_i$ shrinks to zero volume precisely over the points $p_i$ where $df_i = 0$. To every gradient trajectory $\gamma(f_i)$ starting at a point $p_i$ we can associate a 3-chain, which is given by tracing out the corresponding $\sph_i$ in the ALE-fibration. Given three sufficiently generic Morse functions $f_i$, there will be finitely many gradient flow trees connecting the three critical points $p_i$ (see figure \ref{fig:3cycle}). Adding the associated three-chains produces a three-cycle, the boundary of which is given by $\sum_i \alpha_i$ in the ALE fiber. We may produce a closed three-cycle with the topology of a three-sphere by adding a three-cycle $\beta$ such that $\del\beta= \sph_1+\sph_2+\sph_3$. Moreover, this $S^3$ is in fact associative, since it projects to the tree of gradient trajectories and hence minimizes the volume among all the three-cycles which project down to trees connecting $p_1$, $p_2$ and $p_3$. Wrapping an M2-brane on such a cycle gives rise to Yukawa couplings between modes localized at the critical points of $f_1$, $f_2$ and $f_3$. Consequently, the overlap integral \eqref{eq:Yukawa} vanishes if there exists no trivalent gradient flow tree connecting the critical points\footnote{Massless chiral multiplets are found when expanding the 7d action in true zero modes. These are in general linear combinations of the localised perturbative profiles used in \eqref{eq:Yukawa}. The relevant linear combinations are determined by the Morse-Witten complex. The overlap integral determining the Yukawa couplings between the massless modes are thereby linear combinations of \eqref{eq:Yukawa}.}.

\begin{figure}
	\centering
	\includegraphics[width=7cm]{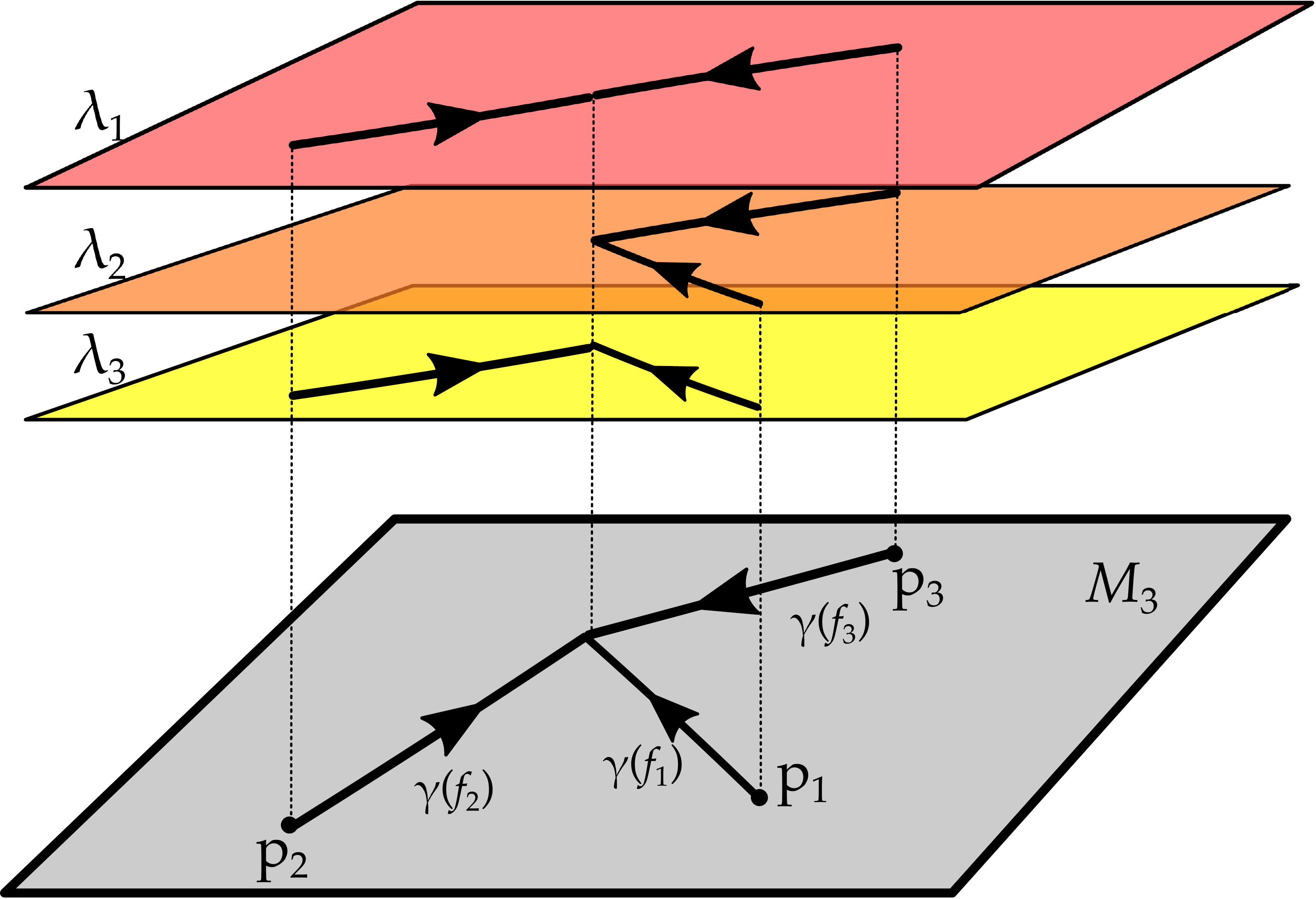}
	\caption{Construction of three-cycle that gives rise to the Yukawa couplings in the spectral Cover picture. The critical loci $p_i$ correspond to the loci where two of the weights $\lambda_j$ are equal, i.e. the corresponding sheets of the spectral cover meet. The uplift of the gradient flow lines $\gamma(f_i)$ sweeps out the associative three-cycle $S^3$ that can then be wrapped by an M2-instanton. This gives rise to the coupling between the three matter states localized at $\lambda_i=0$. The combined flow lines give rise to the gradient flow tree $\gamma (f_1, f_2, f_3)$.}
	\label{fig:SCYuk}
\end{figure}

Similarly, in the spectral cover description, the Yukawa coupling is modeled in terms of a three-sheeted cover, whoch is determined by the graph of $df_i$. The segments of the gradient flow trees determined by the function $f_i$ thus lift to paths on the corresponding sheets; see figure \ref{fig:SCYuk}. The paths connect the points where two sheets pairwise intersect. One can think of the 2-cycles $\alpha_j$ in the ALE-fibration as being stretched between the sheets and the corresponding cycle collapses precisely at points where two sheets meet.

The strength of these interactions is governed by the choice of functions $f_i$. The three-sphere giving rise to the Yukawa coupling is a supersymmetric rigid homology sphere within the $G_2$-manifold and its contribution to the superpotential is again given by \eqref{eq:S3SuperPotContribution}. The sign $n_\gamma=\pm1$ arises in the same manner and is given by an orientation on the moduli space of gradient flow trees. As the Higgs field $\phi_i$ and the gauge field $W_i$ are identified with the periods of the supergravity 3-form $C$ and associative 3-form $\Phi$ the integral is evaluated as
\be\label{eq:VolumeM2}
\int_{S^3_\gamma}(C+i\Phi)= \sum_{j=1}^3\int_{\gamma(f_j)}\int_{\alpha_j} (C+i\Phi) =\sum_{j=1}^3 \int_{\gamma(f_j)} \lb W_j+i\phi_j \rb = i\sum_{j=1}^3 \int_{\gamma(f_j)} tdf_{Q_j}\,,
\ee
Here, we have used that we can gauge the background for the gauge field $W_i$ to zero. Evaluating the final integrals and using that the homological relation between the $\alpha$ implies $\sum_i^3 f_i = 0$ we find
\be\label{eq:SubPotCont}
\Delta W = n_\gamma\exp \lb -\sum_{i=1}^{3} tf_{Q_i}(p_i) \rb\,.
\ee

\subsection{Associatives and Gradient Flow Trees}
\label{sec:AssociativesGFT}

The generation of Yukawa couplings and mass terms from associative three-cycles which project to flow trees on $M_3$ has a natural generalization  \cite{Fukaya_morsehomotopy}, which in the effective theory realizes higher point couplings. 

\begin{figure}
	\centering
	\includegraphics[width= 10cm]{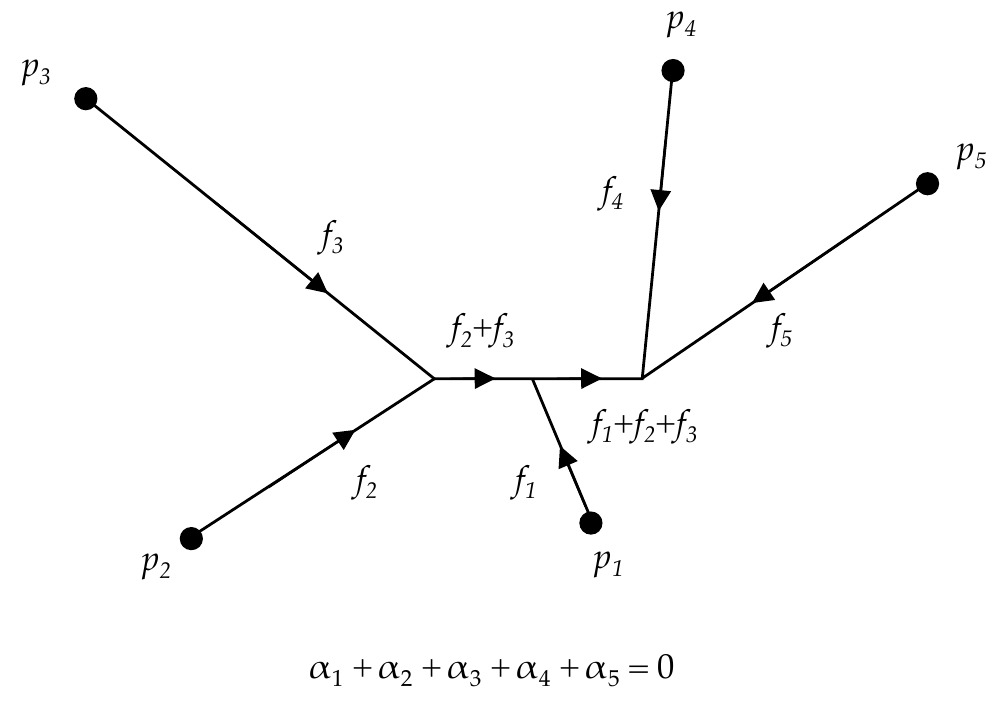}
	\caption{A gradient flow tree 		with 5 external vertices of Morse index 1.}
	\label{fig:GFT}
\end{figure}

We consider a setup in which $G_\perp = S[U(1)^k]$, so that the Higgs background is described by $k$ smooth Morse functions $f_i$. As the associated two-spheres $\alpha_i$ in the ALE fiber sum to zero in homology, the same must be true of the functions $f_i$. Choosing a critical point $p_i$ of each $f_i$ with Morse index $\mu(p_i)$, one can define the moduli space of gradient flow trees
\be\label{eq:ModuliSpace}
\CM(M;f_1,\dots,f_{k\,};p_1,\dots,p_k)\,,
\ee
as the set of gradient flow trees with external vertices $p_1,\dots,p_k$ such that the lines emanating from $p_i$ are ascending gradient flow lines of $f_i$. These form the external edges of the gradient flow tree. Of course we also allow for internal vertices and edges. The flow of these is governed by 
the associated integral linear combinations of the $f_i$, which are in turn determined by a charge conservation constraint. This moduli space $\CM$ has dimension 
\be\label{eq:dimMod}
\dim \CM(M;f_1,\dots,f_{k\,};p_1,\dots,p_k) = k-\sum_{i=1}^k\mu(p_i)\,,
\ee
and there are thus finitely many gradient flow trees connecting $k$ points of Morse index 1. An example of a gradient flow tree for the case of $k=5$ is shown in figure \ref{fig:GFT}.

As before, we can construct a three-cycle by fibering the two-sphere $\alpha$ associated with the Morse function $f$ over each segment $\gamma(f)$. This both guarantees that we end up with an associative, and also that $\alpha$ collapses at the end-points of the flow tree. Furthermore, the fact that we have a tree in $M_3$ implies that the resulting associative three-cycle has the topology of a three-sphere, so that it contributes to the effective superpotential. Using the same manipulations as in \eqref{eq:VolumeM2}, we can compute the volume of such a 3-sphere $\gamma$ as 
\be\label{eq:Volume3Sphere}
\tn{Vol}\,\gamma (f_1,\dots,f_k;p_1,\dots,p_k)=\sum_{i=1}^kf_i(p_i)\,.
\ee
so that the resulting contribution to the superpotential is
\be\label{eq:Const}
\Delta W = \frac{1}{\Mgut^{k-3}}\sum_\gamma n_\gamma e^{-\sum_{i=1}^kf_{i}(p_{a_i})}\,. 
\ee
The scale $\Mgut$ is set by the vev of $\phi$ (see our discussion of this in section \ref{sec:scales}).
Note that there can in general exist several flow trees connecting matter localized at the same loci $p_i$, which can cancel out. 

\begin{figure}
	\centering
	\includegraphics[scale=0.8]{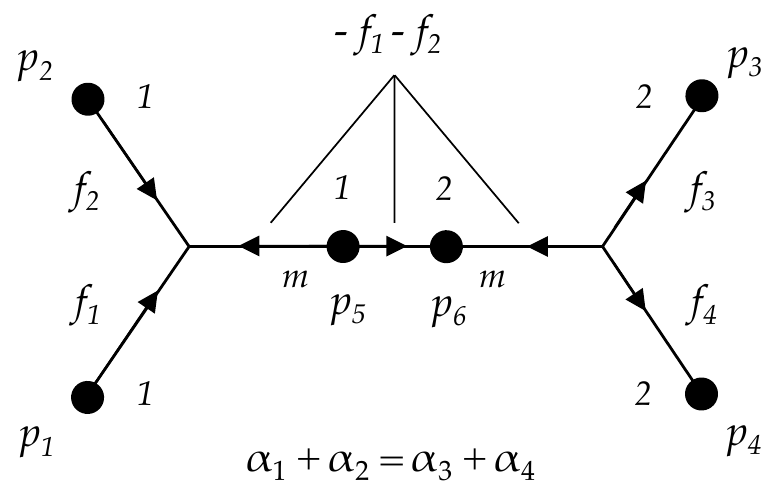}
	\caption{{The figure shows how perturbative ground states participate in 4-point interaction between states localized at $p_1,p_2,p_3,p_4$. The gradient flow tree consists of two trivalent trees connected by a gradient flow line between $p_5$ and $p_6$. We have indicated the relevant Morse functions and Morse indices in the picture. The states localized at $p_5$ and $p_6$ are lifted by instanton corrections and develop a mass $m\sim M_{\tn{Inst}}$. In the 4d effective field theory this gives rise to a 4-point interaction.}}
	\label{fig:IntegrateOut}
\end{figure}

The modes participating in the Yukawa (and higher) couplings are not just the massless states, but in fact all perturbative ground states of the SQM. Below the mass scale 
\be
M_{\tn{Inst}}\sim  \Mgut e^{-tV}\,
\ee
induced by associates over flow lines between two points, we may integrate out the corresponding massive fields, thereby generating higher-dimensional operators in the effective field theory. As $M_{\rm Inst}\ll \Mgut$ these corrections are dominant compared to the couplings induced between the same fields by associatives. {An example is shown in figure \ref{fig:IntegrateOut}.} 

For non-generic choices of the charge distribution the moduli space of gradient flow lines may increase and \eqref{eq:dimMod} is no longer valid. In this case the moduli space of gradient flow lines is not discrete but of dimension 1 and isomorphic to a circle. In the ALE geometry this corresponds to a continuous family of associative submanifolds. In \cite{Beasley:2003fx} it is shown that the contribution of such a family $\mathcal{C}$ of associatives is proportional to $\chi(\mathcal{C})$. 

In the more generic set-up of unfactored spectral cover of rank $n$ the Higgs background can only be diagonalised locally as in \eqref{eq:GeneralBackgroundLocal}. The source terms $\rho$ in the BPS equations are now oriented arbitrarily along $G_\perp$ breaking the gauge symmetry to its commutant $\gut$ in $\gau$. Assuming we can diagonalising the Higgs field in a tubular neighbourhood $T$ of the singularity as
\be
U(x)\phi(x) U^{-1}(x)=\diag(\lambda_1(x),\dots,\lambda_n(x))\quad x\in T\supset \tn{supp}\,(\rho)\,,
\ee
we can impose boundary conditions on our field content as in section \ref{sec:U1SymMod} and proceed with a local analysis. Chiral multiplets still localize at the vanishing points of the Higgs field and the boundary conditions again preclude perturbative zero modes from localising at the boundary. Due to the mixing of the sheets of the spectral cover the background is no longer determined by a set of globally defined functions and we can not relate the cohomologies of $\CD$ counting the zero modes to de Rham cohomologies. The local geometric picture however persists, all interactions are determined by three-cycles of the ALE geometry as in the previous sections with strengths determined by their volumes as in \eqref{eq:SubPotCont}.

Finally, let us briefly comment on the case in which the critical loci are circles, i.e. we are allowing $f_Q$ to be Morse-Bott. Perturbing the set-up slightly we return to the case of Morse theory. The ground states of \eqref{eq:MorseBottGS} now decompose into multiple perturbative ground states
\be
\alpha\wedge \psi_{\perp}\quad \rightarrow \quad \frac{1}{\sqrt{n}}\, \eta_i\quad i=1,\dots,n\,,
\ee
where we have assumed that the circle decomposes into $2n$ critical point of which $n$ have Morse index $1$ and $n$ have a Morse index of $2$. The forms $\eta_i$ are $1,2$-forms depending on whether $\alpha$ is a  $0,1$-form and localize at these critical points of Morse index $1,2$ respectively. We are further assuming that the states $\alpha\wedge \psi_\perp$ and $\eta_i$ are of unit norm. After this perturbation, the  previous analysis applies. The true ground state corresponding to $\alpha\wedge \psi_{\perp}$ is
\be\label{eq:TrueGS}
\frac{1}{\sqrt{n}}\sum_{i=1}^n \eta_i\,. 
\ee

Finally, let us apply these observations to the TCS constructions.
In \cite{Braun:2018fdp} a chain of string dualities was used to argue for the existence of infinitely many associatives on a class of TCS $G_2$-manifolds, and this result was recovered in an orbifold limit in \cite{Acharya:2018nbo}. These associatives are furthermore in one-to-one correspondence with elements of the lattice $E_8 \oplus E_8$. The local limits of these models must be such that $\tilde{G} = G_\perp = E_8 \times E_8$, and the associatives argued for in \cite{Braun:2018fdp,Acharya:2018nbo} must correspond to flow trees in these local models. In fact, the description of the associatives in terms of string junctions in \cite{Braun:2018fdp,Acharya:2018nbo} is already deceptively close to our description in terms of flow trees. It would certainly be interesting to flesh out this correspondence in detail. 

%%%%%%%%%%%%%%%%%%%%%%%%%%%%%%%%%%%%%%%%%%%%%%%%%%%%%%%%%%%%%%%%%%%%%%%%%%%%%%%%%%%%%%%%%%%%%%%%%%%%%%%%%%%%%%%%%%%%%%%%%%%%%%%%%%%%%%%%%%

\section{Higgs Bundles and Twisted Connected Sum $G_2$-manifolds}
\label{sec:Snickers}

In this section we consider local models associated with twisted connected sum (TCS) $G_2$-manifolds, which form the largest known class of examples of compact $G_2$-manifolds \cite{MR2024648,Corti:2012kd}. The TCS construction has by now been covered extensively in the literature, so we will only briefly recapitulate the main points and refer the reader to \cite{MR2024648,Corti:2012kd,Braun:2017uku} for further details. In a nutshell, the power of the TCS construction is that it shows how compact $G_2$-manifolds can be glued from simpler building blocks, which can in turn be constructed using algebraic geometry. Although this makes finding examples relatively straight-forward, TCS $G_2$-manifolds appear to be a rather special class within the set of all $G_2$-manifolds \cite{Braun:2017uku}. Our analysis of local models for $G_2$-manifolds allows us to move away (at least in local models) from the TCS description and explore how to connect TCS $G_2$-manifolds to $G_2$-manifolds giving rise to chiral spectra.

\subsection{TCS $G_2$-Manifolds}

\begin{figure}
	\begin{center}
		\includegraphics[width=8cm]{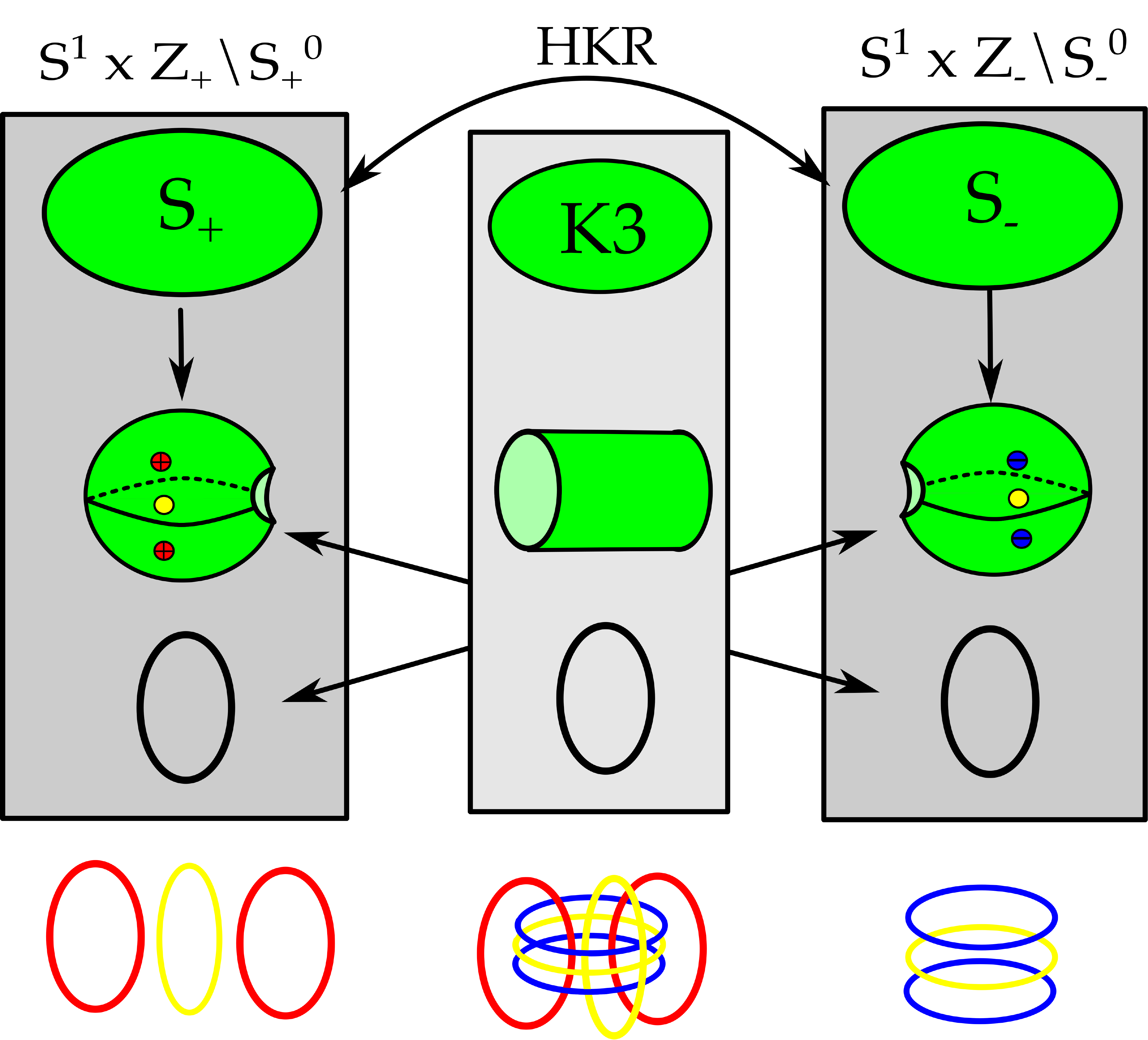}
		\caption{TCS construction of $G_2$-manifolds. Top: Building blocks that are aCyl Calabi-Yau and hyper-K\"ahler rotation (HKR) in asymptotic cyclindrical region. Bottom: Higgs bundle data.
			The critical loci of the Morse-Bott function $f$ (in yellow) and the charge configuration $\rho$ (in red and blue) corresponding to the local limit of a TCS $G_2$-manifold. The figure on the top shows the decomposition of $S^3$ into $\C_\pm \times \s^1_\pm$ and the figure on the bottom shows the location of the same critical loci and charges in a patch $\R^3$ of $S^3$. Every every circle in $X_+ \times \s_+^1$ has linking number one with any of the circles in $X_+ \times \s_+^1$. Note that charge conservation requires that not all loops carry identical charges in this example. }
		\label{fig:TCS_kovalev_config}   
	\end{center}
\end{figure}

The basic ingredient for the twisted connected sum construction is a pair of algebraic threefolds $Z_\pm$, which each admit a K3 fibration
\be
\ba
S_\pm\longrightarrow &  Z_\pm \cr 
& \downarrow {\pi_\pm} \\
&\,  \P_\pm^1
\ea
\ee  
with generic K3 fiber $S_\pm$. The manifolds $Z_\pm$ have to satisfy
\be
c_1(Z_\pm) = [S_\pm]\,,
\ee
i.e. the first Chern class of $Z_\pm$ must be equal to the class of a generic K3-fiber. With some further assumptions on the topology (see \cite[Definition 3.5]{Corti:2012kd}) $Z_\pm$ are then called the building blocks. Excising a generic fiber $ S^0_\pm$ from $Z_\pm$ one obtains a pair of non-compact threefolds $X_\pm = Z_\pm\setminus S^0_\pm$, fibered over a punctured Riemann sphere,
\be
\ba
S_\pm\longrightarrow &  X_\pm \cr 
&\downarrow {\pi_\pm} \cr 
& \C_\pm 
\ea\,,
\ee
which are asymptotically cylindrical (aCyl) Calabi-Yau threefolds. Away from a compact submanifold, the $X_\pm$ have the topology of the cylinder $\R^+\times \s_{b,\pm}^1\times S^0_\pm$ and the Ricci-flat metrics on $X_\pm$ asymptote to the Ricci-flat product metric on this cylinder. The situation is sketched in figure \ref{fig:TCS_kovalev_config}.

%%%%%%%%%

%%%%%%%%%

To form a compact $G_2$-manifold, the aCyl Calabi-Yaus $X_\pm$ are then multiplied by an extra circle $\s^1_{e,\pm}$ and glued together along a their cylindrical regions. The diffeomorphism used for the gluing exchanges the `internal' circles $\s_{b,\pm}^1$ with the `external' circles $\s_{e,\mp}^1$ and identifies the K3 surfaces $S^0_\pm$ by a diffeomorphism which induces a \hk rotation, or Donaldson matching,
\begin{equation}\label{eq:matchball_Donaldson}
\begin{aligned}
\Re(\Omega_\pm^{2,0}) &= \omega_\mp \\  
\Im(\Omega_+^{2,0}) &= -\Im(\Omega_+^{2,0}) \, .
\end{aligned}
\end{equation}
Here, $\Omega_\pm^{2,0}$ and $\omega_\pm$ are the holomorphic $(2,0)$ forms and K\"ahler forms on $S^0_\pm$ which are induced by the complex structures on $X_\pm$.  The compact topological manifold $J$ resulting from this gluing then admits a metric with holonomy $G_2$, which is close to the Ricci-flat metrics on $X_\pm \times \s_{e,\pm}^1$. More precisely, there exists a limit, which we will call `Kovalev limit', in which the cylindrical region becomes arbitrarily long and in this limit the Ricci flat $G_2$-holonomy metric approaches the Calabi-Yau metrics on $X_\pm$. In compactifications of M-theory, modes localized only on $X_+ \times \s^1_{e,+}$ or $X_- \times \s^1_{e,-}$ give rise to subsectors with enhanced $\mathcal{N} = 2$ supersymmetry in four dimensions. These subsectors are coupled such that they mutually only preserve $\mathcal{N} =1$ supersymmetry, and we may think of the length of the cylindrical region as the inverse of their coupling \cite{Guio:2017zfn,Braun:2017csz}.

As both $X_\pm \times \s_{e,\pm}^1$ are fibered by K3 surfaces and the gluing acts separately on the fiber and base, $J$ is (topologically) fibered by K3 surfaces as well. The base of this fibration is a three-sphere $S^3$ glued together from two solid tori. We can hence think of the local models associated with TCS $G_2$-manifolds as describing an ALE space which is cut out from the K3 fiber over a base space $M_3$ which is $S^3$. To engineer non-abelian gauge groups, every ALE fiber of the local geometry and hence every K3 fiber of the associated compact $G_2$-manifold must be singular. It is straightforward to construct acyl Calabi-Yau threefolds in which every K3 fiber has a singularity of ADE type and the work of \cite{Braun:2017csz,Braun:2017uku} suggests that gluing such singular three-folds indeed results in a singular $G_2$-manifold. 

Let us consider this in more detail. Denote the image of 
\be\label{TwoFormGlue_rho}
\rho_\pm: H^2(Z_\pm,\Z)\rightarrow H^2(S_\pm^0,\Z)
\ee
by $N_\pm$. The Donaldson matching implies an idenfication of $H^2(S_+^0,\Z) \simeq H^2(S_-^0,\Z)$. Using this map, every element of 
\be\label{TwoFormGlue}
\mathfrak{g} =   N_+ \cap N_-\, ,
\ee
gives rise to an associated harmonic two-form on $J$: the Poincar\'e dual cycle to such a form is algebraic for both $S_+$ and $S_-$, so that its fibration over the whole base $S^3$ of $J$ is trivial and it sweeps out a five-cycle, which is Poincar\'e dual to a two-form on $J$. The number of independent such two-forms on $J$ is simply given by the rank of $\mathfrak{g}$ \cite{Corti:2012kd}. In compactifications of M-theory on $J$, there are hence $|\mathfrak{g}|$ massless $U(1)$ vectors from the Kaluza-Klein reduction of the three-form $C_3$\footnote{There are in general further massless $U(1)$ vectors associated with classes in the kernel of $\rho_\pm$ \cite{Corti:2012kd}, which associated with the irreducible components of reducible fibers of the K3 fibrations on $Z_\pm$.}. 

The hyper-K\"ahler structure on $S_\pm^0$ is forced by the Donaldson matching to be such that the integral of both $\Omega^{2,0}_\pm$ and $\omega_\pm$ vanishes for every cycle contained in $\mathfrak{g}$. This means that whenever there is a root, i.e. a lattice vector of length $-2$, contained in $\mathfrak{g}$, the K3 fibers $S_\pm^0$ are singular. As $\mathfrak{g}$ sits inside of the polarizing lattices\footnote{The polarizing lattice of a family of K3 surfaces is the sublattice of $H^2(K3,\mathbb{Z})$ which is orthogonal to $\Omega^{2,0}$ for all members of the family.} of the algebraic families $X_\pm$, this implies that every single K3 fiber has a singularity. The type of singularity can be read of by finding the sublattice $\mathfrak{g}_{\rm root} \subset \mathfrak{g}$ generated by the roots of $\mathfrak{g}$. This sublattice must be a (sum of) ADE root lattice(s) and its type determines the corresponding singularity and the resulting simply-laced\footnote{While this data is sufficient to find the singularities associated with simply-laced gauge groups, it is slightly more tricky to find non-simply laced gauge groups. Their emergence in TCS $G_2$-manifolds parallels their emergence in F-theory \cite{Bershadsky:1996nh} in that the exceptional divisors of resolutions of ADE singularities of $S_\pm$ may globally become a single divisor in $X_\pm$ \cite{Braun:2017uku}. In terms of lattice data, this can be expressed by saying that a cycle of self-intersection $n < -2$ contained in $\mathfrak{g}$ can force an ADE singularity in every fiber if it is a linear combination of $-2$ curves in $S_+$ or $S_-$ which are all in the polarizing lattices of the families $S_+$ and $S_-$. The difference between the polarizing lattices and $N_\pm$ determines the `folding' of the ADE Dynkin diagram.} non-abelian gauge group upon compactification of M-theory. 

The matter loci in these models arise as the degeneration loci of the singular K3-fibration i.e. where the singularity worsens. This happens over points in $\P^1_\pm$, each of which gets multiplied with a circle in the TCS construction. This implies that in M-theory compactification on a TCS manifold $J$, matter is localized along circles. This is true at least in the Kovalev (stretched neck) limit in which the metric on the $J$ is well approximated by the metrics on each of the building blocks, which can be thought of as being contained inside $J$ (more precisely, the products $X_\pm\times \s^1_\pm$ are in $J$).

\subsection{Higgs Bundles of TCS $G_2$-manifolds}

We start by considering the local models of the two building blocks individually. As the discussion is the same for both sides, we will drop the $\pm$ subscripts. The first step is to replace the K3 fibration with a local ALE model. The precise details of this local limit depends on the ADE group corresponding to the type of ADE singularity, and are well known in the literature \cite{Lerche:1996an,Billo:1998yr}. Besides an ADE singularity, every ALE fiber contains a number of compact cycles, the volumes of which vary over the base. Such cycles may collapse over points in the base $\C$. At these loci the singularity present in the generic fiber is enhanced and matter is localized. Let 
\be
\sigma \in H_2(S,\Z)
\ee
be such cycle which vanishes at (some) of the corresponding points in the base $\C$. Let us denote the \hk triple by $\Theta = (\omega_{I},\omega_{J},\omega_{K})$. In terms of \eqref{eq:matchball_Donaldson} the \hk structure is simply
\be
\ba
\omega &= \omega_{I}\\
\Omega^{2,0} &= \omega_{J} + i \omega_{K}.
\ea
\ee
After taking the local limit and integrating over $\sigma$ we get a meromorphic function $\phi$ on $\C$:
\be
\phi = \int_\sigma \Theta,
\ee
with zeros precisely where $\sigma$ shrinks to zero volume. The poles and zeros of $\phi$ are located away from $\infty$. Moreover, we can identify $\phi$ with the meromorphic (1,0)-form as $\phi\, dz$. Since the base $\C$ is contractible $\phi = df$, where $f$ is now a Morse function with critical points of index 1 and singular loci corresponding to the poles of $\phi$. After taking a product with the circle we trivially get a Morse-Bott function. 

If (unit) charges are placed at points $a_i\in \C$, the function $\phi$ will be of the form
\be
\phi(z) = \sum_{i=1}^n \frac{1}{z-a_i}\,.
\ee 
Therefore $\phi$ can have at most $n-1$ critical loci, which is generically the case. If we impose charge conservation on each side, there can be at most $n-2$ critical loci of $\phi$.

With this information let us now consider how the Higgs bundle for TCS manifolds. After gluing $\C_+ \times \s^1_{e,+}$ with $\C_- \times \s^1_{e,-}$, the base manifold is $M_3=S^3$. In the Kovalev limit, the critical locus of the harmonic Morse-Bott function $f$ consists of a disjoint union of $m$ circles of Morse index 1. As before, we may engineer such an $f$ by an appropriate configuration of charges $\Graph$ on $S^3$. On $\C_\pm \times \s_\pm^1$, these charges will simply be given by a collection of points on $\C_\pm$ times the circle $ \s_\pm^1$.

From the above discussion we only need the simple observation that matter loci in TCS $G_2$-manifold, at least in the Kovalev limit, are circles. Suppose that there are $m$ matter circles and no points. Using the results of section \ref{sec:1dmatter} we see that the Morse-Bott complex is
\be
\qquad C^1=\Omega^0(\s^1)^m\,, \qquad C^2=\Omega^1(\s^1)^m\,,
\ee
and the cohomology gives just
\be
\label{eq:morsebottresult}
H^1(\tM3, \Sboundary_-)\cong \R^m\,, \qquad H^2(\tM3, \Sboundary_-)\cong \R^m\,.
\ee
We find that every perturbative ground state constitutes a true ground state, the Morse-Bott function $f$ is thus perfect. As each circle gives rise to a pair of chiral and conjugate-chiral zero modes upon Kaluza-Klein reduction, the spectrum associated to this Higgs field configuration $\phi=df$ is non-chiral
\be
\chi(\tM3,{\bf R}_q)=0\,.
\ee 
We can use this result to derive constraints on the function $f$. By the above results the relative Euler characteristic $\chi(\tM3,\Sboundary_-)=0$ vanishes and by Lefschetz duality we find that this implies $\chi(\tM3,\Sboundary_+)=0$. We obtain the topological constraint
\be
\label{eq:topconstraint}
\chi(\tM3)=\chi(\Sboundary_+)=\chi(\Sboundary_-) =0\,.
\ee

There has been a recent attempt to modify the TCS construction to yield singular $G_2$-manifolds with codimension 6 singularities by Chen \cite{Chen}. Instead of smooth bulding blocks Chen takes the $Z_+$ building block to be a threefold with isolated nodal singularities, which means that the non-compact aCyl $G_2$-manifold $X_+\times\s^1_{e+}$ has singularities in codimension 6. However, the standard TCS gluing argument does not work in this case; rather it is conjectured \cite{Chen} that if circles of nodal singularities are replaced by pairs of isolated conical singularities it is possible to glue to a $G_2$-manifold with conical singularities using a modified version of the connected sum construction. In terms of the local model, the collapse of circles into points corresponds to deforming the Morse-Bott function to a generic Morse function, where the same collapse of critical circles to critical points occurs (recall that critical points correspond precisely to isolated singularities of the total space of the $G_2$-manifold). However, even if this conjecture is true, such $G_2$-manifold will still give rise to a non-chiral spectrum by the arguments above.

Finally, let us discuss the spectral covers for a TCS $G_2$-manifold which is given to us in terms of building blocks $X_\pm$ and a gluing map 
\be
\gamma: \qquad S_+^0 \rightarrow S_-^0\,,
\ee
where $S_\pm$ have ADE singularities over $U_\pm \subset \mathbb{C}_\pm$. This gluing of the K3 fibers in the TCS geometry also implies a consistent gluing map for the ALE-fibrations associated with the local model. In general, to be able to glue two given ALE-fibrations together, the two ALE-spaces need to be of the same type, $\tilde{G}_+ = \tilde{G}_-$, and furthermore the periods of the ALE-fibers must satisfy a matching condition. By Torelli theorem for ALE-spaces \cite{Kronheimer:1989pu}, the structure of an ALE-space is completely determined by the periods of the \hk structure forms over the 2-cycles in the root lattice of the algebra of $\widetilde{G}$. Explicitly, the matching condition is 
\be
\int_{\sigma_j}\omega_{I,+}	= \int_{\sigma_j}\omega_{J,-}\,,\qquad \int_{\sigma_j}\omega_{J,+}	= \int_{\sigma_j}\omega_{J,-}\,, \qquad \int_{\sigma_j}\omega_{K,+}	= -\int_{\sigma_j}\omega_{K,-}\,,
\ee
where $\sigma_j$ are the 2-cycles generating the root lattice. Note that this implies that the non-abelian part of the group $G$, i.e. the type if ADE singularity, must be the same on both sides. 

Each $X_\pm$ furthermore has a local model, which is a Higgs bundle $\phi_{(\pm)}$ over $\mathbb{C}_\pm$, and a corresponding spectral cover $\mathcal{C}_{(\pm)}$. The asymptotic values the Higgs fields $\phi_{(\pm),0}$ are similarly related by 
\be
\phi_{(+),0} = \gamma^\ast \phi_{(-),0} \,,
\ee
which induces a gluing of the spectral covers.  
%%%%%%%%%%%%%%%%%%%
\begin{figure}
	\centering
	\includegraphics[width=8cm]{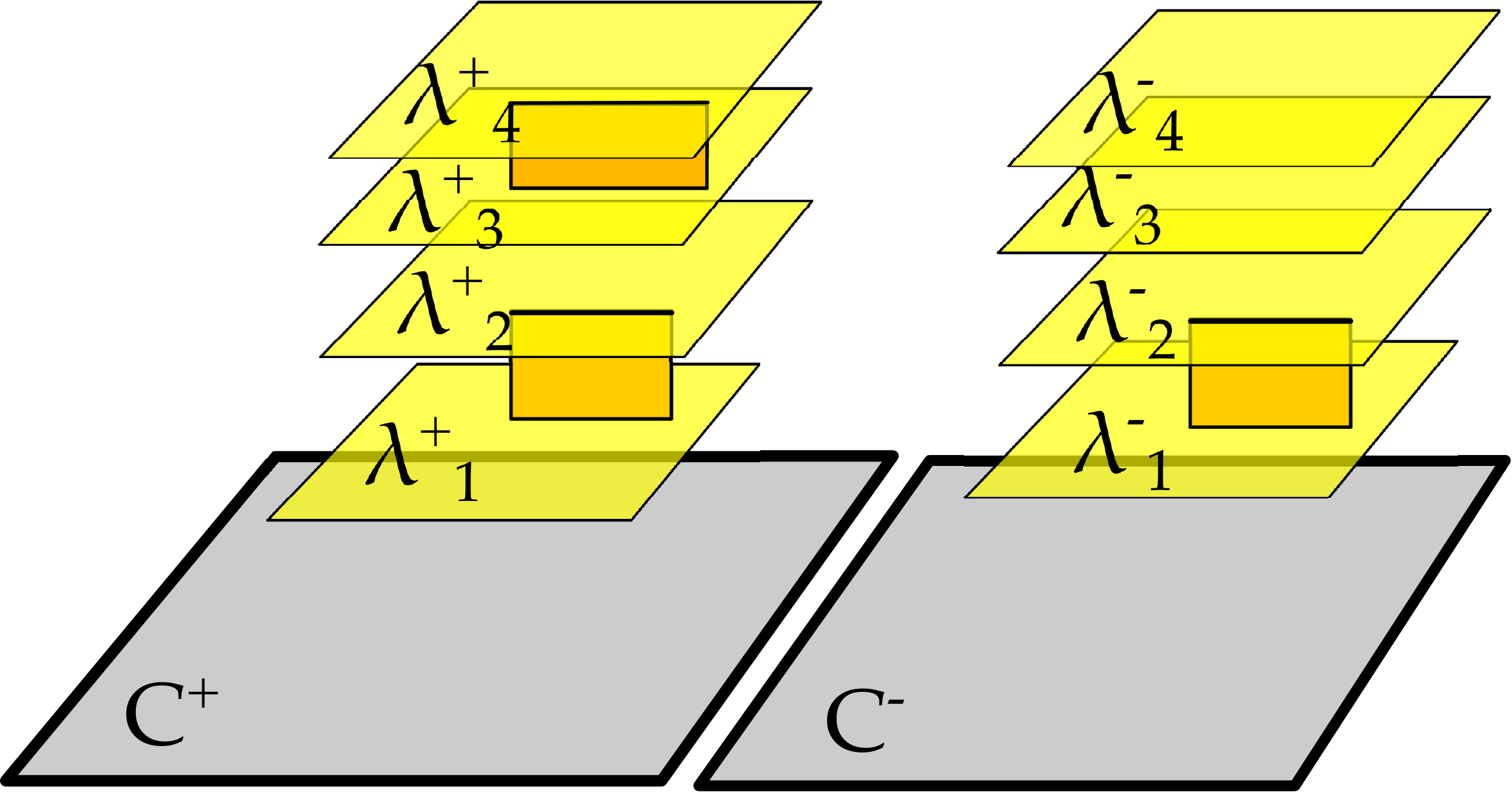}
	\caption{Each of the building blocks $X_\pm$ defines a spectral cover over $\C_\pm$, and these are then glued to a spectral cover over $S^3$. In the example shown here, the cover on $\C_+$ factors into two components and the cover on $\C_-$ factors into three components. These covers are glued such that the resulting spectral cover over $S^3$ has two components with two sheets each. Hence the resulting model has $G_\perp = S[U(2) \times U(2)]$ and there is a single unbroken $U(1)$.}
	\label{fig:SCTCS}
\end{figure}
%%%%%%%%%%%%%%%%%%%

Let us explain the origin of $U(1)$ gauge symmetries in glued spectral covers. Each cover $\mathcal{C}_{(\pm)}$ can have a factorization structure, which defines two-forms (five-cycles) and locally $U(1)$ symmetries. This can be detected by the restriction of the map (\ref{TwoFormGlue_rho}) to the ALE-fibrations over $\C_\pm$. Factorization of the spectral cover $\mathcal{C}$ over $S^3$ after gluing $\mathcal{C}_{(\pm)}$ can likewise be detected by (\ref{TwoFormGlue}), and only those two-forms in the image that lie in the intersection will globally give rise to a two-form and thereby a $U(1)$ symmetry. An example is shown in figure \ref{fig:SCTCS}, where $\mathcal{C}_{(\pm)} \rightarrow \mathbb{C}_\pm$ each is a four-sheeted cover. However $\mathcal{C}_{(\pm)}$ is factored into two (three), and thus locally gives one (two) $U(1)$ symmetries. The gluing is such that the spectral cover $\mathcal{C} \rightarrow S^3$ has only two factors, and thus only gives rise to a single $U(1)$ symmetry. The scale at which the other $U(1)$ is broken is set by the size of the neck region of the TCS-construction.

Besides an analysis via Higgs bundles, the matter spectrum of M-theory on TCS $G_2$-manifolds can also be found using a purely geometric reasoning. The geometry in the vicinity of each matter locus is that of a Calabi-Yau threefold times a circle. The local Calabi-Yau geometry is that of a fibration of an ADE singularity over $\C$ with a further degeneration at a point. Using the usual dictionary between singularities and gauge theory for M-theory or type IIA on Calabi-Yau threefolds, the Cartan generators and weight vectors can be identified with exceptional divisors and curves in the resolved Calabi-Yau geometry \cite{Katz:1996xe,Intriligator:1997pq,Morrison:2011mb,Marsano:2011hv}. Our analysis of Higgs bundles now implies that the multiplicities must be such that each matter locus gives rise to a single vector-like pair of representations. Furthermore, we may determine the $U(1)$ charges by simply integrating the two-forms in $\mathfrak{g}$ which give rise to the $U(1)$s over the exceptional curves of the resolution associated with the matter.

\subsection{Deformation of TCS Higgs Bundles}

Given the local model for TCS $G_2$-manifolds we now consider the behavior of the physics under deformations of the Morse-Bott function. We have seen above that circular critical loci arise in the non-generic $\s^1$-invariant distributions of charges in $S^3$ which are present in the Kovalev limit. The natural question is what happens if this invariance in is broken by a slight deformation. The strategy we will use to describe deformations is to exploit the construction of Morse(-Bott) functions in terms of charge distributions. For every charge distribution $\rho$, there is an associated Morse-Bott function, which in turn lifts to an ALE-fibration, our local model of a $G_2$-manifold. For every deformation of the charge distribution there is hence an associated deformation of the local model. Note that this deformation might be trivial: contrary to the number of deformations of Higgs bundles or the deformations of $G_2$-manifolds, which are finite in number, there are infinitely many deformations of any given charge configuration.

\begin{figure}
	\centering
	\includegraphics[scale=0.4]{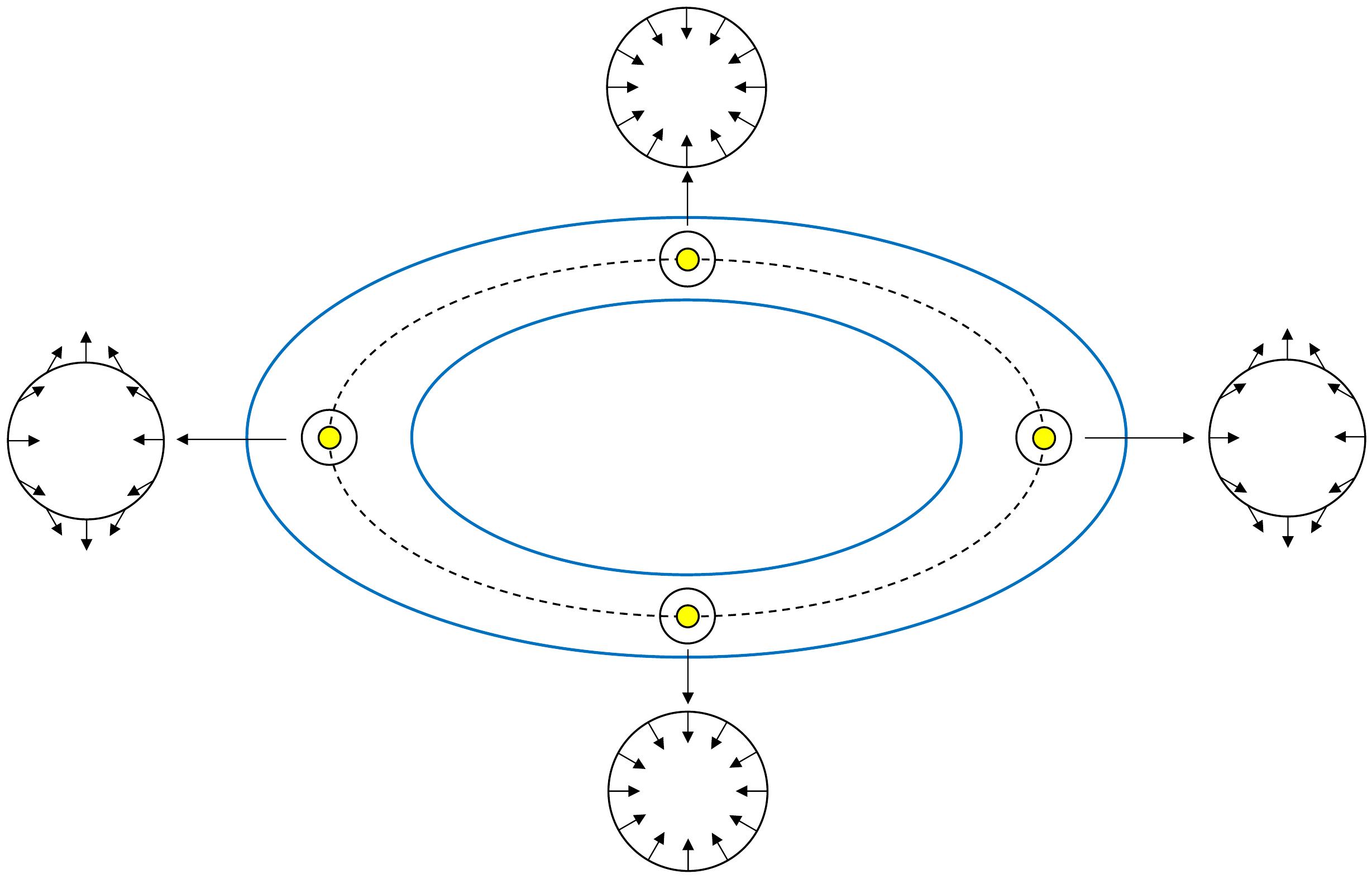} \hspace{1em} 
	\caption{Two homogeneously negatively charged coplanar circles which have been stretched to ellipses. On the ellipse which is equidistant to the two charge loci, the electric field only vanishes at the four points marked as black dots. To find the Morse index, consider the small circles (coplanar to the ellipses) around these points and the restriction of the electric field to these circles. The Morse index of the points near the vertices of the ellipse is 2 and while those near the co-vertices have a Morse index of 1. As before, the critical loci are shown in yellow, and the charges in blue.}
	\label{fig:deformedchargelocus}
\end{figure}

A configuration of charges which produces the Morse-Bott function associated with a TCS $G_2$-manifold in the Kovalev limit must of course be finely tuned, as a generic configuration of charges will always result in critical loci of dimension zero. Let us discuss this in a simple example -- see figure \ref{fig:deformedchargelocus}: consider a charge distribution of two equally charged coplanar and cocentric circles in $\R^3$. This setup has rotational symmetry and correspondingly the critical locus is another coplanar and concentric circle. A generic deformation will destroy the rotational symmetry and lead to critical points instead of a circle. Consider e.g. deforming the charges to ellipses while preserving coplanarity. This collapses the critical locus to two points of Morse index 1 near the vertices of the ellipses and two points of index 2 near the co-vertices.

More generally, the function $f$ will become Morse with isolated critical points for a generic deformation. However, since the topology of $\Sboundary_\pm$ does not change, we still have
\be 
\chi(\tM3,\Sboundary_-) = 0\,.
\ee
Physically this means that any deformation of $f$ will give rise to chiral spectrum  if under the deformation the topology of $\Sboundary_-$ remains unchanged. Denoting the number of points with Morse index $i$ by $m_i$, the Morse inequalities for manifolds with boundary imply
\be
\chi(\tM3,\Sboundary_-) = m_2 - m_1 = 0\, .
\ee
Equally, every deformation of the local model of a TCS $G_2$-manifold that has an associated charge distribution which consists of a number of circles will satisfy $n_\pm = l_\pm$, so that the resulting spectrum is seen to be non-chiral.

\subsection{Chirality and Singular Transitions}

It is not at all surprising that TCS $G_2$-manifolds do not give rise to chiral spectra and that small deformations do not change the chiral index. However the result we have found already has fairly interesting geometrical implications: for a generic small deformation of a TCS $G_2$-manifold, the loci at which matter is localized are no longer one-dimensional but become point-like. This of course implies that the product structure of $X_\pm \times \s^1_{e,\pm}$ must be broken and the periods of the hyper-K\"ahler triplet on the K3 fiber must have a non-trivial dependence along $\s^1_{e,\pm}$. Although such small deformations will not yield $G_2$-manifolds giving rise to chiral spectra, the crucial ingredient, which are point-like singularities, is already present for small deformations of TCS $G_2$-manifolds.

Engineering the ALE-fibration from a Morse function which in turn is determined by a configuration of charges allows us to make the key observation for how to deform TCS $G_2$-manifolds to situation with chiral spectra: we need to make a transition after which $n_\pm = l_\pm$ no longer holds. The simplest way to do so is to bring two circles of equal charge together and then deform them to an object with $l = 2$. For TCS $G_2$-manifolds, there are essentially two different ways to achieve this. 

\begin{figure}
	\centering
	\includegraphics[width=11cm]{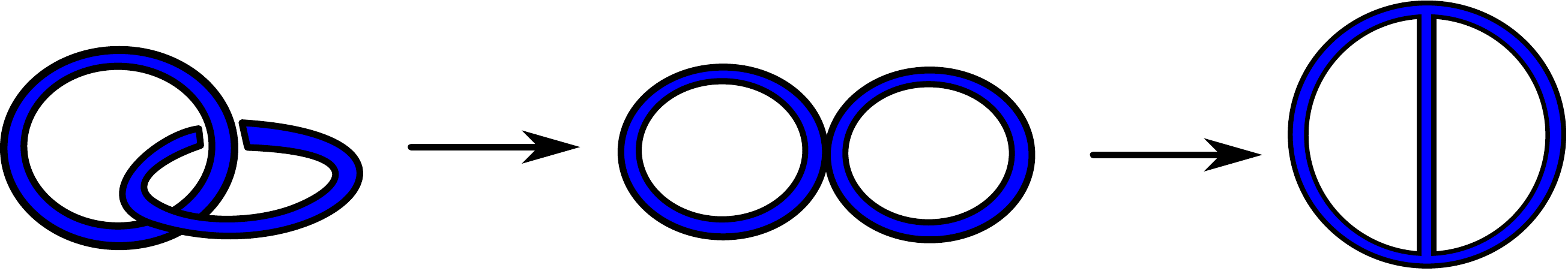} 
	\caption{A transition of the charge configuration which results in a transition between a non-chiral and a chiral spectrum. Starting from a TCS configuration of charges deforming the configuration to one that results in a chiral spectrum.  }
	\label{fig:circle_transition}
\end{figure}

The first option is to take e.g. a positive charge on $\C_+$ and another positive charge on $\C_-$, bring them together, and fuse them as shown in figure \ref{fig:circle_transition}. 
%%%%%%%%%%%%%%%%%
%%%%%%%%%%%%
As now $l_+ - n_+ = 1$ while $l_- - n_-=0$, the resulting spectrum must be chiral. In a generic situation in which $f$ is Morse, i.e. $f$ only has isolated critical points, the critical locus of $f$ hence consists of an odd number of points now. As we started from a non-chiral configuration with an even number of critical points, this implies that some of the critical points must have fused. As the circles of positive charge we have fused originated from different ends of the TCS $G_2$-manifold we started from, the critical points which have fused must likewise originate from different ends. Geometrically, these critical points are nothing but degeneration loci of the K3 fibration of the $G_2$-manifold, so that we have effectively taken specific singular fibers of the K3 fibration into what used to be cylindrical region of the TCS and collided them. As expected from our earlier statement about the absence of chiral spectra in TCS $G_2$-manifolds, this signifies a definite departure from the TCS set-up, where the K3-fibration must be constant in the cylinder region. 

In fact, the type of transition we have just sketched can also be anticipated from the heterotic duals of TCS $G_2$-manifolds, which are given by compactifications on the Schoen Calabi-Yau threefold with different vector bundles \cite{Braun:2017uku}. Such models always have non-chiral spectra and a singular transition connecting the Schoen Calabi-Yau threefold to a different Calabi-Yau threefold (together with appropriate vector bundles) is needed to find a chiral spectrum. The Schoen Calabi-Yau threefold can be described as a fiber product of two $dP_9$s, and it allows singular transitions in which a singular fiber of one $dP_9$ is collided with a singular fiber of the other $dP_9$. As discussed in \cite{Braun:2017uku,Braun:2018fdp}, the duality to a TCS $G_2$-manifold implies that the singular fibers of these two $dP_9$s are separated into disjoint regions of the common $\P^1$ base. A collison between singular fibers from both ends translates to a collision of singular K3 fibers coming from the two separate ends $X_+$ and $X_-$ of the dual TCS $G_2$-manifold. 

The second option is to change the charge configuration corresponding to a TCS $G_2$-manifold by colliding two circles of equal charge which are both located in the same building block. The picture of such a deformation will be similar to the one in figure \ref{fig:circle_transition}, however initially the charged circles will be unlinked. Again, it is clear that this signals a departure from a TCS $G_2$-manifold (and must result in a singular transition on the heterotic side as well): after the transition e.g. $X_+ \times \s^1_{e,+}$ must become a non-compact $G_2$-manifold without the structure of a product.

%%%%%%%%%%%%%%%%%%%%%%%%%%%%%%%%%%%%%%%%%%%
%%%%%%%%%%%%%%%%%%%%%%%%%%%%%%%%%%%%%%%%%%%

%%%%%%%%%%%%%%%%%%%%%%%%%%%%%%%%%%%%%%%%%%%
%%%%%%%%%%%%%%%%%%%%%%%%%%%%%%%%%%%%%%%%%%%

\section{Higgs Bundles for $G_2$s: A User's Manual}
\label{sec:Manual}

We will now give a user-friendly summary of how to build local Higgs bundle models for $G_2$ manifolds, stripping off most of the mathematical baggage and condensing it to the essentials, which might be useful for the practitioners in the field. 

\subsection{Scales}
\label{sec:scales}
Let us briefly discuss the mass scales in the problem, and specify what scale separation gives rise to the decoupling of gravity. For this purpose consider the gauge theory on the associative three-cycle $M_3$ on which the gauge degrees of freedom are localized. The compactification geometry determines the size of the cycle $M_3$ and the volume of the $G_2$-manifold $J$
\be
\hbox{Vol}(M_3) \sim R_{\rm M_3}^3 \,,\qquad 
\hbox{Vol}(J) \sim  R_{G_2}^7 \,,
\ee
and the characteristic size $R_\perp$ of the directions transverse to $M_3$ in $J$ is hence
\be
R_\perp^4 \sim R_{G_2}^7/ R_{\rm M_3}^3 \, .
\ee
In terms of these length scales, the Plank masses $M_{\rm 4}$ in 4D and $M_{\rm 7}$ in 7D are given by
\be
M_{\rm 4}^2 \sim M_{11d}^9  R_{G_2}^7 \,,\qquad M_{\rm 7}^5 \sim M_{11d}^9  R_\perp^4 \, .
\ee
The scale $\Mgut$ at which $\tilde{G}$ is broken to $G$ is set by the an appropriate average of the Higgs background. As the volumes of the compact cycles in the ALE fiber over $M_3$ are set by $\langle \phi \rangle$, the local limit corresponds to the limit in which we can decouple gravity from the gauge theory degrees of freedom:
\be
\Mgut \ll M_{\rm 7} \, .
\ee
Approximating the effective physics by the gauge degrees of freedom is only valid if the two scales $\Mgut$ and $M_{\rm 7}$ can be decoupled and permit the limit $M_{\rm 7}\rightarrow \infty$ while keeping $\Mgut$ constant. Keeping $\phi$ fixed, this limit is equivalent to shrinking $M_3$ inside of $J$. 
Finally, the coupling of the 4D gauge theory $G$ is given in terms of 
\be
{1\over \alpha_{\rm GUT}} \sim M_{11d}^3 R_{M_3}^3 \,.
\ee

\subsection{Matter Content and Interactions}

Here we summarise the construction of some simple backgrounds and the resulting matter content and interactions. 

\begin{enumerate}
	\item Choose the rank $n$ of the Higgs bundle, which is equal to the number of Cartan generators  $\mathfrak{t}^i$ along which a Higgs field background $df_i$ has been turned on. Each of these $n$ abelian directions is sourced by a charge distribution $\rho_i$ of different type which determines the background function $f_i:M_3 \rightarrow \R$ completely, and must integrate to zero on $M_3$. The Higgs bundle is therefore given by
	\be
	i=1,\dots,n\,:\qquad\phi=\mathfrak{t}^idf_i\,,\qquad\rho=\mathfrak{t}^i\rho_i\,, \qquad \Delta f_i=\rho_i\,,\qquad \int_{M_3}\rho_i=0\,. 
	\ee	
	\item This background breaks the gauge symmetry $\gau\rightarrow \gut \times U(1)^n$ and determines the count of representations in 
	$\oplus_Q {\bf R}_Q$, where $Q=(q_1,\dots,q_n)$. To count the zero modes in ${\bf R}_Q$ we need the effective charge distribution and its corresponding potential
	\be
	\rho_Q=\sum_{i=1}^n q_i\rho_i\,,\qquad f_Q=\sum_{i=1}^n q_if_i\,. 
	\ee 
	At every point in $M_3$ where $df_Q = 0$, there is a localized chiral multiplet transforming in ${\bf R}_Q$. For every flow line governed by $f_Q$ between a pair of such points, there is a mass term for the associated chiral multiplets. Given a charge configuration $\rho_Q$, the resulting massless spectrum can be described in terms of the numbers $n_\pm^Q$ of positively and negatively charged component, and the total number $\ell_\pm^Q$ of loops. 
	
	The massless spectrum is counted by
	\be\ba
    \#\, \tn{chiral multiplets in\ }{\bf R}_Q &= \ell_+^Q+n_-^Q- r^Q -1\,,\\
	\#\, \tn{chiral multiplets in }\overline{{\bf R}}_{-Q}& = \ell_-^Q+n_+^Q- r^Q -1\,,\\
	\chi_Q&=(l_-^Q-l_+^Q)+(n_+^Q-n_-^Q)\,.
	\ea\ee
	Here $r^Q$ denotes the number of negatively charged loops which are independent in homology when embedded into $\tM3\setminus \rho_Q^+$\,.
	\item Interactions between three 4d chiral fields localized at points $p_s$ transforming in ${\bf R}_{Q_s}$ can only arise if $Q_1+Q_2+Q_3=0$ and if there exists a trivalent gradient flow tree between them. In general, there can be several such flow trees and cancellations between them can occur. Furthermore, there are mass terms for some of the localized zero modes and one needs to integrate out those massive fields to find the Yukawa couplings between the massless fields.
	
\end{enumerate}

\subsection{Retro-Model-Building 1: Top Yukawa}
\label{sec:TopYuk}

We close this section with two retro-inspired model building applications. First we will consider the top Yukawa coupling in an $\mathcal{N}=1$ $SU(5)$ toy model. We take $M_3=S^3$ and $\gau=E_6$, i.e. two abelian directions parametrised by $f_a$ and $f_b$ are turned on. The corresponding decompositions read
\be\ba
E_6 \quad \rightarrow \quad &SU(5)\times U(1)_a\times U(1)_b\,,\\
{\bf 78}  \quad \rightarrow \quad &{\bf 1}_{0,0} \oplus {\bf 1}_{0,0} \oplus {\bf 1}_{-5,-3} \oplus {\bf 1}_{5,3} \oplus {\bf 24}_{0,0} \\ &\oplus {\bf 5}_{-3,3} \oplus \overline{\bf 5}_{3,-3} \oplus {\bf 10}_{-1,-3} \oplus \overline{\bf 10}_{1,3} \oplus {\bf 10}_{4,0} \oplus \overline{\bf 10}_{-4,0}\,. 
\ea\ee
The effective Morse functions are
\be\ba\label{eq:EffectiveMorseFunctions}
{\bf 5}_{-3,3}:&\qquad f_{5}=-3f_a+3f_b \,, \\
{\bf 10}_{-1,-3}:&\qquad f_{10}^{(1)}=-f_a-3f_b \,, \\
{\bf 10}_{4,0}:&\qquad f_{10}^{(2)}=4f_a \,. \\
\ea\ee 
and they are determined by two independent charge distributions $\rho_a$ and $\rho_b$. Using the linear combinations $\rho_1=-\rho_a-3\rho_b$ and $\rho_2=4\rho_a$ we can write the charge vectors $Q_s$ as
\be\ba\label{eq:ChoiceOfCharge}
Q_1=Q_{f_{10}^{(1)}}=(1,0) \,, \qquad
Q_2=Q_{f_{10}^{(2)}}=(0,1) \,. \qquad 
Q_3=Q_{f_{5}}=(-1,-1) \,.
\ea\ee 

In our model, we distribute $n_\pm^{(1)}+1$ and $n_\pm^{(2)}+1$ negative and positive point charges of types $1$ and $2$ throughout $M_3$. This yields generically $n_\pm^{(1)}$ and $n_\pm^{(2)}$ critical points of the functions $f_{Q_1}$ and $f_{Q_2}$ as seen in example \ref{sec:Ex2}. Of these $n_-^{(1)},n_-^{(2)}$ have Morse index  1, and $n_+^{(1)},n_+^{(2)}$ have Morse index 2 with respect to $f_{Q_1}$ and $f_{Q_2}$. 

\begin{figure}
	\centering
	\includegraphics[width= 7cm]{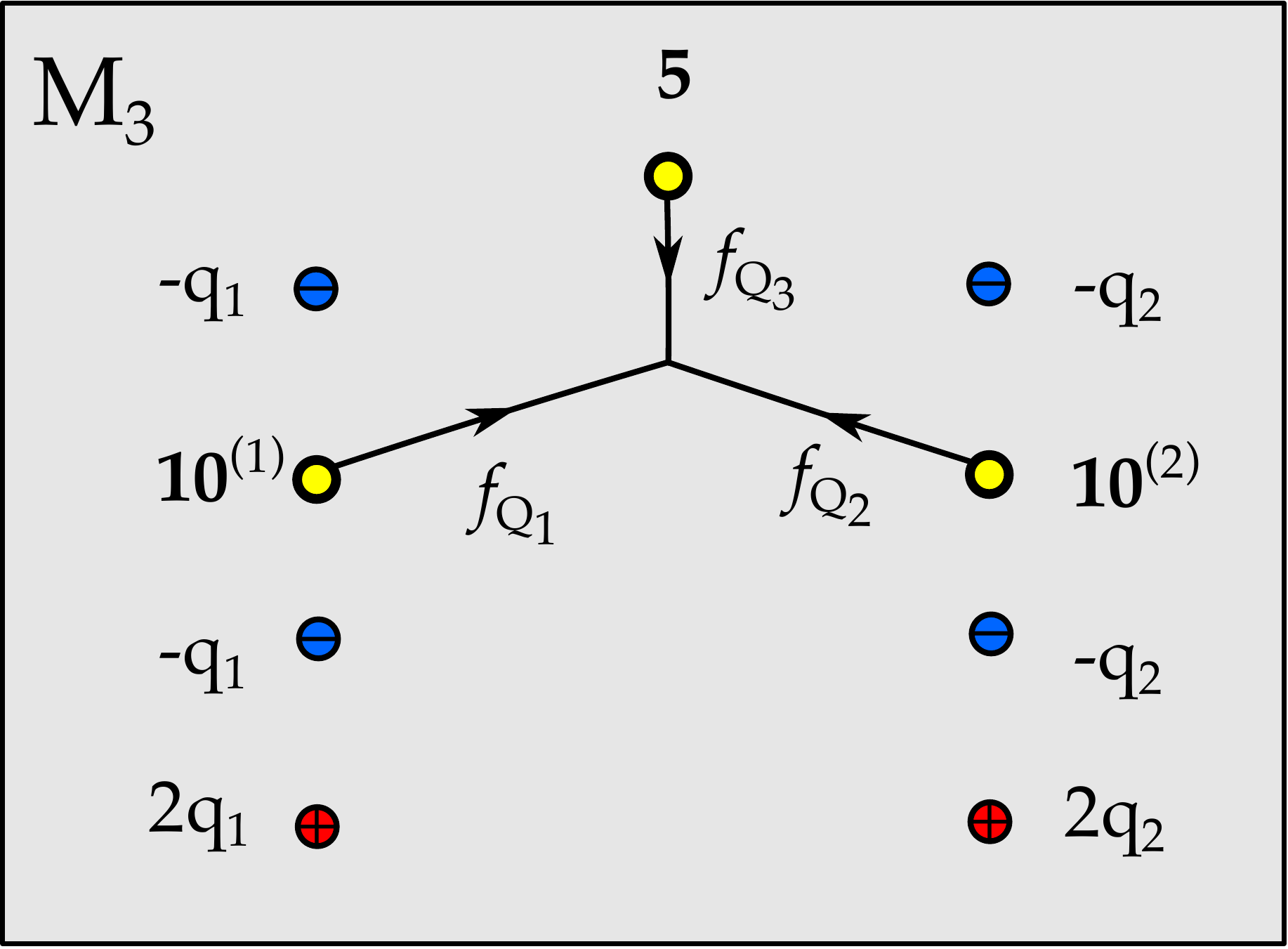}
	\caption{The point charge distributions $\rho_1$ and $\rho_2$ with two types of $U(1)$ charges $(1)$ and $(2)$. Negative charges (blue) and positive charges (red) add to zero for both configurations and they each give rise to critical points (yellow) at which matter transforming in ${\bf 10}^{(1)}$ and ${\bf 10}^{(2)}$ is localized. The charge distribution $\rho_3 = -\rho_1 - \rho_2$ gives rise to a third critical point (yellow) at which matter transforming in ${\bf 5}$ resides. A flow tree between these $3$ critical points gives rise to a Yukawa coupling of type ${\bf 10}^{(1)}\,{\bf 10}^{(2)}\,{\bf 5}$.}
	\label{fig:TopYukawa}
\end{figure}

An example is shown in figure \ref{fig:TopYukawa}. Here we have embedded 3 point charges of type 1 and 2 in $M_3$ such that their total charge vanishes. The charge configurations of type 1 and 2 exhibit a single critical point each (yellow), while their linear combination has a critical point (yellow) in the patch of $M_3$ depicted. Their interaction is determined by the gradient flow tree connecting the critical points and is of type ${\bf 10}^{(1)}\,{\bf 10}^{(2)}\,{\bf 5}$. Denoting the critical points at which the ${\bf 10}^{(1,2)}$ matter localizes by $p_1,p_2$ and the critical point at which the ${\bf 5}$ matter localizes by $p_3$ we have to leading order in $1/t$
\be
\lambda_{({\bf10},{\bf 10},{\bf 5})}\sim e^{-t(f_{Q_1}(p_1)+f_{Q_2}(p_2)+f_{Q_3}(p_3))}\,,\qquad df_{Q_i}(p_i)=0\,.
\ee
So far the analysis has only concentrated on a local patch in $\tM3$ and in principle there may exist further gradient flow trees connecting the points $p_i$ in a complete model, and therefore further contributions to the ${\bf 10}^{(1)}\,{\bf 10}^{(2)}\,{\bf 5}$ coupling.

\subsection{Retro-Model-Building 2: And $SU(5)$ GUT}

Finally, we provide a full $SU(5)$ GUT type model with all matter and Yukawa couplings. The goal is to construct three generations of chiral matter in ${\bf 10}$ and $\overline{\bf 5}$, as well as a pair of Higgs fields in the ${\bf 5}_H$ and $\overline{\bf 5}_H$, and the top and bottom Yukawa couplings 
\be\label{Yukis}
W_{\rm Yuk} = \lambda_{\rm top} \, {\bf 10} \, {\bf 10} \, {\bf 5}_H + \lambda_{\rm bottom}\, \overline{\bf 5}\,  \overline{\bf 5}_H \, {\bf 10} \,. 
\ee
The top Yukawa was already engineered in the local $G_2$ spectral cover in section \ref{sec:TopYuk}.
To break the GUT group the standard model gauge group can e.g. be achieved by turning on discrete Wilson lines. We will not discuss this here, but it would be interesting to incorporate this into the Higgs bundle framework, see also \cite{Witten:2001bf} for a discussion.  
%
%Another useful model building tool is the presence of additional $U(1)$ symmetries, which e.g. can protect from proton decay, in particular dimension four and five operators
%\be
%W_{\rm dim 4} = \lambda_{\rm dim 4} \overline{\bf 5} \, \overline{\bf 5}\,  {\bf 10} \,,\qquad 
%W_{\rm dim 5} = \lambda_{\rm dim 5} {\bf 10} \, {\bf 10}\,  {\bf 10}\,   \overline{\bf 5} \,.
%\ee
%
%An economic way of achieving this is to require a $U(1)_{PQ}$ symmetry, which has the property that  
%\be
%q_{PQ} ({\bf 5}_{H}) + q_{PQ} (\overline{\bf 5}_{H}) \not =0 \,,
%\ee
%thus forbidding a $\mu$-term $W_{\mu} = \mu {\bf 5}_{H}\, \overline{\bf 5}_{H}$. 

There are numerous ways in generating the chiral matter content of an $SU(5)$ GUT model in the Higgs bundle setup. We will choose one that is minimal and requires only point charges. 
Furthermore to incorporate the superpotential couplings we will consider a Higgs field with $\gau = E_8$ and $\langle\phi\rangle$ taking values in $G_\perp = S[U(1)^5]$. 

The Higgsing that we consider is a special case of 
\be\label{eq:E8generalDecomp}
\ba
E_8 \quad & \rightarrow  \quad SU(5)_{\rm GUT} \times  SU(5)_{\perp} \cr 
{\bf 248}& \rightarrow \quad  ({\bf 24}, {\bf 1}) \oplus ({\bf 1}, {\bf 24})   \oplus {({\bf 10}, {\bf 5})} \oplus (\overline{\bf 5}, {\bf 10})
\oplus (\overline{\bf 10}, \overline{\bf 5}) \oplus ({\bf 5}, \overline{\bf 10}) \,.
\ea
\ee
in which
\be\label{eq:E8breaking}
\ba
E_8 \quad & \rightarrow  \quad SU(5)_{\rm GUT} \times  U(1)^4  \, .
\ea
\ee
The way to parametrize the charges is in terms of the embedding of a $U(1)^4$ into the Cartan subalgebra (CSA) of $SU(5)$, where we have five generators 
$\mathfrak{t}_i$, which satisfy 
\be
\sum_{i=1}^5 \mathfrak{t}_i =0\,.
\ee
Any $U(1)_{\alpha}$ can be parametrized now as a linear combination
\be
\mathfrak{t}_\alpha = \sum m_\alpha^i \mathfrak{t}_i \,.
\ee
The charges under the CSA generators $(\mathfrak{t}_1, \cdots, \mathfrak{t}_5)$ of the matter fields of the $SU(5)_{\rm GUT}$ are 
\be
\begin{array}{c|c|c}
	\hbox{Matter Field} \ &\  \hbox{$U(1)$-Charges under CSA} \ &\  \hbox{Spectral Cover}\cr \hline 
	{\bf 10}_i  & (\delta_{i,n})_{n=1, \cdots, 5} & \lambda_i=0 \cr 
	\overline{\bf 5}_{ij} \,,\ i>j&(\delta_{i,n} + \delta_{j, n})_{n=1, \cdots, 5}  & \lambda_i + \lambda_j =0\cr 
	{\bf 1}_{ij} \,,\ i>j& (\delta_{i,n} -\delta_{j, n})_{n=1, \cdots, 5} \,, \ i> j  & \lambda_i - \lambda_j =0 
\end{array}
\ee
We also include the loci where these matter fields are localized in $\tM3$ in terms of the $\lambda_i$, $i=1, \cdots, 5$, with $\sum_i \lambda_i =0$, which are the weights of the fundamental representation of the $SU(5)_\perp$.
For each  $\lambda_i$, we define a corresponding Morse function $f_i$ with 
\be\label{eq:Morsefunction}
\lambda_i = df_i \,.
\ee
Let us assign the matter points as follows
\be\label{eq:MatterContent}
\begin{array}{c|c|c}
	\hbox{GUT Multiplet} & \hbox{SC realization} & \hbox{Locus}\cr \hline 
	{\bf 10}_{M}^{(1)} & {\bf 10}_{1}  & \lambda_1 =0   \cr 
	{\bf 10}_{M}^{(2)} & {\bf 10}_{2}  & \lambda_2 =0   \cr 
	{\bf 10}_{M}^{(2)} & {\bf 10}_{3}  & \lambda_3 =0  \cr 
	{\bf 5}_{H} & {\bf 5}_{23} & -(\lambda_2 + \lambda_3)=0  \cr 
	{\bf 5}_{\rm ex} & {\bf 5}_{45} & -(\lambda_4 + \lambda_5)=0  \cr 
	\overline{\bf 5}_{M}^{(1)} & {\bf 5}_{24} & \lambda_2 + \lambda_4=0  \cr 
	\overline{\bf 5}_{M}^{(2)} & {\bf 5}_{15} & \lambda_1 + \lambda_5=0  \cr 
	\overline{\bf 5}_{M}^{(3)} & {\bf 5}_{35} & \lambda_3 + \lambda_5=0  
\end{array}
\ee
The top Yukawa coupling takes the form ${\bf 10}_{M}^{(2)} \times {\bf 10}_{M}^{(3)} \times {\bf 5}_{H}$ as all other combinations are forbidden by the $U(1)$ symmetries, i.e. there is no bottom Yukawa coupling, this must be generated beyond the set-up. There is one additional multiplet valued in ${\bf 5}$ beyond the required matter for the GUT model. 
This example has one extra matter multiplet, and it would be interesting to see whether different charge configurations give rise to exactly the GUT spectrum. 

\begin{figure}
	\centering
	\includegraphics[width=15cm]{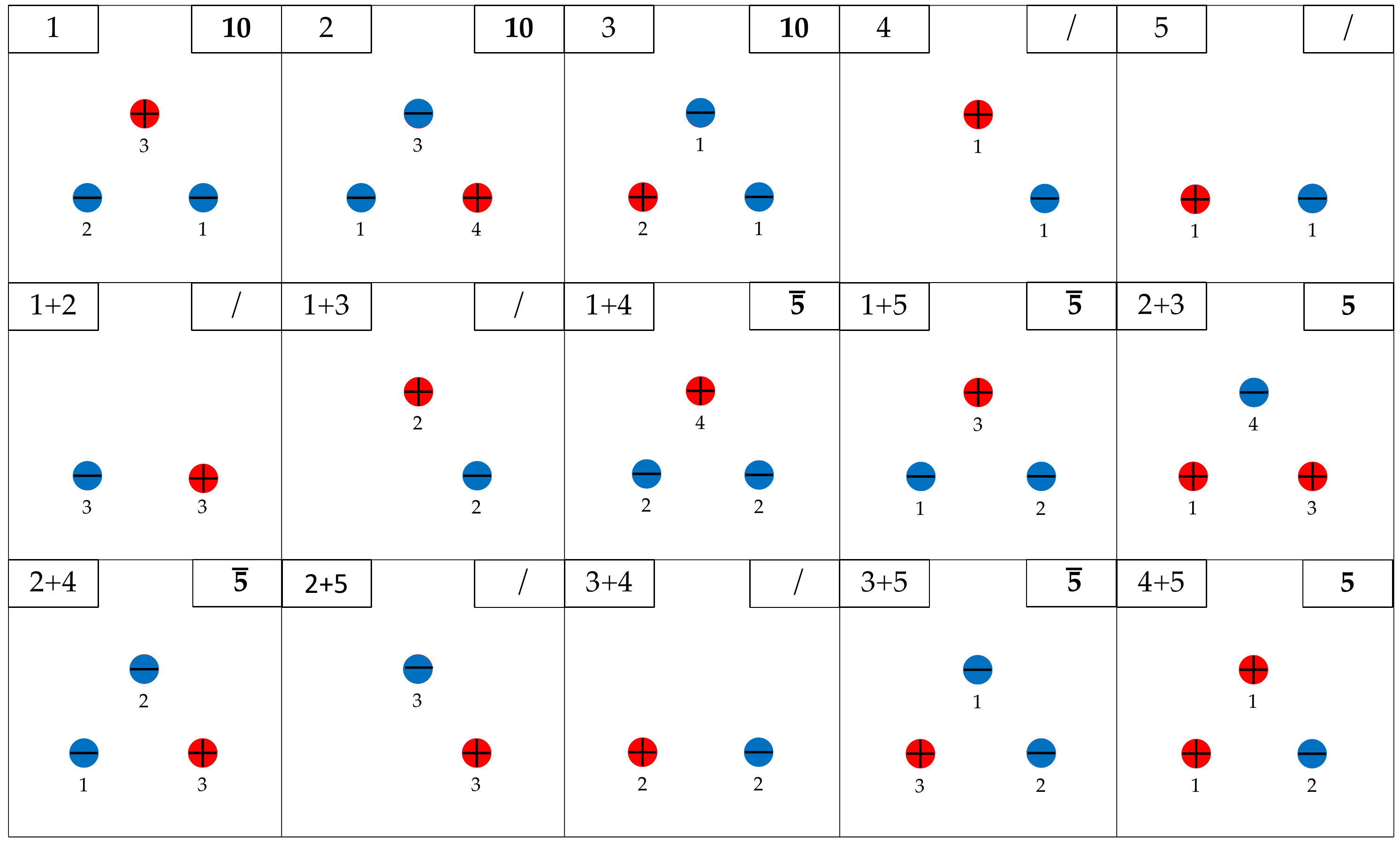}
	\caption{A charge configuration leading to a chiral spectrum with 3 multiplets transforming in ${\bf 10}$, 4 in $\overline{\bf 5}$ and 2 in ${\bf 5}$. There are further multiplets transforming in ${\bf 1}$ which are not depicted. Each box shows the same 3 points in $M_3$. The first row shows 5 fundamental charge distributions with the units of charge denoted as subscripts. The total charge of each box vanishes and also adding the first row of pictures yields a vanishing charge distribution. This reflects \eqref{eq:ChargeConstraint}. The bottom two rows show superpositions of the fundamental charge distributions as noted in the top left corner. Each box contributes a single chiral multiplet, denoted in the top right corner of each box, if it depicts 3 charged points. This realizes the matter content as in \eqref{eq:MatterContent}. }
	\label{fig:10105}
\end{figure}

To realise this spectrum via electrostatic charge distributions we translate the above into the language of section \ref{sec:BPSConfig} and its higher rank generalisations. Note from \eqref{eq:E8breaking} that $4$ factors of $U(1)$ have been broken off, this gives us 4 types of charge with which to build the model. Due to this special abelian background the decomposition in \eqref{eq:E8generalDecomp} decomposes further and the fundamental weights of $SU(5)_\perp$ are now associated with fundamental charge vectors $Q_i^F$ such that \eqref{eq:Morsefunction} now becomes
\be\label{eq:ChargeConstraint}
\lambda_i=df_i=df_{Q_i^F}\,, \qquad \sum_{i=1}^5 Q_i^F=0\,.
\ee
Upon a redefinition of $U(1)$ generators we may take these charge vectors to be
\be
Q_5^F=(-1,-1,-1,-1)\,, \qquad Q_i^F=\delta_{ik} \quad k=1,\dots,4\,. 
\ee
Placing $n_i^-$ negative and $n_i^+$ positive point charges of type $i=1,\dots,4$ we obtain by \eqref{eq:Chiralcount} 
\be\ba
&n_i^--1 \tn{  chiral multiplets transforming in }{\bf 10}_{Q^F_i}\,, \\
&n_i^+-1 \tn{  chiral multiplets transforming in }\overline{\bf 10}_{-Q^F_i}\,, \\
\ea\ee
generically. The number of chiral multiplets valued in $\bar {\bf 5}_M$ and ${\bf 5}_H$ is fixed by these choices and computed by taking the relevant linear combinations of the charges as listed in \eqref{eq:MatterContent}.

The simplest charge configuration possible is obtain by collecting the four types of different charge at three distinct points in $M_3$ where $i=1,2,3$. The only constraint on each charge configuration is that it must be of vanishing total charge. A fifth charge configuration is generated via the last relation in \eqref{eq:ChargeConstraint}. We depict a possible distribution of point charges in figure \ref{fig:10105}. The charge distribution of this example yields a chiral spectrum with $3,4,2$ chiral multiplets transforming in the representations ${\bf 10},{\bf \bar{5}},{\bf 5}$ respectively. These multiplets reside at the critical points of the relevant combinations of the charge. The set-up allows for a single top Yukawa coupling.
Clearly there is a lot of room to extend these models and improve them and it would be interesting to see the full extent of the phenomenological implications of this framework.

%%%%%%%%%%%%%%%%%%%%%%%%%%%%%%%%%%%%%%%%%%%
%%%%%%%%%%%%%%%%%%%%%%%%%%%%%%%%%%%%%%%%%%%

\section{Conclusions and Outlook}
\label{sec:CD}

The main result of this paper is a study of the gauge sector of M-theory compactifications on $G_2$-holonomy manifolds to 4d $\mathcal{N}=1$ supersymmetric gauge theories. The structure that governs this theory is a Higgs bundle on an associative three-cycle $M_3$, i.e. a gauge field $W$ and a one-form Higgs field $\phi$ on $M_3$, satisfying the BPS equations (\ref{BPS}). We have focused exclusively on the case of $W=0$ and $\phi$ Higgs field, and have given a detailed description on how to engineer and analyse backgrounds satisfying the BPS equations for abelian Higgs fields. Furthermore, we have shown how to apply this formalism to the case of TCS $G_2$-manifolds. Although these are not interesting for phenomenological applications, we have qualitatively shown under which conditions singular transitions of such compactifications can give rise to chiral 4d spectra and thus bring us somewhat closer to the main open question in this field, i.e. the construction of compact $G_2$ manifolds with codimension 7 singularities.  There are various concrete directions to build on the present work: 
\begin{enumerate}
\item Using the analogy with electrostatics not only allows to (implicitely) construct abelian Higgs field backgrounds, but can furthermore be used to find the zero mode spectrum. In the case of genuinly non-Abelian Higgs field, i.e. in situations in which the spectral cover does not factor completely, this method can not be straightforwardly applied. For model building applications one hence needs both a concise way to specify such solutions and a efficient way to determine the resulting spectrum of zero modes. 
\item `T-branes': Once the non-factored spectral covers are understood, there is also of course the extension to `T-branes' \cite{Cecotti:2010bp}, i.e. non-diagonalizable Higgs vevs. Clearly these would be interesting to study in the present context as well. Similar to the state of affairs in F-Theory, such backgrounds require to supply extra data on top the geometry of a $G_2$-manifold. 
\item In terms of applications to model building, we have given examples of $SU(5)$ GUT models, we have not discussed mechanisms of GUT breaking. A thorough investigation of such mechanisms includes investigating the effect of flat gauge field for $\pi_1(M_3)\not= 1$. Furthermore it would be interesting to give a comprehensive analysis of the possible charge destributions, that give rise to semi-realistic GUT models. Can these e.g. be systematically analyzed as in e.g. F-theory?
In particular, since  $U(1)$ symmetries are paramount here, what type of constraints are there on $U(1)$-charges. Much progress on this has appeared in F-theory (see e.g. \cite{Cvetic:2018bni} for a review), which would be interesting to complement with a $G_2$-type analysis. 
\item M-theory/Heterotic duality: M-theory on K3 is dual to heterotic on $T^3$. Applied fiberwise to the ALE-fibrations (and the associated Higgs bundles) that we have studied in this paper, one can ask what this entails for the dual heterotic models on Calabi-Yau three-folds. On the heterotic side, the $T^3$ becomes the fiber of the SYZ fibration of the Calabi-Yau threefold and it would be interesting to understand how an application of this duality gives rise to holomorphic vector bundles which are specified by varying flat bundles on the SYZ fiber as a generalization of \cite{Friedman:1997yq}. For TCS $G_2$-manifolds this has been done in \cite{Braun:2017uku}. 
\end{enumerate}

%%%%%%%%%%%%%%%%%%%%%%%%%%%%%%%%%%%%%%%%%%%
%%%%%%%%%%%%%%%%%%%%%%%%%%%%%%%%%%%%%%%%%%%

%%%%%%%%%%%%%%%%%%%%%%%%%%%%%%%%%%%
%%%%%%%%%%%%%%%%%%%%%%%%%%%%%%%%%%%
\subsection*{Acknowledgments}

We thank David Bosticco, Xenia de la Ossa, Julius Eckhard, Jim Halverson,  Mark Haskins, Heeyeon Kim, Dave Morrison, James Sparks for discussions. 
AB, SC and SSN are supported by the ERC Consolidator Grant 682608 ``Higgs bundles: Supersymmetric Gauge Theories and Geometry (HIGGSBNDL)''. MH is supported by the Studienstiftung des Deutschen Volkes.
%%%%%%%%%%%%%%%%%%%%%%%%%%%%%%%%%%%
%%%%%%%%%%%%%%%%%%%%%%%%%%%%%%%%%%%

\appendix

%%%%%%%%%%%%%%%%%%%%%%%%%%%%%%%%%%%%%%%%%%%
%%%%%%%%%%%%%%%%%%%%%%%%%%%%%%%%%%%%%%%%%%%

\section{Conventions}

\subsection{Glossary}
\label{app:Gloss}

{\footnotesize
\begin{tabular}{c|l}
Label & Meaning \cr \hline
$M_3$ & Associative three-cycle  \cr 
$\gau$ & Unhiggsed gauge group \cr 
$G$ & Gauge group in 4d (arising from Higgsing from $\gau \rightarrow G \times G_\perp$)\cr 
$\phi$ & One-form Higgs field in $\Omega^1 (\tM3) \otimes \hbox{Ad} (G_\perp)$\cr 
$f$ & Morse-Bott function or electrostatic potential with $\phi =df$ \cr
$\rho$ & Charge distribution on $M_3$ supported on $\Gamma$\cr  
$Q$ & Vector of $U(1)$ charges \cr
$f_Q$ & Charge weighted sum of Morse-Bott functions \cr 
$\Gamma$ & Subspace of $M_3$ where electrostatic charge distribution is localized\cr
$\tM3$  & 	 $M_3\setminus T(\Gamma)$, where  $T(\Gamma)$ is a tubular neighborhood of $\Gamma$.  \cr 
$\Sigma$ & $\partial \tM3$\cr 
$\gamma (f_1, \cdots, f_n)$ & Gradient flow tree specified by Morse-Bott functions $f_i$\cr 
\end{tabular}
}

\subsection{Spinors}
\label{app:Spinors}

The Clifford-algebras in $4,7,10$ dimension are denoted by 
    \be 
4d \leftrightarrow \gamma\,,\qquad 7d \leftrightarrow \hat\gamma\,,\qquad 10d \leftrightarrow \Gamma\,.
    \ee
    We realise the gamma matrices as{\footnotesize
\begin{eq}{ExplicitGamma}
    \Gamma^0 &=\sigma^1 \otimes \hat\gamma^0 \otimes \sigma^0=\sigma^1 \otimes \gamma^0 \otimes \sigma^0 \otimes \sigma^0 \cr 
    \Gamma^1 &=\sigma^1 \otimes \hat\gamma^1 \otimes \sigma^0=\sigma^1 \otimes \gamma^1 \otimes \sigma^0 \otimes \sigma^0 \cr 
    \Gamma^2 &=\sigma^1 \otimes \hat\gamma^2 \otimes \sigma^0=\sigma^1 \otimes \gamma^2 \otimes \sigma^0 \otimes \sigma^0 \cr 
    \Gamma^3 &=\sigma^1 \otimes \hat\gamma^3 \otimes \sigma^0=\sigma^1 \otimes \gamma^3 \otimes \sigma^0 \otimes \sigma^0 \cr 
    \Gamma^4 &=\sigma^1 \otimes \hat\gamma^4 \otimes \sigma^0=\sigma^1 \otimes \gamma_5 \otimes \sigma^1 \otimes \sigma^0 \cr 
    \Gamma^5 &=\sigma^1 \otimes \hat\gamma^5 \otimes \sigma^0=\sigma^1 \otimes \gamma_5 \otimes \sigma^2 \otimes \sigma^0  \cr 
    \Gamma^6 &=\sigma^1 \otimes \hat\gamma^6 \otimes \sigma^0 =\sigma^1 \otimes \gamma_5 \otimes \sigma^3 \otimes \sigma^0 \cr 
    \Gamma^7 &=\sigma^2 \otimes I_8 \otimes \sigma^1=\sigma^2 \otimes I_4 \otimes \sigma^0 \otimes \sigma^1 \cr 
    \Gamma^8 &=\sigma^2 \otimes I_8 \otimes \sigma^2=\sigma^2 \otimes I_4 \otimes \sigma^0 \otimes \sigma^2 \cr 
    \Gamma^9 &=\sigma^2 \otimes I_8 \otimes \sigma^3=\sigma^2 \otimes I_4 \otimes \sigma^0 \otimes \sigma^3 \,.
\end{eq}}
 Here $\sigma^0= \hbox{Id}_2$ and  the 4d gamma matrices are 
\begin{eq}{Explicit 4d Dirac matrices}
\gamma^\mu = \lb \begin{array}{cc} 0 & \sigma^\mu \\ \bar{\sigma}^\mu & 0 \end{array}\rb\,,\qquad (\sigma^\mu)=(-\sigma^0,\sigma^i)\,,\qquad (\bar{\sigma}^\mu)=(-\sigma^0,-\sigma^i)\,,
\end{eq}
where 4d signature is $\R^{3,1}$.
The chirality, $B$-matrices, charge conjugation matrices and Lorentz-generators will be denoted by:
\begin{eq}{Related matrix notation}
\gamma_5\,, \Gamma_c\,,\qquad B_4\,, B_{10}\,, \qquad C_4\,,C_{10}\,, \qquad \Sigma_4\,,\Sigma_7\,, \Sigma_{10}\,,
\end{eq}
respectively for $4d ,7d ,10d$ gamma matrices. 
\begin{alignat}{3}\label{Explicitrelatedmatrices1}
\gamma_5&=i \gamma^0 \cdots \gamma^3\,,\qquad &&\Gamma_c && =- \Gamma^0 \cdots \Gamma^9\cr 
B_4 &= \gamma_5 \gamma^0\gamma^1\gamma^3\,,\qquad &&B_{10}&&= - \Gamma^0\Gamma^1\Gamma^3\Gamma^5\Gamma^7\Gamma^9 \,,  \\
C_4 &= B_4 \gamma^0\,,\qquad &&C_{10}&&= - B_{10}\Gamma^0 \,,
\end{alignat}
and
\begin{alignat}{5}\label{Explicitrelated matrices2}
    \Sigma_4^{\mu\nu}&=-\frac{i}{4} [\gamma^\mu,\gamma^\nu]\,,\qquad &\Sigma_7^{\mu\nu}& =-\frac{i}{4} [\hat\gamma^\mu,\hat\gamma^\nu]\,,\qquad &\Sigma_{10}^{\mu\nu}&=-\frac{i}{4} [\Gamma^\mu,\Gamma^\nu]\,.
\end{alignat}
As one of the defining properties of the $B$-matrices is $B^*B=1$ there strictly speaking does not exist a matrix $B_7$. However the definition of the $B$-matrices as product of all imaginary gamma matrices can be extended to odd dimensions. We thus define $B_7$ and its corresponding charge conjugation matrix $C_7$ by
\begin{eq}{B7}
    B_7=\hat\gamma^0\hat\gamma^1\hat\gamma^3\hat\gamma^5\,, \qquad C_7=B_7 \hat{\gamma}^0\,.
\end{eq}
The three $B$-matrices fit together as
    \begin{eq}{3B}
    B_{10}=\sigma^0\otimes B_7 \otimes i\sigma_2 = \sigma^0\otimes B_4 \otimes (-\sigma^2) \otimes i\sigma_2 \,.
    \end{eq}
We collect the relations satisfied by the above matrices in table \ref{Table of matrix relations}. Finally we list anti-symmetric combinations needed to specify the Lorentz generators $\Sigma_{10},\Sigma_7,\Sigma_4$  
\begin{alignat}{3}
\label{Antisymmetric Dirac matrix combination}
\Gamma^{\mu\nu}&=\sigma^0 \otimes \gamma^{\mu\nu}\otimes I_4 \,,\qquad &&\Gamma^{\mu \underline{k}}&&=\sigma^0\otimes \gamma^\mu\gamma_5 \otimes \sigma^{\underline{k}} \otimes \sigma^0  \,,\\
\Gamma^{\underline{k}\underline{l}}&= I_8 \otimes \sigma^{\underline{k}\underline{l}} \otimes \sigma^0 \,,\qquad &&\Gamma^{\mu \hat\imath}&&=i\sigma^3\otimes \gamma^\mu \otimes \sigma^0 \otimes \sigma^{\hat\imath} \,,\\
\Gamma^{\hat\imath\hat\jmath}&=I_{16} \otimes \sigma^{\hat\imath\hat\jmath}\,,\qquad &&\Gamma^{\underline{k}\hat\imath}&&=i\sigma^3\otimes \gamma_5 \otimes \sigma^{\underline{k}} \otimes \sigma^{\hat\imath} \,,
\end{alignat}
where indices run as $\mu=0, \dots ,3$ and $\underline{k},\underline{l}=1,2,3$ and $\hat\imath,\hat\jmath=1,2,3$.

\begin{table}
    \label{Table of matrix relations}
    \begin{center}
   {\footnotesize      \begin{tabular}{||c|| c c c ||} 
         \hline
         & Chirality & B-matrix & Charge  Conj.   \\ [0.5ex] 
         \hline\hline
         4d & $\begin{array}{c} \gamma_5 \gamma_5 = 1 \\ \{\gamma_5,\gamma\}=0 \\ {[}\gamma_5,\Sigma_4{]}=0 \\ \gamma_5=\gamma_5^*~ (\dagger) \\
         \gamma_5=\gamma_5^T~ (\dagger) \end{array}$ & $\begin{array}{c} B_4^*B_4=1 \\ B_{4} \gamma B_{4}^{-1}=+ \gamma^* \\ B_4 \gamma_5 B_4^{-1}=-\gamma_5^* \\
         B_4 \Sigma_4 B_4^{-1}=-\Sigma_4^* \\ B_4=B_4^T~ (\dagger) \end{array}$ & $\begin{array}{c} C_{4} \gamma C_{4}^{-1}=+ \gamma^T \\ C_{4} \Sigma_4 C_{4}^{-1}=- \Sigma_4^T \\ C_4 C_4=-1~ (\dagger) \\ C_4=C_4^*~ (\dagger)  \\ C_4=-C_4^T~ (\dagger) \end{array}$  \\ 
         \hline
         7d & $\begin{array}{c} \tn{} \\ \tn{does not exist} \\  \tn{} \end{array}$ & $\begin{array}{c} B_7^*B_7=-1 \\ B_{7} \hat\gamma B_{7}^{-1}=+ \hat\gamma^*\\ B_7 \Sigma_7 B_7^{-1}=-\Sigma_7^* \\ B_7=-B_7^T~ (\dagger) \end{array}$ & $\begin{array}{c}  C_{7} \Sigma_7 C_{7}^{-1}=- \Sigma_7^T \\  C_7 C_7=-1~ (\dagger) \\ C_7=-C_7^*~ (\dagger) \\ C_7=C_7^T~ (\dagger) \end{array}$  \\
         \hline
         10d & $\begin{array}{c} \Gamma_c \Gamma_c = 1 \\ \{\gamma_5,\Gamma\}=0 \\ {[}\Gamma_c,\Sigma_{10}{]}=0 \\  \Gamma_c=\Gamma_c^*~ (\dagger) \\
         \Gamma_c=\Gamma_c^T~ (\dagger)  \end{array}$ & $\begin{array}{c} B_{10}^*B_{10}=1 \\ B_{10} \Gamma B_{10}^{-1}=+ \Gamma^* \\ B_{10} \Gamma_c B_{10}^{-1}=+\Gamma_c^* \\  B_{10} \Sigma_{10} B_{10}^{-1}=-\Sigma_{10}^* \\ B_{10}=B_{10}^T~ (\dagger) \end{array}$ & $\begin{array}{c} C_{10} \Gamma C_{10}^{-1}=- \Gamma^T \\ C_{10} \Sigma_{10} C_{10}^{-1}=- \Sigma_{10}^T  \\  C_{10} C_{10}=1~ (\dagger) \\ C_{10}=-C_{10}^*~ (\dagger) \\ C_{10}=-C_{10}^T~ (\dagger)  \end{array}$  \\ [0.5ex] \hline
        \end{tabular}}
    \end{center}
\caption{List of matrix relations. The unmarked relations are fundamental and necessary to the definition of the chiral, B and charge conjugation matrices. The daggered relations are a consequence of the explicit realisation of the gamma matrices. The $B_7$ is defined in analogy to the $B_4,B_{10}$ matrices but cannot be used to implement a Majorana condition as $B_7^*B_7=-1$. In odd dimensions there exists no notion of chirality as the representation of the Clifford-algbra is already irreducible. We suppress space-time indices.}
\end{table}

The 10d Majorana-condition $B_{10}\lambda=\lambda^*$ leads to a symplectic Majorana-constraint on the 7d spinors and a Majorana-constraints on the 4d spinors. We trace through the decomposition of the spinor representation as detailed in \eqref{DimRed}, \eqref{UntwistedReduction} and \eqref{TwistedFieldcontent} and make these constraints explicit.

By (\ref{eq: 3B}) the constraint inherited by the 7d spinors $\lambda_{\alpha\hat\alpha}$ is
\be\label{SymMajo}
\lambda_{\alpha\hat\alpha}=(i\sigma^2)_{\hat\alpha}^{~\hat\beta}\lambda_{\alpha\hat\beta}^*\,,
\ee
which is a symplectic Majorana-condition, \cite{VanProeyen:1999ni}. Here the indices run as $\alpha=1,\dots,8$ and $\hat\alpha=1,2$. The 7d spinors satisfy no further constraints.

We next turn to the spinors $\lambda_{\alpha\underline{\alpha}\hat\alpha}$ in 4d of the untwisted symmetry group. By (\ref{eq: 3B}) these are required to satisfy the Majorana-constraint
\be\label{untwistedMajo}
\lambda_{\alpha\underline{\alpha}\hat\alpha}=(-\sigma^2)_{\underline{\alpha}}^{~\underline{\beta}}(i\sigma^2)_{\hat\alpha}^{~\hat\beta}\lambda_{\alpha\underline{\beta}\hat\beta}^*\,,
\ee
where the indices run as $\alpha,\underline{\alpha},\hat\alpha=1,2$. There are no further constraints on the spinors.

After performing the twist we find the Dirac spinors $(\lambda_0,\lambda_i)$, which carry twisted indices, to be constrained as
	\begin{eq}{Twisted Majorana}
    iB_4 \lambda_0 = \lambda_0^*\,,\qquad   iB_4 \lambda_i = -\lambda_i^*\,,
    \end{eq}
with $i=1,2,3$. Decomposing these Dirac spinors into Weyl spinors $\lambda_0=(i\chi_\alpha,i\bar{\xi}^{\dot\alpha})$ and $\lambda_i=(\psi_{i\alpha},\bar{\zeta}_i^{~\dot\alpha})$ the conditions are rewritten explicitly with the charge conjugation matrix as
\be\begin{split}
(i\bar{\sigma}^2)^{\dot\alpha \alpha} \chi_\alpha =  -\lb \bar{\xi}^{\dot\alpha}\rb^*\,, \qquad
 (i\bar{\sigma}^2)^{\dot\alpha \alpha} \psi_{i\alpha} = - \lb \bar{\zeta}_i^{~\dot\alpha}\rb^*\,,
	\end{split}\ee
which are, due to the introduction of a factor of $i$, nothing but two canonical Majorana-conditions
\be\label{MajoranaCons}
\chi_\alpha=\xi_{\alpha}\,,\qquad \psi_{i\alpha}=\zeta_{i\alpha}\,. 
\ee
Using this we can rewrite the 4d dimensionally reduced action of 10d SYM interms of 1+3 unconstrained Weyl spinors $\chi,\psi_i$ in 4d. The resulting action is given in \eqref{TwistedActionMain}.

%%%%%%%%%%%%%%%%%%%%%
     
\section{Derivation of 4d Effective Theory}
\label{app:SUSY}

\subsection{Off-shell action for 7d SYM}
\label{app:OffShell}

Following \cite{Beasley:2008dc} we rewrite the twisted 7d SYM action into a mostly off-shell form. We introduce complex auxiliary fields $H_i$ packaged in 3 chiral multiplets
\be\ba
\Phi_i=\varphi_i +\sqrt{2} \theta \psi_i + \theta \theta H_i + \cdots\,, 
\ea\ee
and a real auxiliary scalar field $D$ contained in a vector multiplet in WZ gauge
\be
V=-\theta \sigma^\mu \bar{\theta} A_\mu + i \theta\theta\bar{\theta}\bar{\chi}-\bar{\theta}\bar{\theta}\theta\chi+\frac{1}{2}\theta \theta \bar{\theta} \bar{\theta} D \,.
\ee
The $\CN=1$ supersymmetry algebra
\be\label{SUSYAlg}
\{\delta_\alpha,\delta_\beta\}=\{\bar{\delta}_{\dot\alpha},\bar{\delta}_{\dot\beta}\}=0\,,\qquad \{\delta_\alpha,\bar{\delta}_{\dot\alpha}\}=2i(\sigma^\mu)_{\alpha\dot\alpha}D_\mu \,,
\ee
can not be implemented off-shell for a vector multiplet in WZ gauge. Either manifest gauge invariance or manifest supersymmetry need to be demitted. It is possible to maintain the reduced algebra
\be\label{constraint1}
\{\delta_\alpha,\delta_\beta\}=\{\bar{\delta}_{\dot\alpha},\bar{\delta}_{\dot\beta}\}=0\,,
\ee
and manifest gauge invariance at the same time. This is achieved if we choose the bosonic variations to be of the form 
\be\ba
\delta_\alpha A^\mu&=+i\sigma^\mu_{\alpha\dot\alpha}\bar{\chi}^{\dot\alpha} \,, \qquad &&\bar{\delta}_{\dot\alpha} A^\mu&&=-i\chi^\alpha \sigma^\mu_{\alpha\dot\alpha} \cr 
\delta_\alpha \varphi_i&=\sqrt{2} \psi_{i\alpha} \,, \qquad &&\bar{\delta}_{\dot\alpha} \varphi_i&&=0 \cr 
\delta_\alpha \bar{\varphi}_i&=0 \,, \qquad &&\bar{\delta}_{\dot\alpha} \bar{\varphi}_i&&=\sqrt{2} \bar{\psi}_{i\dot\alpha} \,,
\ea\ee
and fermionic variations \eqref{fermionicVar} to be of the form
\be\ba
\delta_\alpha \chi_\beta &=+ F_{\mu\nu}\sigma^{\mu\nu}_{\alpha\beta}-iD \epsilon_{\alpha\beta} \,, \qquad &&\bar{\delta}_{\dot\alpha} \chi_\beta &&=0 \cr 
\delta_{\alpha} \bar{\chi}_{\dot\beta} &=0  \,, \qquad &&\bar{\delta}_{\dot\alpha} \bar{\chi}_{\dot\beta} &&=+ F_{\mu\nu}\bar{\sigma}^{\mu\nu}_{\dot\alpha\dot\beta}+iD \epsilon_{\dot\alpha\dot\beta} \cr 
\delta_\alpha \psi^m_\beta &=-\sqrt{2} H^m \epsilon_{\alpha\beta} \,, \qquad &&\bar{\delta}_{\dot\alpha} \psi^m_{\beta} &&=+i\sqrt{2}(F_{\varphi})_\mu^{~k} \sigma^\mu_{\beta\dot\alpha}  \cr 
\delta_\alpha \bar{\psi}^m_{\dot\beta} &=+i\sqrt{2}(F_{\bar\varphi})_\mu^{~k} \sigma^\mu_{\beta\dot\alpha} \,, \qquad &&\bar{\delta}_{\dot\alpha} \bar{\psi}^m_{\dot\beta} &&= -\sqrt{2} \bar{H}^m \epsilon_{\dot\alpha\dot\beta} \,,
\ea\ee
where $\delta_\alpha(\,\cdot\,)=[Q_\alpha,\cdot\,\}$ and $\bar{\delta}_{\dot\alpha}(\cdot)=[\bar{Q}_{\dot\alpha},\cdot\,\}$. The variations of the auxilary fields $H_i,D$ are then given by
\be\begin{alignedat}{3}
	\delta_\alpha D&=-(\sigma^\mu)_{\alpha\dot\alpha}D_\mu\chi^\alpha\,,\qquad && \bar{\delta}_{\dot\alpha} D&&=-(\sigma^\mu)_{\alpha\dot\alpha}D_\mu\bar{\chi}^{\dot\alpha}\cr 
	\delta_{\alpha} H_i&=0\,, \quad && \bar{\delta}_{\dot\alpha} H_i&&=i\sqrt{2}\sigma^\mu_{\alpha\dot\alpha}D_\mu\psi_i^\alpha-2i\CD_i\bar\chi_{\dot\alpha}\cr 
	\delta_{\alpha} \bar{H}_i&=i\sqrt{2}\sigma^\mu_{\alpha\dot\alpha}D_\mu\bar{\psi}_i^{\dot\alpha}+2i\bar\CD_i\chi_{\alpha} \,, \quad && \bar{\delta}_{\dot\alpha} \bar{H}_i&&=0 \,.
\end{alignedat}\ee

We next derive the action invariant under the above transformations which after integrating out the auxiliary fields reduces to the on-shell action previously obtained via dimensional reduction and twisting \eqref{TwistedActionMain}. To obtain an action annihilated by $\delta_{\alpha},\bar{\delta}_{\dot\alpha}$ it suffices to construct a real action which is annihilated by either of the two operators. The other operator will annihilate the action as the operators are conjugate to another. A real action must thus be built from exact and closed terms with respect to $\bar{\delta}^2$ or $\bar{\delta}_{\dot\alpha}$ which are annihilated by  $\bar{\delta}_{\dot\alpha}$. 

The operators
\be\ba
\mathcal{O}&=\int_{M_3}d^3x\,\tn{Tr}\lbb\frac{1}{8}\bar{\chi}\bar{\chi}\rbb\cr 
\mathcal{O}^{(1)}_{\dot\beta}&=\int_{M_3}d^3x\,\tn{Tr}\lbb\frac{i}{4}(\CF_{\bar{\varphi}})_{ij}\epsilon^{ijk}\bar{\psi}_{k\dot\beta}\rbb\cr 
\mathcal{O}^{(2)}_{\dot\beta}&=\int_{M_3}d^3x\,\tn{Tr}\lbb\frac{1}{2\sqrt{2}}\bar\psi_{\dot\beta}^kH_k+\frac{i}{2\sqrt{2}}\lb F_{\bar\varphi}\rb_{\mu k}\sigma^\mu_{\alpha\dot\beta}\psi^{\alpha k}-\frac{i}{2} I_{\varphi,\bar\varphi}\bar\chi_{\dot\beta}\rbb\cr 
\ea\ee
give rise to $\bar{Q}$-exact and closed terms
\be\ba\label{eq:OffShellAction}
I_1=\bar{\delta}^2\mathcal{O}&=\int_{M_3}d^3x\,\tn{Tr}\lbb +i D_\mu \chi^\alpha (\sigma^\mu)_{\alpha\dot\alpha}\bar{\chi}^{\dot\alpha}-\frac{1}{4} F_{\mu\nu}F^{\mu\nu}-\frac{i}{8}\epsilon^{\mu\nu\rho\sigma}F_{\mu\nu}F_{\rho\sigma}+\frac{1}{2}D^2 \rbb\,,\\
I_2=\epsilon^{\dot\alpha\dot\beta} \bar{\delta}_{\dot\alpha}\mathcal{O}^{(1)}_{\dot\beta}&=\int_{M_3}d^3x\,\tn{Tr}\lbb\frac{i}{\sqrt{2}} \bar{\CD}_i\bar{\psi}_{j\dot\alpha }\epsilon^{ijk}\bar{\psi}^{\dot\alpha}_k +\frac{i}{\sqrt{2}} (F_{\bar{\varphi}})_{ij}\epsilon^{ijk}\bar{H}_k\rbb\,,\\
I_3=\epsilon^{\dot\alpha\dot\beta}\bar{\delta}_{\dot\alpha}\mathcal{O}^{(2)}_{\dot\beta}&=\int_{M_3}d^3x\,\tn{Tr}\,\bigg[ H^k H_k-\frac{i}{2}D_\mu\psi_k\sigma^\mu \bar\psi^k+\frac{i}{\sqrt{2}}\bar\chi\CD_k\bar\psi^k \\ &~~~~~~~~~~~~~~~~~~~ -\frac{i}{2} D_\mu \psi^k\sigma^\mu \bar\psi_k-i\sqrt{2}\chi \bar\CD_k\psi^k-\lb F_{\bar\varphi}\rb_{\mu k}\lb F_{\varphi}\rb^{\mu k} \\ 
&~~~~~~~~~~~~~~~~~~~+\frac{i}{\sqrt{2}}\bar\chi\CD_k\bar\psi^k+I_{\varphi,\bar\varphi}D \bigg]\,.
\ea\ee
Note that
\be
\bar{\delta}_{\dot\alpha}\mathcal{O}^{(1)}_{\dot\beta}=-\bar{\delta}_{\dot\beta}\mathcal{O}^{(1)}_{\dot\alpha}\,, \qquad \bar{\delta}_{\dot\alpha}\mathcal{O}^{(2)}_{\dot\beta}=-\bar{\delta}_{\dot\beta}\mathcal{O}^{(2)}_{\dot\alpha}\,, 
\ee
which is necessary for $\bar{\delta}_{\dot\alpha}$ to annihilate $I_2$ and $I_3$. The action \eqref{TwistedActionMain} taken off-shell becomes
\be
S_B+S_F=\frac{1}{g_7^2}(I_1+I_2+I_2^*+I_3)\,.
\ee
In more detail the lagrangian reads
\be\ba\label{OffShellBulk}
\CL_{7d}=&\frac{1}{g_7^2}\tn{Tr}\Big[ -\frac{1}{4}F_{\mu\nu}F^{\mu\nu}-D_\mu \bar{\varphi}_k D^\mu\varphi^k+i D_\mu \chi \sigma^\mu\bar{\chi} -i D_\mu \psi_k \sigma^\mu\bar{\psi}^k\\
& +\frac{1}{2}D^2+H^k\bar{H}_k+DI_{\varphi, \bar{\varphi}}+\frac{i}{\sqrt{2}}(F_{\bar{\varphi}})_{ij}\epsilon^{ijk}\bar{H}_k-\frac{i}{\sqrt{2}}(F_{\varphi})_{ij}\epsilon^{ijk}H_k \\
&-\frac{i}{\sqrt{2}}\epsilon^{ijk}\psi_i\CD_j\psi_{k}+\frac{i}{\sqrt{2}}\epsilon^{ijk}\bar{\psi}_i\bar{\CD}_j\bar{\psi}_{k}-\sqrt{2}i\chi \bar{\CD}_i \psi^i+\sqrt{2}i\bar{\chi} \CD_i \bar{\psi}^i\Big] \,.
\ea\ee
Note that the term
\be
F_{\mu\nu}F_{\rho\sigma}\epsilon^{\mu\nu\rho\sigma}\,,
\ee
giving the Pontryagin density can be dropped as it is topological, therefore supersymmetric on its own and thus not necessary for the supersymmetric invariance of the action. The auxiliary fields are eliminated via their equations of motion
\be\label{AuxiliaryEOM}
D=-I_{\varphi,\bar\varphi}\,,\qquad H_k=-\frac{i}{\sqrt{2}}(F_{\bar{\varphi}})_{ij}\epsilon^{ijk}\,, \qquad \bar{H}_k=+\frac{i}{\sqrt{2}}(F_{\varphi})_{ij}\epsilon^{ijk}\,.
\ee
This returns the on-shell action \eqref{TwistedActionMain} and the corresponding supersymmetric variations \eqref{bosonicVar},\eqref{fermionicVar}.

\subsection{Effective 4d Action}
\label{sec:EffAction}

In section \ref{sec:ZeroModes} we discussed the zero modes upon reduction to 4d and enumerated these in section \ref{sec:Haribo}. In section \ref{sec:MatterInteractions} we studied the 4d $\CN=1$ superpotential terms governing the interaction of localized modes of the 4d $\CN=1$. In this appendix we give more details on the reduction to 4d. We begin by reproducing the assumptions on the Higgs background of section \ref{sec:MatterInteractions}.

We take the Higgs background to be turned on along $n$ abelian given by the Cartan generators $\mathfrak{t}^i$. Its profile is the parametrised parametrised by $n$ singular harmonic Morse functions $f_i:M_3\rightarrow \R$ as
\be\ba
\phi= \sum_{i=1}^n\mathfrak{t}^idf_i\,.
\ea\ee
As a consequence the gauge symmetry breaks as $\gau\rightarrow \gut \times U(1)^n$ and fields are repackaged into irreducible representations of the remnant gauge symmetry as
\be\label{eq:DecompOfReps}
\tn{Ad}\, \gau \quad \rightarrow \quad \tn{Ad}\, \gut \oplus \tn{Ad} (U(1)^n)\oplus \bigoplus_{Q= (q_1,\dots, q_n)} \tn{\textbf{R}}_{Q  }\,,
\ee
where $Q$ denotes a vector of $U(1)$ charges.

Next we introduce 4d $\CN=1$ multiplets valued in the irreducible representations of $\gut\times U(1)^n$ of \eqref{eq:DecompOfReps}. We begin with those valued in the representations $\tn{Ad}\, \gut$ and $ \tn{Ad} (U(1)^n)$ 
\be\ba
V'&=-\theta \sigma^\mu \bar{\theta} A_\mu' + i \theta\theta\bar{\theta}\bar{\chi}'-\bar{\theta}\bar{\theta}\theta\chi'+\frac{1}{2}\theta \theta \bar{\theta} \bar{\theta} D' \,, \\
V''&=-\theta \sigma^\mu \bar{\theta} \tilde A_\mu'' + i \theta\theta\bar{\theta}\bar{\chi}''-\bar{\theta}\bar{\theta}\theta\chi''+\frac{1}{2}\theta \theta \bar{\theta} \bar{\theta} D'' \,.
\ea\ee  
which are uncharged under all factors of $U(1)$ and will be referred to as uncharged or bulk fields. 

Then there are $b^1(\tM3)$  $\tn{Ad}\, \gut$ and  $\tn{Ad}\,U(1)^n $  chiral multiplets respectively denoted as
\be\ba
\Phi_{(i)}'&=\varphi_{(i)} +\sqrt{2} \theta \psi_{(i)} + \theta \theta H_{(i)} + \cdots\,, \\
\tilde\Phi_{(i)}&=\tilde\varphi_{(i)} +\sqrt{2} \theta \tilde\psi_{(i)} + \theta \theta \tilde H_{(i)} + \cdots\,.
\ea\ee

We further have chiral multiplets valued in every representation ${\bf R}_{Q_r}$ which we will referred to as charged or localized fields. These we denote as
\be\ba\label{eq:PerturbativeZeromodes}
\Phi_{(a,r)}=\varphi_{(a,r)} +\sqrt{2} \theta \psi_{(a,r)} + \theta \theta H_{(a,r)} + \cdots\,, 
\ea\ee
where the index denotes the representation via the corresponding charge vector $Q_r$ the multiplet transforms in and $a$ denotes the critical point of Morse index 1 of the function $f_{Q_r}$ to which the multiplet is associated. The index $a$ omits critical points of Morse index 2. We think of all manipulations in the gauge algebra as embedded in $\tn{Ad}\,\gau$.

The 4d Lagrangian organizes itself into three parts. The first part are the standard kinetic terms in 4d $\CN=1$ superspace, the second are the superpotential terms to which bulk fields contribute and the third are the superpotential terms to which only the localized fields contribute. The latter two were focus of section \ref{sec:MatterInteractions}.

Expanding the action \eqref{OffShellBulk} in zero modes the first part of the 4d Lagrangian is found to be
\be\label{eq:4dLagrangian1}
\CL_{\tn{kin}}=\CL_{\tn{gauge}}^{\tn{bulk}}+\CL_{\tn{ch}}^{\tn{bulk}}+\CL_{\tn{ch}}^{\tn{loc}} \,,
\ee
with the individual contributions to the Lagrangian being given by
\be\ba
\CL_{\tn{gauge}}^{\tn{bulk}}&=\frac{1}{g_4^2}\lb \CL_{\tn{gauge}}+\sum_m\CL^{(m)}_{\tn{gauge}}\rb\,,
\\ \CL_{\tn{gauge}}&=\frac{1}{4}WW|_{\theta\theta}+\frac{1}{4}\bar{W}\bar{W}|_{\bar{\theta}\bar{\theta}} \,,\\
\CL^{(m)}_{\tn{gauge}}&=\frac{1}{4} W^{(m)}W^{(m)}|_{\theta\theta}+\frac{1}{4}\bar{W}^{(m)}\bar{W}^{(m)}|_{\bar{\theta}\bar{\theta}} \,,
\ea\ee
and
\be\ba
\CL_{\tn{ch}}^{\tn{bulk}}&= \sum_{i}\lb \CL_{\tn{kin}}^{(i)}+\sum_{m}\CL_{\tn{kin}}^{(i,m)}\rb\,,\\
\CL_{\tn{kin}}^{(i)}&= \lb \Phi_{(i)}' \rb^\dagger e^{2V'}\Phi_{(i)}'\Big|_{\theta\theta\bar\theta\bar\theta}\,,\\
\CL_{\tn{kin}}^{(i)}&=\lb \tilde\Phi_{(i)} \rb^\dagger  \tilde \Phi_{(i)}\Big|_{\theta\theta\bar\theta\bar\theta}\,,\\
\ea\ee
and
\be\ba\label{eq:KineticTermsLoc}
\CL_{\tn{ch}}^{\tn{loc}}&= \sum_{a,r,\mu(p_a)=1  }\CL_{\tn{kin},(a,r)}+\sum_{m,a,r,\mu(p_a)=1}\CL_{\tn{kin},(a,r)}^{(m)}\,,\\
\CL_{\tn{kin},(a,p)}&= \lb \Phi_{(a,p)} \rb^\dagger e^{2V'}\Phi_{(a,p)}\Big|_{\theta\theta\bar\theta\bar\theta}\,,\\
\CL_{\tn{kin},(a,p)}^{(m)}&=\lb \Phi_{(a,p)} \rb^\dagger e^{2QV''}\Phi_{(a,p)}\Big|_{\theta\theta\bar\theta\bar\theta}\,.
\ea\ee
The limits of the sum \eqref{eq:KineticTermsLoc} avoid an over counting as points of Morse index $2$ of $f_Q$ contribute a chiral multiplet of charge $-Q$ which are already included as the sum over $p$ runs over all charge vectors $Q_p$\,. The notation $QV''$ is short for the charge weighted sum $q_iV^{(i)}$ of the individual abelian gauge fields contained in $V''$.

The second part of the 4d action can be inferred from the terms
\be
H^k\bar{H}_k+\frac{i}{\sqrt{2}}(F_{\bar{\varphi}})_{ij}\epsilon^{ijk}\bar{H}_k-\frac{i}{\sqrt{2}}(F_{\varphi})_{ij}\epsilon^{ijk}H_k
\ee
and its Lagrangian is found to take the form
\be\ba\label{eq:ZukawaInteraction}
\CL_{\tn{Zuk}}&= \frac{1}{3}\sum_{i,p}\sum_{ab}g^{(i)}_{(ab,p)}\tilde\Phi_{(i)}\Phi_{(a,p)}\Phi_{(b,-p)}\Big|_{\theta\theta}+\tn{h.c.}\\
&~~~\:~\,+\sum_{i,p}\sum_{ab}g_{(ab,p)}^{(i)}\Phi_{(i)}' \Phi_{(a,p)}\Phi_{(b,-p)}\Big|_{\theta\theta}+\tn{h.c.}\,.
\ea\ee
The sums over $a,b$ are restricted to points of Morse index 1 and $a\neq b$ . Here the Yukawa interaction $g^{(i)}_{(ab,p)}$ is determined by overlap integrals of zero modes. The relevant integral is
\be
Z^{(a,p),(b,-p),(i)}=\int_{\tM3} \varphi^{(a,p)}\wedge \varphi^{(b,-p)}\wedge h^{(i)}\,,
\ee
where $\varphi^{(a,p)}$ are again perturbative zero modes localized at $p_a\in\tM3$ transforming in ${\bf R}_{Q_p}$ and $h^{(i)}$ is the $i$-th harmonic form on $\tM3$ with $i=1,\dots,b^1(\tM3)$. The modes $\varphi^{(a,p)}$ are not to be confused with $\varphi_{(a,p)}$ in \eqref{eq:PerturbativeZeromodes}. They combine to the first term in the KK reduction from 7d to 4d as
\be
\varphi=\varphi_{(a,p)}\varphi^{(a,p)}+\cdots \,.
\ee
The coupling has the value
\be
g^{(i)}_{(ab,p)}=-\frac{i}{\sqrt{2}}Z^{(a,p),(b,-p),(i)}\,.
\ee

The last piece of the effective 4d Lagrangian was determined in section \ref{sec:MatterInteractions} and so we are brief in its discussion. The mass terms and Yukawa couplings between localized modes can be inferred from interaction the term 
\be\label{eq:IntTerms}
\psi\wedge \CD \psi=\psi\wedge \CD^{(0)} \psi+\psi\wedge [\delta\varphi \wedge\,, \psi]\,,
\ee
and its conjugate which we have here split into its background part and dynamical part respectively. More precisely $\CD^{(0)}=d+df_Q\wedge$ and $\psi,\delta\varphi$ display a vanishing background while being dynamical. The background part gives rise to the mass terms whereas the 3-point part returns Yukawa couplings upon reduction to 4d. The corresponding Lagrangians read

\be\ba\label{eq:MassYukawaSupPot}
\CL_{\tn{Mass}}&=\frac{1}{2}\sum_{p}\sum_{ab}m_{(a,p),(b,-p)}\Phi_{(a,p)}\Phi_{(b,-p)}\Big|_{\theta\theta}+\tn{h.c.}\,,\\
\CL_{\tn{Yuk}}&=\frac{1}{3}\sum_{pqr}\sum_{abc}g_{(a,p),(b,q),(c,r)}\Phi_{(a,p)}\Phi_{(b,q)}\Phi_{(c,r)}\Big|_{\theta\theta}+\tn{h.c.}\,,
\ea\ee
where the sums over $a,b,c$ are restricted to points of Morse index 1 and always distinct. The couplings are 
\be\ba
m_{(a,p),(b,q)}&=\frac{i}{\sqrt{2}}M^{ab}_{Q_p}\,, \\
g_{(a,p),(b,q),(c,r)}&=\frac{i}{\sqrt{2}}Y^{abc}_{pqr}\,,
\ea\ee
with the mass matrix as in \eqref{eq:Massmatrix} and the Yukawa coupling as in \eqref{eq:Yukawa}. The complete 4d action descending from the SYM description is thus
\be\label{eq:effLag}
S_{4d}=\int d^{\,4}x \tn{Tr}\lbb \CL_{\tn{kin}} +\CL_{\tn{Zuk}}+\CL_{\tn{Mass}}+\CL_{\tn{Yuk}}\rbb \,.
\ee
We conclude with a comment on scales. The internal zero modes along $\tM3$ are dimensionless as can be read off from the 7d action derived in section \ref{app:OffShell}. The 7d gauge coupling $g_7$ thus contributes the mass scale and we find $M^{ab}_{Q_p}$, which denotes the mass terms for the conjugate representations ${\bf R}_{Q_p},\overline{\bf R}_{-Q_{p}}$ in analogy to \eqref{eq:MEMB}, to be associated to $\Mgut$ (exponentially suppressed by the volume of three-cycles responsible for the mass term). Of course $Y^{abc}_{pqr}$ is dimensionless. The masses $M^{ab}_{Q_p}$ thus contribute a new mass scale $M_{\tn{Inst}}$ which is linked to $\Mgut$ via the average volume $\tn{Vol} \gamma$ of the $3$-cycles in the ALE geometry as
\be
M^{ab}_{Q_p}\sim M_{\tn{Inst}}\sim \Mgut e^{-\tn{Vol} \gamma}\,.
\ee
To obtain the true effective theory we need to integrate out all massive modes. As $M_{\rm Inst}$ is in general much lighter than the lightest state of the KK tower we can indeed constrain considerations to modes whose masses are induced by M2-branes as these will yield the largest corrections. Thus integrating these states out will alter the Lagrangian \eqref{eq:effLag} to 
\be
S_{4d}=\int d^{\,4}x \tn{Tr}\lbb \CL_{\tn{kin}} +\CL_{\tn{Zuk}}+\CL_{\tn{Yuk}}+ \dots +\frac{1}{M_{\tn{Inst}}^k}\CL^{k}_{\tn{Int}}+\dots\rbb \,.
\ee
where we have made the mass scale $M_{\tn{Inst}}$ suppressing the higher point interactions explicit.

\section{Boundary Conditions}
\label{sec:BoundaryAndHodge}

In this appendix we provide some details underpinning the computation of the $\CD$-cohomology groups in section \ref{sec:MilkyWay}. The argument was originally introduced by Witten \cite{Witten:1982im} for closed manifolds. The extension to the manifolds with boundary includes the additional subtlety of choosing the appropriate boundary conditions, which allow us to apply the Hodge theory arguments on manifolds with boundary. 
	
	Following \cite{chang1995} we consider the case where we have a single boundary component. We first focus on defining the appropriate domains of the standard operators $d$ and $d^\dagger$ and then show that the cohomology groups are invariant under the deformation by $f$. We define $d_1$ and $d_1^\dagger$ as operators with the domains given by
	\be\ba
	D(d_1) &= \Omega^p(\tM3),\\
	D(d^\dagger_1) &= \left\{\alpha\in\Omega^p(\tM3)\,\left|\,\ast\alpha_n|_{\Sboundary}= 0\right. \right\}.
	\ea\ee
	This is the Neumann boundary condition. To obtain a self-adjoint Laplacian for these boundary conditions we have to further restrict the domain of the Laplace operator. 
	We define $\Delta_1$ to be the standard metric Laplacian, with the domain
	\be\ba
	D(\Delta_1) = \{\alpha\in\Omega^p(M)\,\left|\, \ast\alpha_n|_{\Sboundary}= \ast(d\alpha)_n|_{\Sboundary}=0 \right.\}.
	\ea\ee
	Note that $d_1$ is simply the standard de Rham differential, so the cohomology of the resulting complex is the (absolute) de Rham cohomology $H^\ast(\tM3)$. For the Dirichlet boundary conditions we define the operators $d_2$ and $d^\dagger_2$, with domains
	\be\ba
	D(d_2) &= \left\{\alpha\in\Omega^p(\tM3)\,\left|\,\alpha_t|_{\Sboundary}= 0\right. \right\},\\
	D(d^\dagger_2) &= \Omega^p(\tM3), 
	\ea\ee
	and $\Delta_2$, with the domain
	\be\ba
	D(\Delta_2) = \left\{\alpha\in\Omega^p(\tM3)\,\left|\, \alpha_t|_{\Sboundary}= (d^\dagger\alpha)_t|_{\Sboundary}=0 \right.\right\}.
	\ea\ee
	We use the indices to formally distinguish between the two operators, based on their different domains. With these domains, the two Laplace operators are self-adjoint and we have the Hodge decompositions \cite{chang1995,SchwarzHodge}
	\be
	\label{eq:HodgeDecomposition}
	\ba
	\Omega^p(\tM3) &= R(d^{p-1}_i)\oplus R((d^\dagger_i)^{p+1})\oplus N(\Delta^p_i),\\
	N(d^{p}_i) &= R(d^{p-1}_i)\oplus N(\Delta^p_i),\\
	N((d^\dagger_i)^{p-1}) &= R((d^\dagger_i)^{p})\oplus N(\Delta^p_i).
	\ea
	\ee
	Here, $N$ and $R$ denote the nullspace and range respectively. Superscripts denote the degree of forms on which the operators act. Observe that this implies 
	\be
	N(\Delta^p_i) = N(d^{p}_i)/R(d^{p-1}_i).
	\ee
	The quotient on the right side is precisely the $p$-th cohomology group of the operator $d_i$. As already noted above, the cohomology of operator $d_1$ is the de Rham cohomology $H^p(\tM3)$ by definition. To identify the cohomology groups of the operator $d_2$, observe that $\alpha_t|_\Graph = 0$ is the same as saying $\alpha|_{\Sboundary}= 0$. This implies that the cohomology of $d_2$ is isomorphic to the relative cohomology  $H^p(\tM3,\Sboundary)$ \cite{ElementsTopAlg}. 
	
	Let us now consider how the deformation of the complex by a smooth function $f$ relates to the above. Recall that $\CD$ is in fact the deformed de Rham differential $\CD = e^{-qf}d e^{qf}$. Denote $\CD_i = e^{-qf}d_i e^{qf}$ and $\CD_i^\dagger = e^{qf}d_i^\dagger e^{-qf}$ i.e. we impose the same boundary conditions on $\CD_i$ (resp $\CD_i^\dagger$) as we did on $d_i$ (resp. $d_i^\dagger$). Let $\Delta_{f,i} = \CD_i\CD_i^\dagger + \CD_i^\dagger\CD_i$ denote the twisted Laplacian from \eqref{eq:WittenLap}. We set $D(\Delta_{f,i})= D(\Delta_i)$. We have the same Hodge decomposition for the operators $\CD_i, \CD_i^\dagger, \Delta_{f,i}$ as in \eqref{eq:HodgeDecomposition} as well as the identification
	\be
	H_{\CD_i}^p (\tM3) = N(\Delta_{f,i}^p).
	\ee
	Finally, the map $\alpha \mapsto e^{-q f}\alpha$ induces an isomorphism between the $d_i$-complex and $\CD_i$-complex (multiplication by smooth functions preserves the boundary conditions). This in turn induces isomorphisms of the cohomology groups. Combining the arguments above, we see that
	\be
	N(\Delta_{f,i}^p)= N(\Delta_i^p),
	\ee
	so the number of ground states of $\Delta_{f,i}$ is independent of $f$. Moreover, 
	\be
	\ba
	H^p_{\CD_1}(\tM3) &= H^p(\tM3)\\
	H^p_{\CD_2}(\tM3) &= H^p(\tM3,\Sboundary).
	\ea
	\ee 
	This provides the simplest example where the cohomology of $\CD$ is computed directly in terms of the cohomology of the underlying manifold with boundary. However, the case with single boundary component is not interesting in our context, as it violates the charge conservation condition we obtain from the electrostatics problem. However, the above discussion clearly outlines the general structure, which persists in the case of mixed boundary conditions, which we consider in \ref{sec:MilkyWay}.

%%%%%%%%%%%%%%%%%%%%%%%%

%%%%%%%%%%%%%%%%%%%%%%%%%%%%%%%%%%%%%%%%%%%
%%%         APPENDICES END HERE
%%%%%%%%%%%%%%%%%%%%%%%%%%%%%%%%%%%%%%%%%%%

%%
%\bibliographystyle{JHEP}
%\bibliography{G2}
\providecommand{\href}[2]{#2}\begingroup\raggedright\endgroup

%%%%%%%%%%%%%%%%%%%%%%%%%%%%%%%%%%%%%%%%%
%%%%%%%%%%%%%%%%%%%%%%%%%%%%%%%%%%%%%%%%%

%%%%%%%%%%%%%%%%%%%%%%%%%%%%%%%%%%%%
%%%%%%%%%%%%%%%%%%%%%%%%%%%%%%%%%%%%
\end{document}